\documentclass[final,5p,times,twocolumn,authoryear]{elsarticle}

\usepackage{amssymb}
\usepackage{lipsum}
\usepackage{listings}
\usepackage{CJKutf8}
\usepackage{empheq}
\usepackage{comment}
\usepackage{xcolor}
\usepackage[utf8]{inputenc}
\usepackage{graphicx}
\usepackage{amsmath}
\usepackage{mathtools}
\usepackage{url}
\usepackage{hyperref}
\usepackage{subcaption}
\usepackage{cuted}
\journal{Icarus}
\begin{document}
\begin{frontmatter}

\title{Cosmic Hydrogen and Ice Loss Lines\\ \small \begin{CJK*}{UTF8}{bsmi}
人能常清靜，天地悉皆歸
\end{CJK*}}

\author[1]{Li Zeng\corref{cor1}}
\ead{astrozeng@gmail.com}

\author[1]{Stein B. Jacobsen}
\ead{jacobsen@neodymium.harvard.edu}

\cortext[cor1]{Corresponding author}
\affiliation[1]{organization={Department of Earth and Planetary Sciences, Harvard University},
            addressline={20 Oxford Street}, 
            city={Cambridge},
            postcode={02138}, 
            state={MA},
            country={USA}}




\begin{abstract}
We explain the overall equilibrium-temperature-dependent trend in the exoplanet mass-radius diagram, using the escape mechanisms of hydrogen and relevant volatiles, and the chemical equilibrium calculation of molecular hydrogen (H$_2$) break-up into atomic hydrogen (H). We identify two \emph{Cosmic Hydrogen and Ice Loss Lines} (CHILLs) in the mass-radius diagram. Gas disks are well known to disperse in ten million years. However, gas-rich planets may lose some or almost all gas on a much longer timescale. We thus hypothesize that most planets that are born out of a hydrogen-gas-dominated nebular disk begin by possessing a primordial H$_2$-envelope. This envelope is gradually lost due to escape processes caused by host-stellar radiation.
\end{abstract}


\end{frontmatter}
\section{Introduction} \label{sec:intro}


Planetary systems are thought to form in a residual disk after the formation of a star. This residual disk is thought to be of the same composition as the star: primarily hydrogen and helium, but a small component (1\%) can form ices (H$_2$O, NH$_3$ and CH$_4$) and an even smaller component (0.5\%) can form silicate rock and metals. One possible outcome would be that planets of roughly stellar composition (gas giants) formed at various distances in the residual protoplanetary disk. However, we know from exoplanet observations that this is not the case, in fact gas giants makes up a relatively small portion ($\lesssim$10\%) of the observed exoplanet inventory. With recent observations of exoplanets around small M-dwarf host stars~\citep{Luque2022, Cherubim2022, Piaulet2022, Diamond-Lowe2022, Madhusudhan2023, Roy2023}, as well as atmospheric characterization of puffy exoplanets by JWST~\citep{OCallaghan2022,Alderson2023,Rustamkulov2023,Feinstein2023,Ahrer2023}, the water world hypothesis becomes viable. Likely, the most common exoplanet ($\gtrsim$60\%) appears be what we call water worlds, consisting dominantly of H$_2$O, NH$_3$ and CH$_4$ ice components~\citep{PNAS:Zeng2019}, which is also consistent with the statistics on the planet radius versus flux (Rp-f) diagram{\color{black}, since there are planets of $\sim$3R$_{\oplus}$ existing at high flux where simple H$_2$-He envelope should no longer hold but water can.} Earth-like rock-metal planets are the second most common ($\sim$30\%) of observed exoplanets. This raises the question of how almost all of the H$_2$-He component was lost from most planets and how in addition the ice components were lost from Earth-like rock-metal planets. We thus infer that there must be processes resulting in the \emph{Cosmic Hydrogen and Ice Loss Lines} that we call CHILLs. In this paper we make use of the exoplanet systematics in the mass-radius diagram to infer, discuss, and model these processes. 


We thus hypothesize that the formation of most planets greater than about one Earth mass (1 M$_{\oplus}$) would start by acquiring a thick primordial hydrogen-helium envelope while being embedded in the residual gaseous nebular disk. Then, this primordial hydrogen-helium envelope is eroded away over time by various mechanisms primarily caused by the host star, at various escape rates depending on the host-stellar irradiation level and other factors. Therefore, when we look at the distribution of exoplanets in the mass-radius diagram, we see two Cosmic Hydrogen and Ice Loss Lines (CHILLs) of hydrogen and related volatile escape. 

The CHILL\#1 may correspond to the complete removal of the primordial hydrogen-helium envelopes from smaller planets including super-Earths and mini-Neptunes. This removal helps to reveal the nature of their core materials which are made of elements heavier than H/He. This can explain the observed bi-modal radius distribution of small exoplanets separated by the \emph{small planet radius gap} as core materials dominated by the elements of oxygen, carbon, nitrogen, and their chemical compounds, versus core materials dominated by magnesium, silicon, iron, and their chemical compounds~\citep{Zeng2021}. 

On the other hand, the CHILL\#2 may correspond to the complete loss of the hydrogen-helium envelope from the more massive gas giant exoplanets, which are dominated by hydrogen and helium in their composition. Their masses typically range from a few tens to a few hundred Earth masses (M$_{\oplus}$). The current study of their atmospheric composition may add in more constraints and inputs in their models. 

{\color{black}
\section{The Hydrogen EOS} \label{sec:H2EOS}

The H$_2$ EOS is needed to understand the hydrogen-dominated envelopes and atmospheres on planets.

\begin{figure}[!ht]
    \centering
    \includegraphics[width=0.45\textwidth, angle=0]{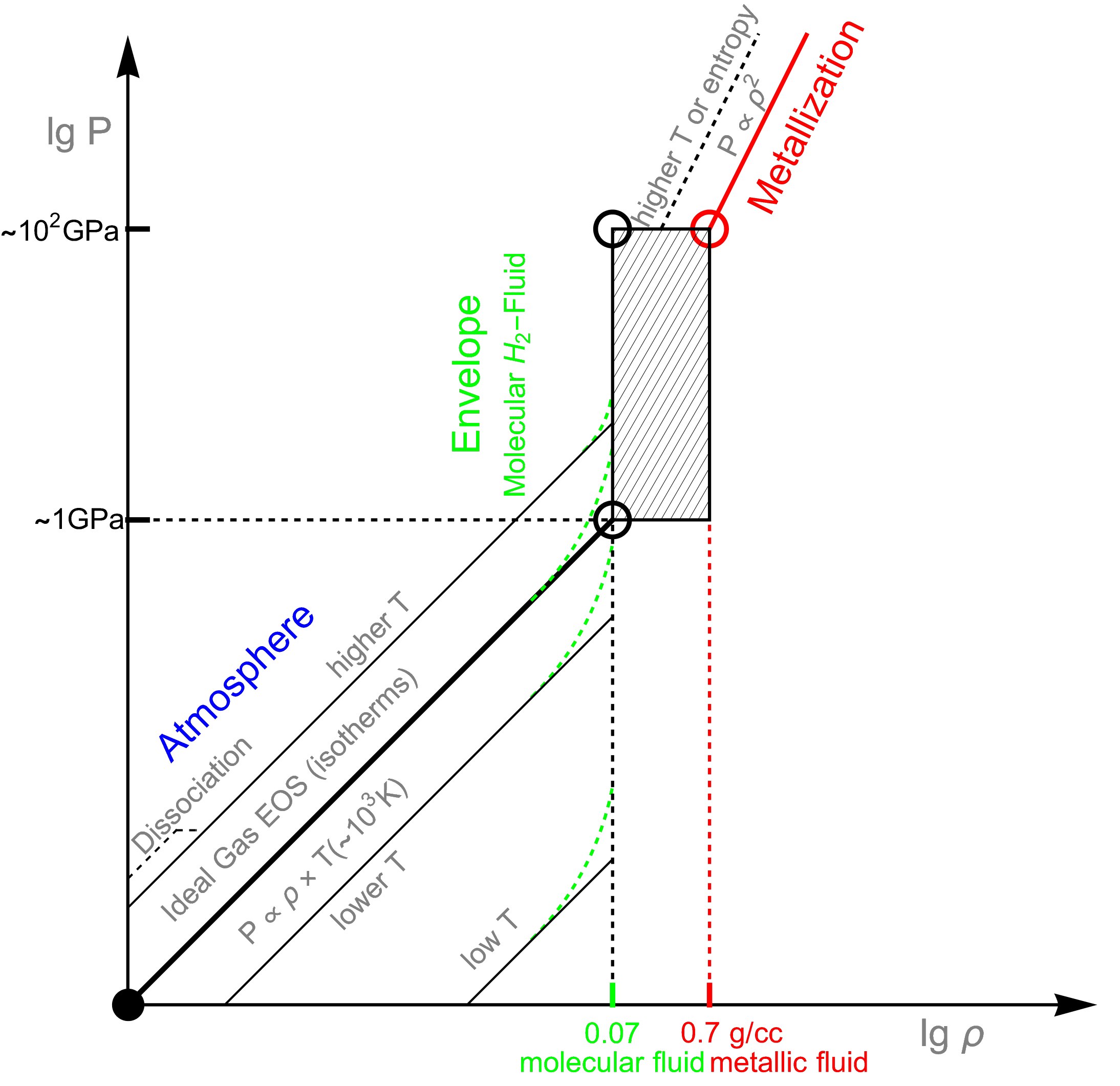}
    \caption{Schematic Representation of Hydrogen EOS in three regimes: (1) Gaseous, (2) Dense (super-critical) Molecular Fluid, and (3) Metallic Fluid. The hydrogen EOS is commonly encountered in exoplanet interior modelling, because hydrogen is the most abundant element available in the universe. The (1) gaseous EOS can be first approximated by an ideal gas law, and then with a modification term taking into account inter-molecular collisions and molecular dimensions. The (3) metallic EOS can be first approximately by a power law with power-index of two which takes into account the energy of de-localized electrons. The (2) molecular fluid EOS bears most of the complexity (grey-shaded region) because molecules strongly interact with neighbors that their average distance is shorter than the sum of their van der Waal radii and the pressure effect and temperature effect are of equal importance.}
    \label{fig:H2eos}
\end{figure}

The EOS is best-described by considering three different regimes from the structure of matter perspective (see Figure~\ref{fig:H2eos}): (1) gaseous, (2) high-density molecular (super-critical) fluid, (3) high-density metallic fluid.

(1) Gaseous EOS. This is applicable when the distance of neighboring molecules is large compared to the linear dimension of molecule ($\sim10^{-8}$cm). The ideal gas law is the simplest EOS to describe such a state ($\rho \ll 0.07$g/cc), while modification of which gives better approximation to the EOS at relatively high density taking into account the interactions of neighboring molecules. One convinient form of such a real gas EOS is given by~\citep{Tsien1965}, by defining a characteristic temperature $\Theta_1 \equiv \epsilon/k_B$, where $\epsilon$ is the characteristic energy of inter-molecular potential, $v \equiv \mu/\rho$ is the average total volume assigned to each molecule, and $v^* \equiv D^3$ is the volume occupied by each molecule's physical dimension (effectively the reach of their outermost occupied electronic orbitals), that is, the cubic power of the equivalent molecular diameter $D$,

\begin{equation}
    \underbrace{\frac{P\cdot v}{k_B T} = 1}_{\text{ideal gas EOS}} + \underbrace{\left( \frac{1}{0.278 \cdot \left(T/\Theta_1 \right)^{1/6} \cdot \left(v/v^* \right) - 0.177} \right)}_{\text{non-ideal correction: up to $\sim$few to ten, but reduces to 0 at high T}}
\end{equation}

Notice that the first part of such an EOS is simply the ideal gas EOS, while the second part gives the non-ideal correction primarily due to occupied volume of molecule. This correction becomes smaller at higher temperatures because the effective size of a molecule in collision becomes smaller due to higher kinetic energy of the molecules involved in a collision. The effective collisional cross-section $\sigma$ for molecules decreases with increasing temperature or increasing collisional kinetic energy, because molecules are not rigid spheres but are \emph{squeezy}. In other words, the effective size of each molecule which makes its way into the correction term decreases.

(2) High-density molecular (super-critical) fluid hydrogen EOS. This EOS is generally complicated because molecules strongly interact with each other with partially-overlapping outer electronic orbitals with neighbors in this physical regime. The general feature is that this EOS quickly stiffens when a critical density is reached. This critcal density for hydrogen fluid is about 1/16-th that of water at ambient conditions. Then, this fluid EOS follows the general compression pattern of other materials, with similar slopes and curve shapes on the mass-radius diagram. The EOS eventually softens when the matter metallizes by further compression for a factor of about ten-fold in volume (from $\sim0.07$g/cc compresses to $\sim0.7$g/cc), or equivalently, a reduction of a factor of two in each linear dimension. At this point, the electronic orbitals of neighboring molecules overlap so much that electrons become de-localized.

(3) High-density metallic fluid hydrogen EOS. Ultimately the compression leads to the metallization where the EOS again softens and becomes simple due to the presence of de-localized electrons. For a significant range of density and pressure this metallic fluid EOS can be approximated by a power-law with power-index of two~\citep{Stevenson2017Ge131:PlanetaryEvolution},

\begin{equation}
    \underbrace{P \approx K \cdot \rho^2 \approx K' \cdot \frac{1}{v^2}}_{\text{characteristic of any metal under extreme compression}}
\end{equation}

where $K$ and $K'$ are proportionality constants. $K \approx 200\text{~GPa} \approx 2\cdot 10^{11}$Pa, where $\rho$ is measured in g/cc. This value of proportionality constant $K$ can change if heavier elements or components with higher mean molecular weights are mixed into this envelope.

Assume \emph{volume additivity}, and define the mass fraction of total heavier elements other than hydrogen-helium in the mixture as $Z$~\citep{Stevenson2017Ge131:PlanetaryEvolution}, we have,

\begin{equation}
    \frac{1}{\rho(P)} = \left( \frac{1-Z}{\sqrt{P/K}} + \frac{Z}{\rho_{\text{heavy}}} \right) \approx \frac{1-Z}{\sqrt{P/K}}
\end{equation}

This simply suggests that we can define an effective $K_{\text{eff}}$ so that,

\begin{equation}
    P = K_{\text{eff}} \cdot \rho^2, \text{~where $K_{\text{eff}} = (1-Z)^2 \cdot K$}
\end{equation}

The radius of a planet in this regime does not vary much with the change of mass and can be approximated as:

\begin{equation}
    R_p \approx \sqrt{\frac{\pi K_{\text{eff}}}{2 G}} \approx \sqrt{\frac{\pi K}{2 G}} \cdot (1-Z)
\end{equation}

where $G$ is the gravitational constant. 

Thus, the planet radius or envelope thickness is reduced by the factor $(1-Z)$. \citep{Stevenson2017Ge131:PlanetaryEvolution} also shows that $R_p$ remains essentially \emph{unchanged} if this much $Z$ heavier elements all concentrate into a core. So $R_p$ is only related to the total amount ($Z$) but \emph{not sensitive} to the exact distribution of heavier elements within the planet interior profile. We will then use this relation in a later discussion of the negative slope of CHILL in the mass-radius plot.
}


\section{Escape Mechanism} \label{sec:mechanism}

\subsection{Geological Evidence}

Nebular gas disks typically remove on the 5-10 million years timescale~\citep{Armitage2010AstrophysicsFormation, Hartmann2008}. On the other hand, the noble gas abundances sourced from the Earth's deep mantle suggest an early existence of a massive H$_2$-He proto-atmosphere/envelope of the order of $\gtrsim10^{3}$bar (0.1 GPa, or equivalently, $10^{3}$ times the mass of the current Earth's N$_2$-O$_2$ atmosphere) in equilibrium with the Earth's early molten magma ocean surface~\citep{Harper1996NobleAccretion}. This massive H$_2$-He proto-atmosphere/envelope presumably has lasted long enough to allow the in-gassing of its helium ($^3$He) and neon ($^{20}$Ne, $^{22}$Ne) content into the Earth's early molten mantle (magma ocean), which later freezes (solidifies) to become the present-day Earth's mantle. Later on, this massive proto-atmosphere is removed. Subsequently, a meteoritic (planetary) atmospheric component is accreted, and is reflected by the abundances of heavier noble gases including argon ($^{36}$Ar) and xenon ($^{130}$Xe). {\color{black}Therefore, the geological evidence points to the existence of two chemical reservoirs available during planet formation: the "\emph{solar}" and the "\emph{planetary}" (="\emph{meteoritic}"), corresponding to the free gas versus the gas trapped in solid phases in the nebula, respectively. The Earth's mantle plume source $^{20}$Ne-$^{22}$Ne ratio and $^{3}$He-$^{20}$Ne ratio preserve the "\emph{solar}"-like relative abundance ratio, while the Earth's mantle plume source $^{36}$Ar/$^{130}$Xe ratio is close to the "\emph{planetary}" ($=$"\emph{meteoritic}") abundance ratio~\citep{Jacobsen2005,Williams2019}. 

In order for significant in-gassing to occur over a limited time duration of planet formation, it requires enough pressure built up at the interface between the molten magma ocean and the overlaying gaseous layer. To the first order approximation, the in-gassing rate is proportional to the number density of gaseous species present, which is in turn proportional to the pressure directly overlying the molten surface. Small bodies do not have enough gravity to build up that pressure required for in-gassing for the limited lifetime of the gas disk.} 

Thus, the overall noble gas abundances of the light versus the heavy noble gas species paint a general two-stage picture of Earth formation: First, a co-accretion of a massive H$_2$-He proto-atmosphere/envelope with the proto-Earth, and Second, the subsequent removal and replacement of such proto-atmosphere/envelope. Likewise, other planets more massive than one Earth mass (1 M$_{\oplus}$) would likely acquire a H$_2$-He proto-atmosphere/envelope during their formation while being embedded in a H$_2$-He gas-rich disk~\citep{Hayashi1979, Mizuno1980FormationPlanets, Sekiya1980, Hartmann2008}. Now, the question is the timescale that this primordial atmosphere/envelope can last on exoplanets under various host-stellar irradiation conditions, resulting from different physical and initial conditions compared to our own solar system planets.

Hydrogen (H and H$_2$) is thereby inferred as the dominant species in the upper atmosphere of many of the close-in exoplanets, due to (1) the availability/abundance of hydrogen and (2) its lightest weight and thus its largest scale height compared to the other atmospheric species. In any atmosphere above a certain critical level (the von Karman line, see~\ref{layering}), the gas density becomes low enough, and the mean free path of molecular collision becomes large enough, so that each gas species separates out according to its own scale height determined by its own molecular weight at a given temperature. For example, this is the case for Earth's atmosphere, which orbits the Sun at 1AU. This physics should also be true for exoplanets which are closer in and simultaneously more H$_2$-rich or H$_2$O-rich. Therefore, we expect the escape and dissociation of hydrogen and hydrogen-bearing species as one of the main physical mechanisms (see~\ref{appendix:H2Dissociation} and~\ref{appendix:H2PhotoDissociation}) \emph{sculpting} the overall distribution of exoplanets on the mass-radius diagram. The escape processes can be powered by the intense host-stellar electromagnetic irradiation at all wavelengths~\citep{Owen2018AtmosphericExoplanets}, or by the planetary cores themselves~\citep{Ginzburg2016Super-EarthRetention, Ginzburg2018}. Here, we propose a new mechanism of shock-wave and phase-transition escape scenario as powered by the absorption of incident stellar bolometric photon flux (mostly in the visible wavelengths).

\subsection{Collisional/Thermal Dissociation}
The dissociation of hydrogen molecules and other hydrogen-bearing species into \emph{free} hydrogen atoms is favored at increasing temperature or decreasing pressure within the assembly of gas species. The relative proportion between the neutral atomic hydrogen (H-atom) and the molecular hydrogen (H$_2$) will change at different heights in the atmosphere. The breakup of H-H covalent bond involves a reversible chemical reaction under the consideration of collisional equilibrium (see~\ref{appendix:H2Dissociation} for details):

\begin{equation}
\text{H}_2 \rightleftharpoons \text{H} + \text{H}
\end{equation}

Likewise, the dissociation of any other hydrogen-bearing species (X-H) such as water (H-O-H) involves the following:

\begin{equation}
\text{X-H} \rightleftharpoons \text{X} + \text{H}
\end{equation}

{\color{black}On one hand, a higher total pressure would favor a shift towards more H$_2$-molecules or more associated hydrogen-bearing molecules (hydrides) X-H within the assembly.}

{\color{black}On the other hand, a higher temperature would favor a shift towards more dissociated molecules, fractals, and single hydrogen atoms within the assembly. The \emph{effective} collisional cross-sections ($\sigma$'s) of all the different species participating in the dissociation reaction network will definitely affect the \emph{rate} of the dissociation reaction in either the forward or the backward direction.

Just as in a \emph{tug-of-war} game in which two teams pull at opposite ends of a rope, if the rope breaks all of a sudden, then the two teams would usually not stay still but would fall apart with extra kinetic energy. 

Likewise, if a bond breaks or fractures, then its constituents would typically not stand still but would fly apart with extra kinetic energy corresponding to the same order of magnitude as the bond strength (Fig.~\ref{fig:h2break}). 

\begin{figure}[!ht]
    \centering
    \includegraphics[width=0.4\textwidth, angle=0]{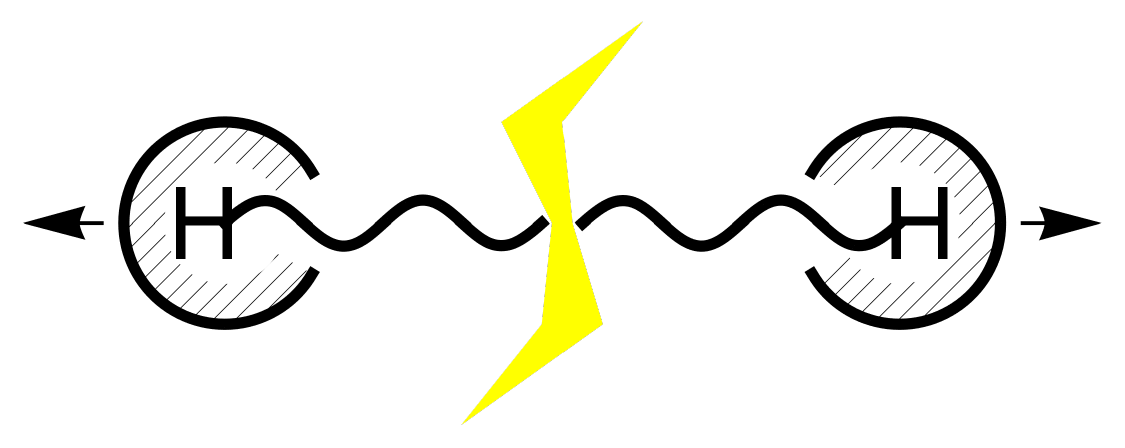}
    \caption{Schematic diagram of a hydrogen molecule dissociation, showing bond breakage and constituents flying apart.}
    \label{fig:h2break}
\end{figure}

If it is the same physical mechanism which breaks the covalent bond that also launches the H-atom into space, then, one can estimate the fly-apart velocity of escaping atom with the following \emph{back-of-envelope calculation} using energy conservation:

\begin{equation}
    m_{\text{H-atom}} \sim m_{\text{proton}} \sim 1.67 \times 10^{-27} \text{~g}
\end{equation}

\begin{equation}
    2 \cdot \left( \frac{1}{2} \cdot m_{\text{H-atom}} \cdot v^2 \right) \sim 4.5 \text{~eV}
\end{equation}

\begin{equation}
    v \sim 20 \text{~km/s}
\end{equation}

This value ($\sim$20 km/s) falls right within the ballpark of the observed range of ($\sim$10-30 km/s) of the high-speed neutral H-atom outflow observed on several exoplanets~\citep{Zhang2022,Zhang2023}. This is a first piece of supportive evidence towards our hypothesis of escape due to \emph{molecular dissociation}. This is a \emph{molecular catapult}.

The discussion here also applies to hydrogen atom sourcing from the dissociation of other hydrides X-H  such as H$_2$O molecule. However, in those scenarios, due to the heavy mass of X compared to H and \emph{conservation of momentum}, the single H-atom after dissociation would carry away most of the kinetic energy.
}


\subsection{Phase Transition: Thermal (entropy-limited) versus Shock-driven (energy-limited) Escape}
\label{phasetranstion}

\begin{figure*}
\centering
\begin{subfigure}{1\columnwidth}
    \includegraphics[width=\textwidth]{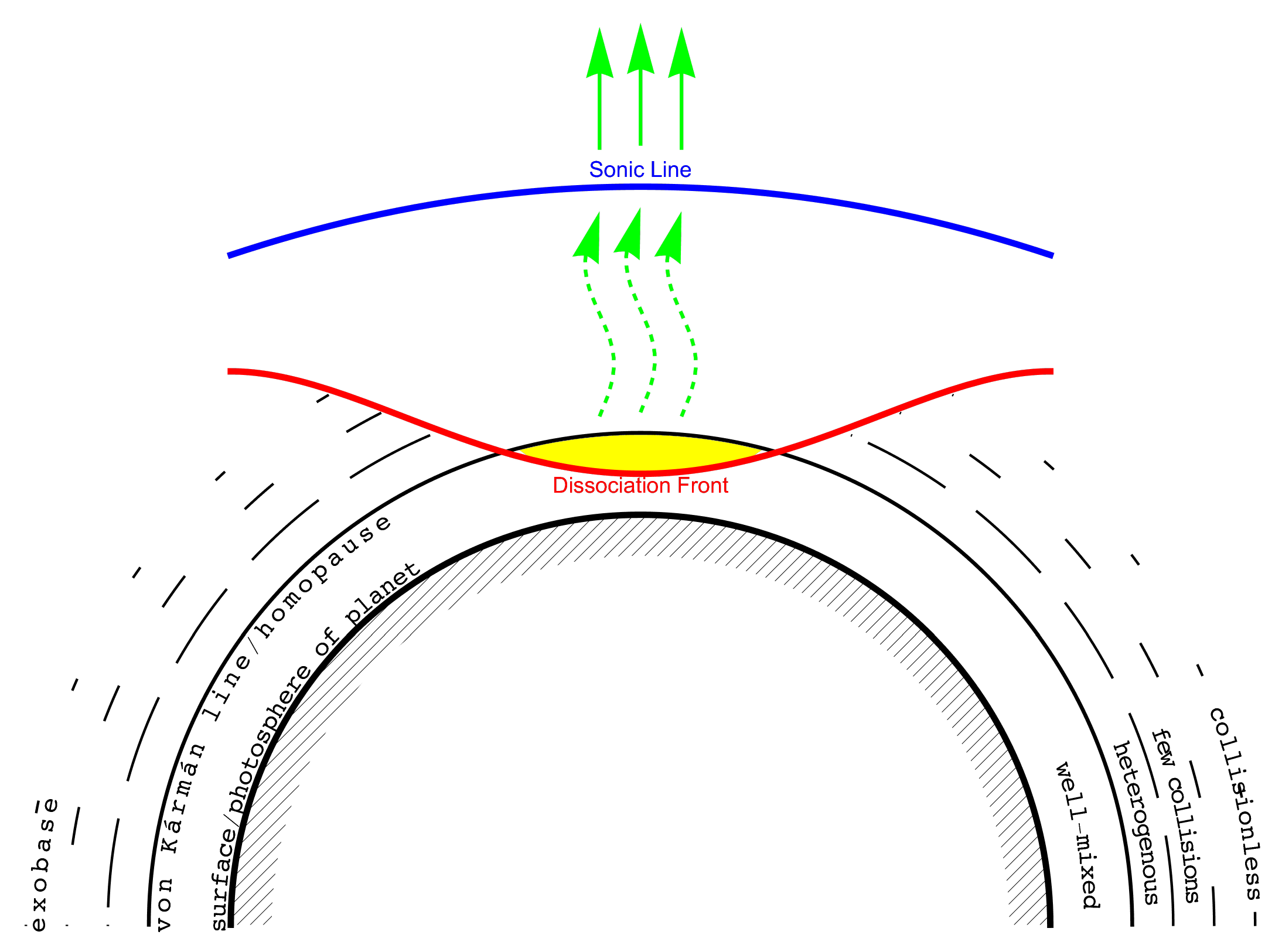}
    \caption{}
    \label{fig:first}
\end{subfigure}
\hfill
\begin{subfigure}{1\columnwidth}
    \includegraphics[width=\textwidth]{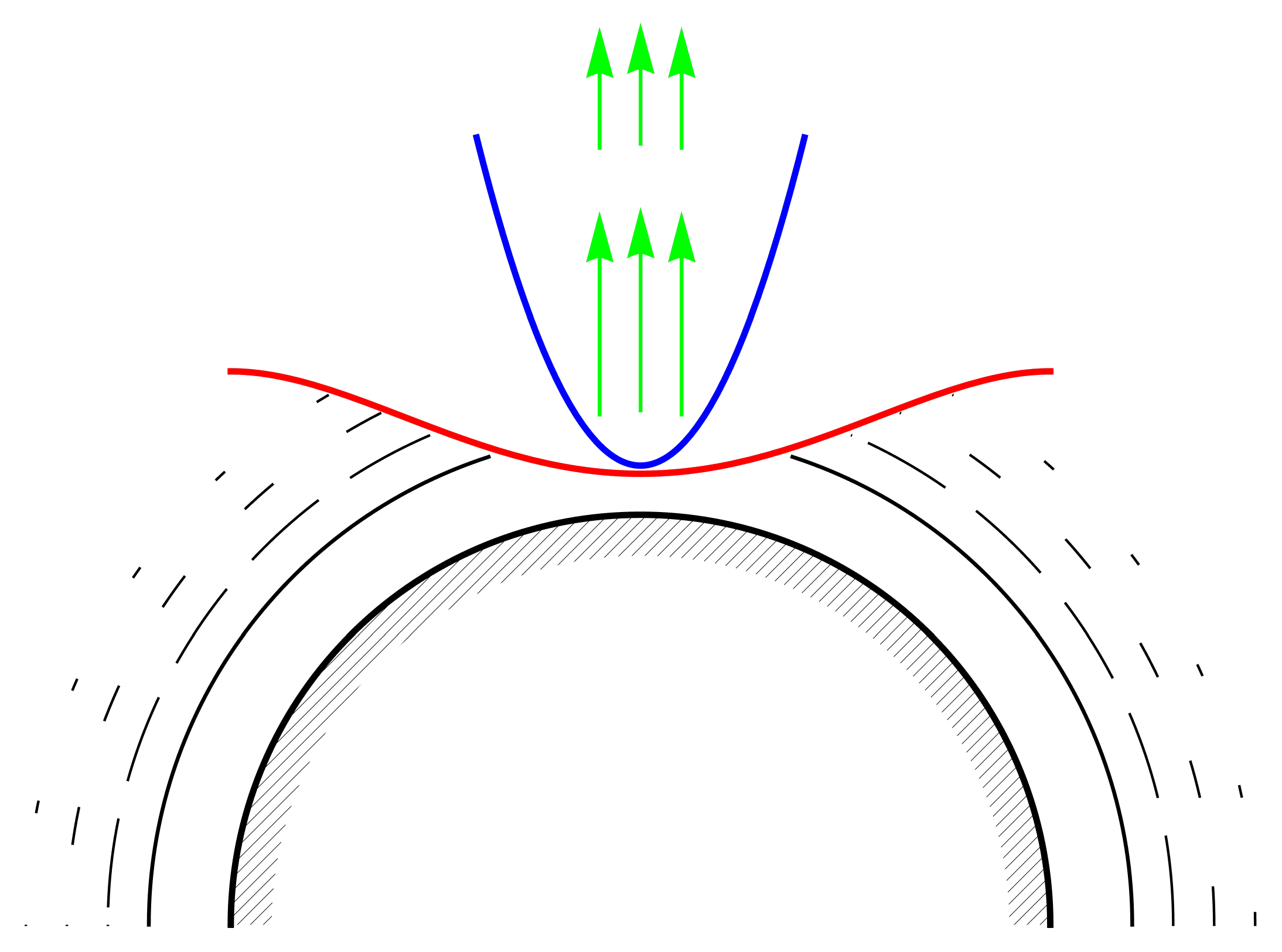}
    \caption{}
    \label{fig:second}
\end{subfigure}
\hfill
\begin{subfigure}{1\columnwidth}
    \includegraphics[width=\textwidth]{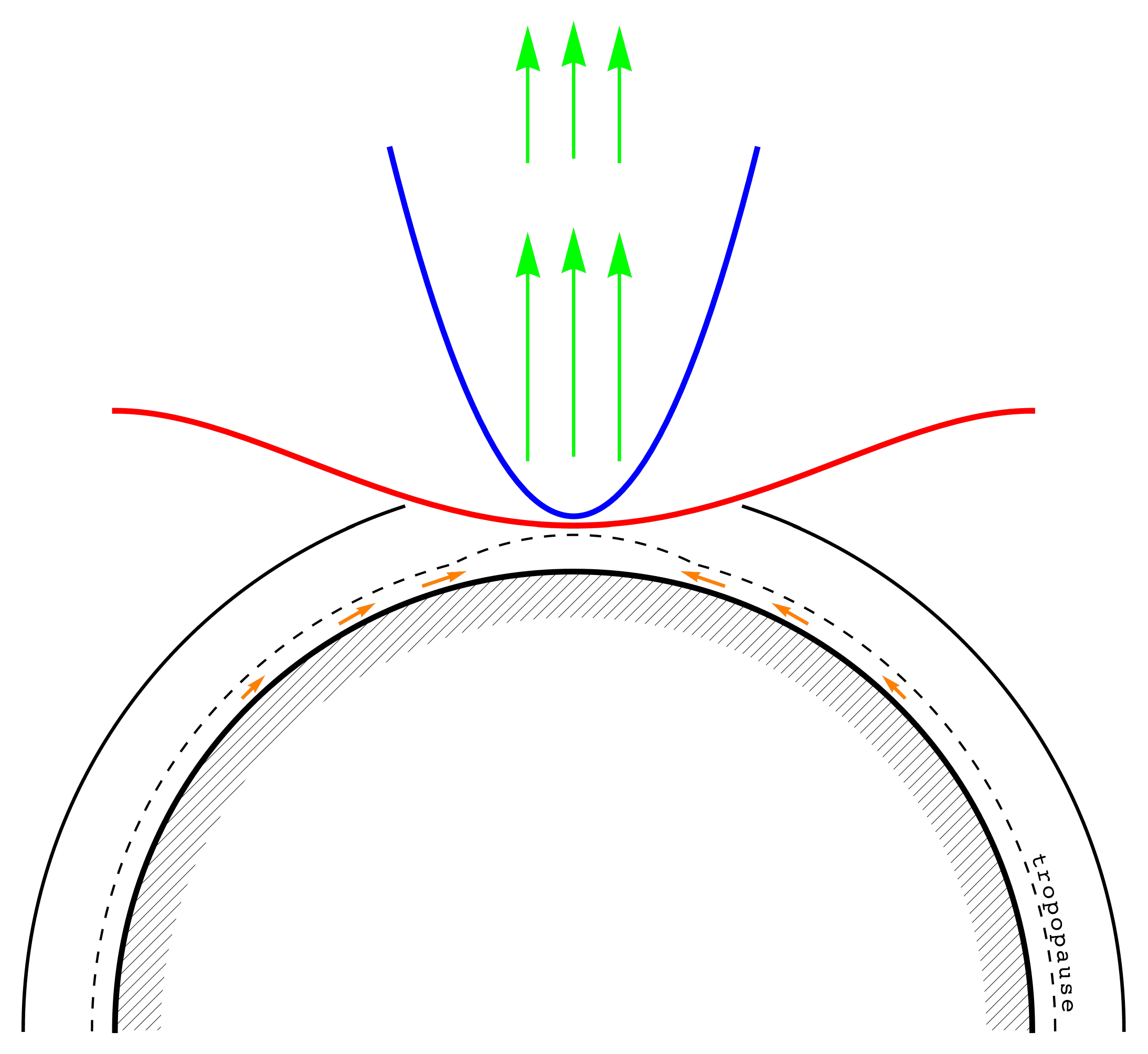}
    \caption{}
    \label{fig:third}
\end{subfigure}
\hfill
\begin{subfigure}{1\columnwidth}
    \includegraphics[width=\textwidth]{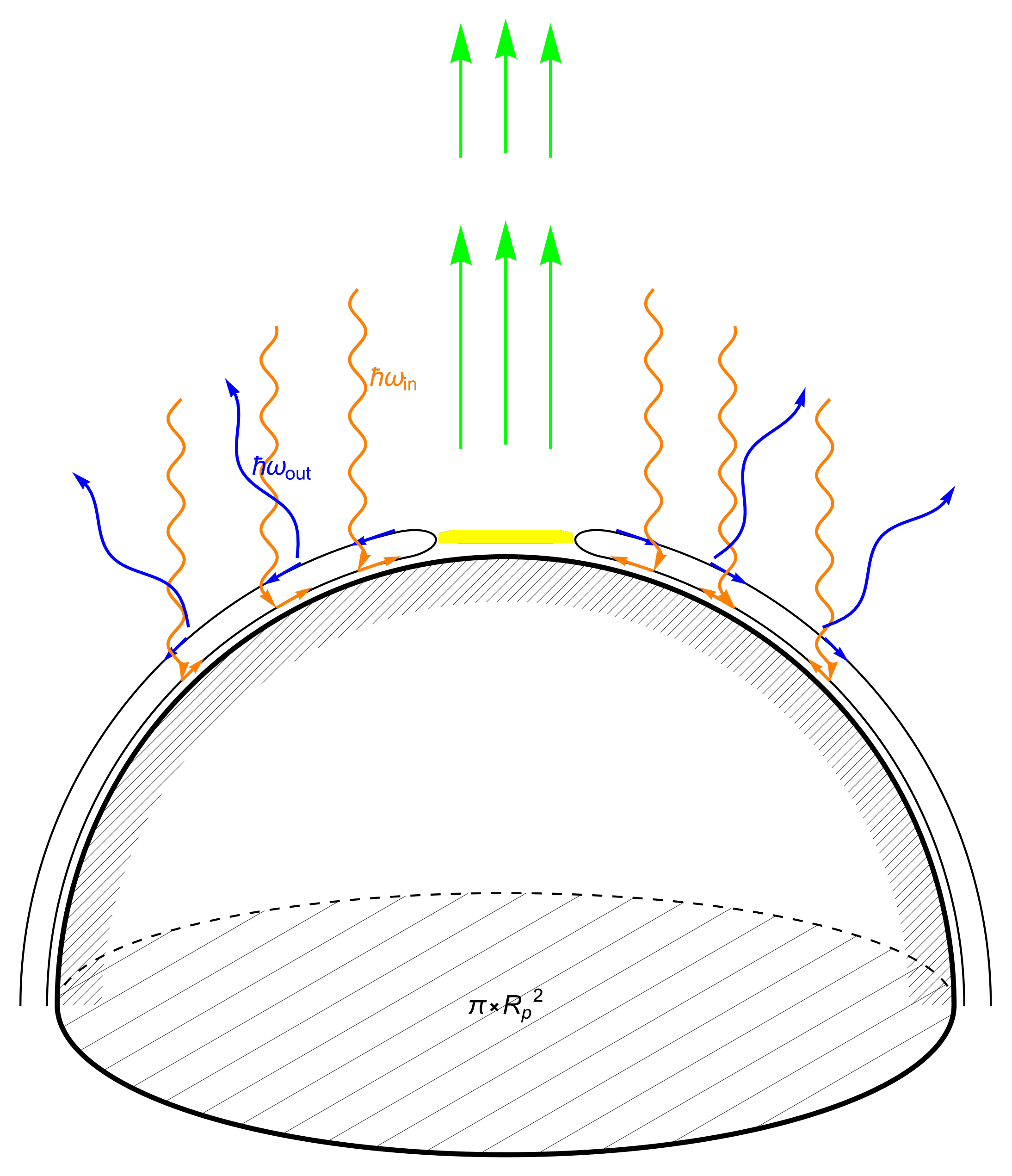}
    \caption{}
    \label{fig:fourth}
\end{subfigure}
        
\caption{Schematic diagrams illustrate the escape process of neutral H-atoms due to the interplay of critical layerings in the planetary atmosphere, in particular, the sonic line and the dissociation line of X-H bond, and the emergence of a shock front and critical phenomenon. In (a), the lenticular body (lens-shaped region colored yellow) is presumably the source region of hot neutral hydrogen atoms prepared ready for thermal escape as powered by the incident host-stellar photons. Subplot (a) is applicable to the low-flux \emph{entropy-limited} escape scenario. Subplot (b)(c)(d) show the transition into the \emph{energy-limited} regime when the flow structure undergoes a \emph{phase transition} and is accelerated by an expansive shock front, at either near the \emph{tropopause}-level ($\sim0.1$-bar) for small exoplanets or near the \emph{photosphere}-level ($\gtrsim1$-bar) for giant exoplanets. Green arrows show the escaping H-atoms (see~\ref{appendix:EnergeticView}). Subplot (b) shows that the \emph{sonic line} bends inward under higher photon flux to catch up with the \emph{dissociation front}, when they touch they form a shock front. Subplot (c) shows that when this shock front touches the \emph{tropopause} (dashed line), it then has gained direct access to the bulk circulating material of planetary body. Subplot (d) shows that the escape process is now coupled to the global planet circulation and is no longer limited by any diffusive process. Matter flowing into this \emph{shock front} is dissociated and accelerated across this thin shock front to high velocity of $\sim$10-30 km/s. The outflow likely becomes beamed also. Effects due to planet rotation and orbital revolution have been ignored in this simplified picture. If included, however, observations show that the hot spot or the source region of escaping flux can either lead ahead or lag behind the direct sub-stellar point as a result of planetary-scale rotation and fluid motion.}
\label{fig:draw}
\end{figure*}


Now, let's visualize a planetary surface rich in H$_2$ or H$_2$O subject to host-stellar photons' bombardment. When we tune up the photon flux, both the \emph{dissociation line} and the \emph{sonic line} would push deeper into the planet fluid body (see~\ref{layering} for the definitions of those lines), but at somewhat different paces. Under low-flux, the \emph{sonic line} always lies higher above. However, under higher flux, \emph{sonic line} would quickly catch up with the the \emph{dissociation line}. When they touch, it would lead to the formation of a \emph{shock front}. {\color{black}Thus, this leads to the emergence of a critical phenomenon at a critical photon flux level (Fig.~\ref{fig:draw}).} Essentially, this creates a planetary-scale rocket engine~\citep{Tsiolkovsky2004}.






{\color{black}Under low-flux (Fig.~\ref{fig:draw}(a))}, the \emph{lens-shaped region} is like a combustion chamber which sources the hot expanding neutral atomic hydrogen gas, and is powered by incoming host-stellar photons. The nozzle of this rocket engine is like the sonic level where the escaping outflow transitions from being sub-sonic to super-sonic due to the conservation of mass flux (continuity equation), and achieves a velocity of the same order-of-magnitude as the planet escape velocity ($v_{\text{esc}} \equiv \sqrt{2GM_p/r_c}$) or even higher. Additional heating of the escaping flux along its course of outflow may occur due to other physical mechanisms which would increase its degree of ionization.

{\color{black}Under high-flux (Fig.~\ref{fig:draw}(b,c,d))}, the \emph{lens-shaped region} in (a) would transform into a shock front. 

This escaping flux of material provides an alternative pathway to cool the planet surface in addition to the escaping flux of thermal infrared photons from the planetary surface/photo-sphere. A detailed energy balance can be shown that the escape flux of material is highly sensitive to the planet equilibrium temperature, which can be calculated from the bolometric flux-level of its host star upon the planet (see~\ref{appendix:EnergeticView} and~\ref{tropopause}). 

{\color{black}The stellar photons which the planet intercepts were in thermal equilibrium with the stellar surface at a much higher temperature (a few thousand Kelvin), while they are far far away from equilibrium with the planetary surface, in terms of their spatial distribution as well as energy distribution when they reach the planetary surface. 

Therefore, it is exactly this dis-equilibrium which drives flows and all kinds of phenomena occurring on the planetary surface. Thus, the entire planet surface with its fluid can be viewed as a heat engine which absorbs the stellar photons from a high temperature heat source and do all kinds of useful work such as moving fluids around and dissociating molecules, while eventually cooling off to space with infrared photons as the low temperature heat sink. One example of such work is to \emph{liberate} certain atoms or certain molecules from the bondage of chemical bond as well as the gravitational bond into space. In particular, we consider the liberation of single neutral hydrogen atom (H-atom), the lightest atom in our universe, from the bondage of hydrogen molecules (H-H) and other hydrides (X-H), and from the planetary surface into space.

Because of that, the escape process is fundamentally driven by an \emph{asymmetric driving force}, and we expect a regime-shift to occur as flux level is increased above a \emph{critical point}, which is analogous to a \emph{phase transition} potentially describable by a theory of \emph{symmetry-breaking}. The geometry of layering under static atmosphere would morph locally in the hot spot region, to change its curvature and even merge to produce discontinuity under more dynamic setting with higher incoming stellar photon flux.

One can visualize that as the incoming photon flux level increases, the \emph{sonic line} would bend inward and retreat backwards towards the planetary surface and eventually merge with the \emph{dissociation front}. When these two fronts merge, then a \emph{shock front} forms. Matter which flows into this shock front are quickly (partially)-dissociated within this very thin shock front and then accelerated to high escaping velocity ($\sim$10-30 km/s) which is much higher than typical sonic speed ($\sim$1-3 km/s) expected on a planetary surface (see~\ref{tropopause}). 

Then, the depth that this \emph{shock front} eats into the planet body depends on the incoming photon flux level. When the \emph{shock front} reaches the \emph{tropopause}, then the escaping neutral H-atom flux is no longer limited by diffusion but is directly coupled to the winds and circulation within the troposphere of the planetary body, and greatly enhanced. This transition may be paralleled with a \emph{phase transition} from an \emph{entropy-limited} regime to an \emph{energy-limited} regime.}



{\color{black}At even higher incoming photon flux level, this \emph{shock front} would touch the \emph{photosphere} directly so that photon energy absorbed there is directed converted into the dissociation energy as well as kinetic energy of escaping atoms. 

Ordinarily, such an \emph{"expansive" shock front} is impossible in nature, meaning that the inflowing material into a shock front is \emph{subsonic} while the outflowing material away from a shock front is \emph{supersonic}, because it violates the \emph{Second Law of Thermodynamics} that there is a net entropy decrease across the shock front if there are no other physical mechanisms involved~\citep{LandauLifshitz2013, Zeldovich1967, VonKarman1956}. Precisely, here the dissociation of molecules and the breakage of H-H or X-H covalent bond provides this physical mechanism that it is the extra \emph{degree-of-freedom} in this system which ultimately generates more entropy by the absorption and conversion of stellar photons into liberating neutral H-atom.}






{\color{black}Yet another alternative way to visualize the physics is that as the incoming photon flux is increased, the planet's atmosphere is trying to catch up with higher and higher wind speed. However, there is a physical limit (sonic speed) in the material property of any fluid which can mobilize and carry energy around so much fast without forming a shock front. So, at a \emph{critical point} of incoming photon flux level, the matter flow can no longer catch up with its limited sonic speed and a \emph{shock front} emerges in the hot spot region. This \emph{shock front} is again interpreted as a \emph{phase transition} which accelerates the exiting outflow from the ordinary thermal escape scenario into strong hydrodynamic escape scenario. Within this shock front, which is very thin compared to the physical dimension of the planet, the hydrogen-molecules and other hydrogen bearing molecules such as water-molecules dissociate and produce hydrogen atoms to escape at high velocity of $\sim$10-30 km/s. 

Now this (1) \emph{phase transition} in the escape geometry directly correlates with the (2) \emph{phase transition} of exoplanet temperature-dependent distribution in the mass-radius diagram, which is ultimately tied to the (3) \emph{phase transition} from hydrogen molecule H-H and hydride molecules X-H into free hydrogen atom. Marvelously, the three \emph{phase transitions} tie together in one physical problem, a trinity! Also, this shock-front driven escape mechanism involves three fronts: transonic front, collisional front (von Karman line), and dissociation front merge into one single shock front (another trinity!) to drive a beamed outflow. Thus, there is double trinity in this physical problem. 

This \emph{phase transition} may help explain the peculiar \emph{negative slopes} of the two observed cut-offs in the mass-radius diagram to be discussed in the next section, as compared to positive slopes expected in the ordinary thermal escape scenario with contours more or less parallel to the gravitational potential contour M/R. 

The winds at the level of \emph{troposphere} and \emph{photosphere} would be first driven by pressure gradient which is induced by the thermal gradient and converge towards the hot spot from all directions (Fig.~\ref{fig:draw}(c,d)). Along the way, the matter in the winds gradually pick up more and more thermal energy and also speed by continuously absorbing host-stellar photons. Of course, the winds also radiate back into space but there is a net increase in their energy as they \emph{sail} towards the hot spot. Eventually, when all these winds converge and collide in the hot spot they dump their thermal energy (characterized by small-scale random motion) as well as kinetic energy (characterized by large-scale uniform motion) entering the \emph{shock front}. Then, this converging energy dissociates molecules and drive escape. Furthermore, the compensating (returning) material flux is radiatively-cooled in a compensating flow away from hot spot region at a somewhat higher level in the atmosphere (Fig.~\ref{fig:draw}(d)). This is a planetary-scale heat engine which uses photosphere and its working material---fluid on planet surface as conveyor-belt to concentrate stellar energy absorbed into a small region, liberate neutral H-atoms, and produce collimated outflow.}



Further investigations of the escape flux and its dependence on (1) the planetary system architecture (2) and the host stellar spectral type are discussed in Section~\ref{sec:PuffyPlanets} and Section~\ref{sec:SmallerHosts}.


\section{Mass-Radius Diagram\label{sec:observation}}

An updated version of the {\color{black}\emph{ManipulatePlanet} Mathematica code} for synthesizing and analyzing the multi-dimensional exoplanets' data~\citep{Zeng2021} (\url{https://community.wolfram.com/groups/-/m/t/2445247}), is utilized to study the well-studied transiting exoplanets in the Transiting Exoplanet Property Catalogue (TEPCat) ~\citep{TEPCat2011JohnSouthworth} (\url{www.astro.keele.ac.uk/jkt/tepcat/}), to interpret the Exoplanet Systematics in the Mass-Radius Diagram.

Here, we consider two sets of mass-radius diagrams: one set for smaller exoplanets which include the super-Earths/Mini-Neptunes with typical mass range of 0-20 Earth masses (M$_{\oplus}$) and radius range of 0-5 Earth radii (R$_{\oplus}$), and the other set for gas giant exoplanets with typical mass range of 50-2000 Earth masses (M$_{\oplus}$) and radius range of 8-20 Earth radii (R$_{\oplus}$). Since we consider hydrogen as one of the main components of the atmosphere and envelope of some exoplanets, we need to exploit the hydrogen Equation-of-States (EOS) in detail, which is done in Section~\ref{sec:H2EOS}. 

The mass-radius distribution of exoplanets is highly sensitive to the planet equilibrium temperature, or equivalently, the flux-level that planet receives from its host star. The actual temperature distribution on planet surface may not be uniform due to geometric factors as well as the efficiency of energy transport on planet surface, and planetary rotation. The details are discussed in Section~\ref{appendix:PlanetTemp}. The average planet equilibrium temperature (T$_{\text{eq}}$) have a simple correlation with the temperature of sub-stellar hot spot---the highest temperature achievable on such a planet surface, and also the source region of the material escape flux considered in our model.

In the process of analyzing the mass-radius diagram, we identify two cut-offs as straight lines with negative slope in the lgM-lgR plot, one cut-off for the small exoplanets which we call the Cosmic Hydrogen and Ice Loss Line \#1 (CHILL \#1), and the other cut-off for the giant exoplanets which we call the Cosmic Hydrogen and Ice Loss Line \#2 (CHILL \#2).


\subsection{CHILL\#1 for small exoplanets}
\begin{figure}[!ht]
    \centering
    \includegraphics[width=0.45\textwidth, angle=0]{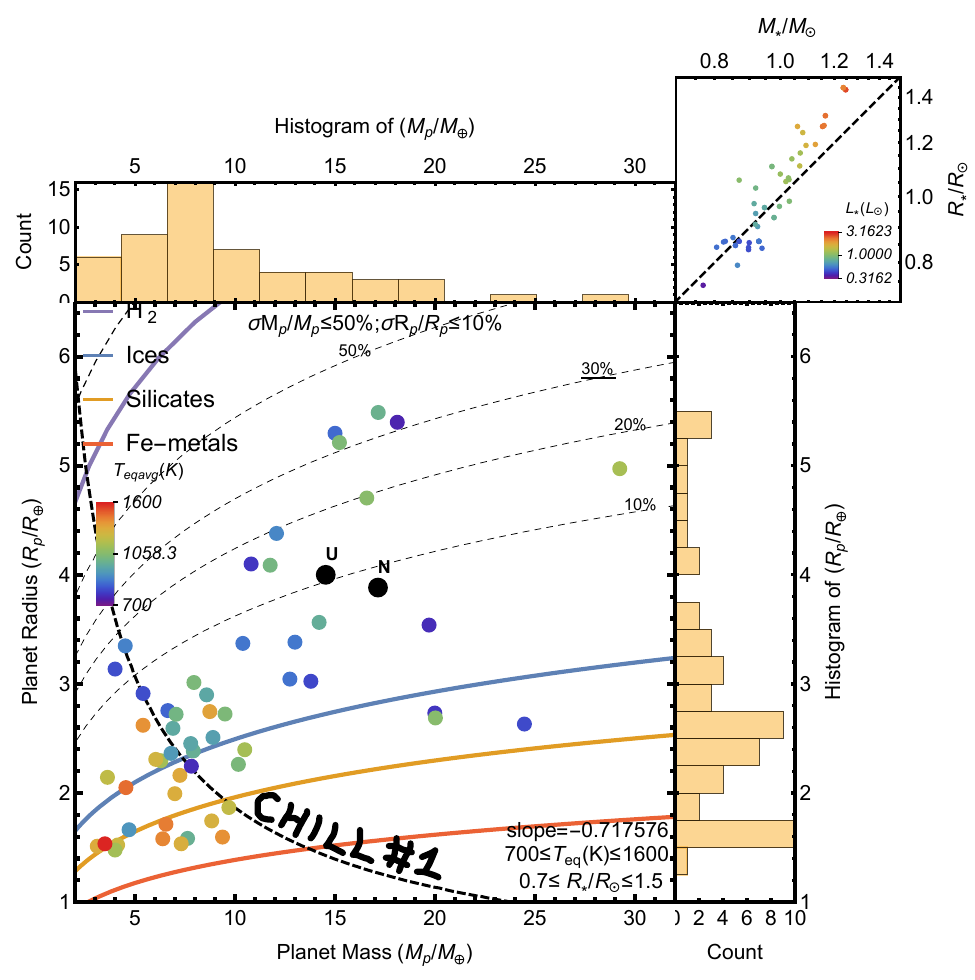}
    \caption{CHILL\#1 (thick dashed line) for Small Exoplanets: Super-Earths/Mini-Neptunes/Cores with H$_2$-Envelopes. Linear-Linear Plot in Planet Masses ($M_p$) and Radii ($R_p$). Planets are color-coded logarithmically according to their average global equilibrium temperatures (T$_{\text{eqavg}}$) assuming bond-albedo = 0 and efficient heat transport and re-distribution. The purple mass-radius curve for H$_2$ is for cold dense molecular fluid which is approximately one-sixteenth the density of cold H$_2$O fluid at similar pressures or planet masses. {\color{black}The four thin dashed mass-radius curves in between that of cold H$_2$ fluid and cold H$_2$O fluid are for the 10\%, 20\%, 30\%, and 50\% mass-mixtures of the two end members, i.e., they represent '10\%', '20\%', '30\%', and '50\%' of (cold) hydrogen envelope by mass fractions. } The subplot at the top-right corner shows the mass-radius relation of the host stars color-coded according to the stellar luminosity. These host-stars are Main-Sequence (MS) stars hotter, bigger, and more massive than M-dwarfs.}
    \label{fig:eb1_1}
\end{figure}

\begin{figure}[!ht]
    \centering
    \includegraphics[width=0.45\textwidth, angle=0]{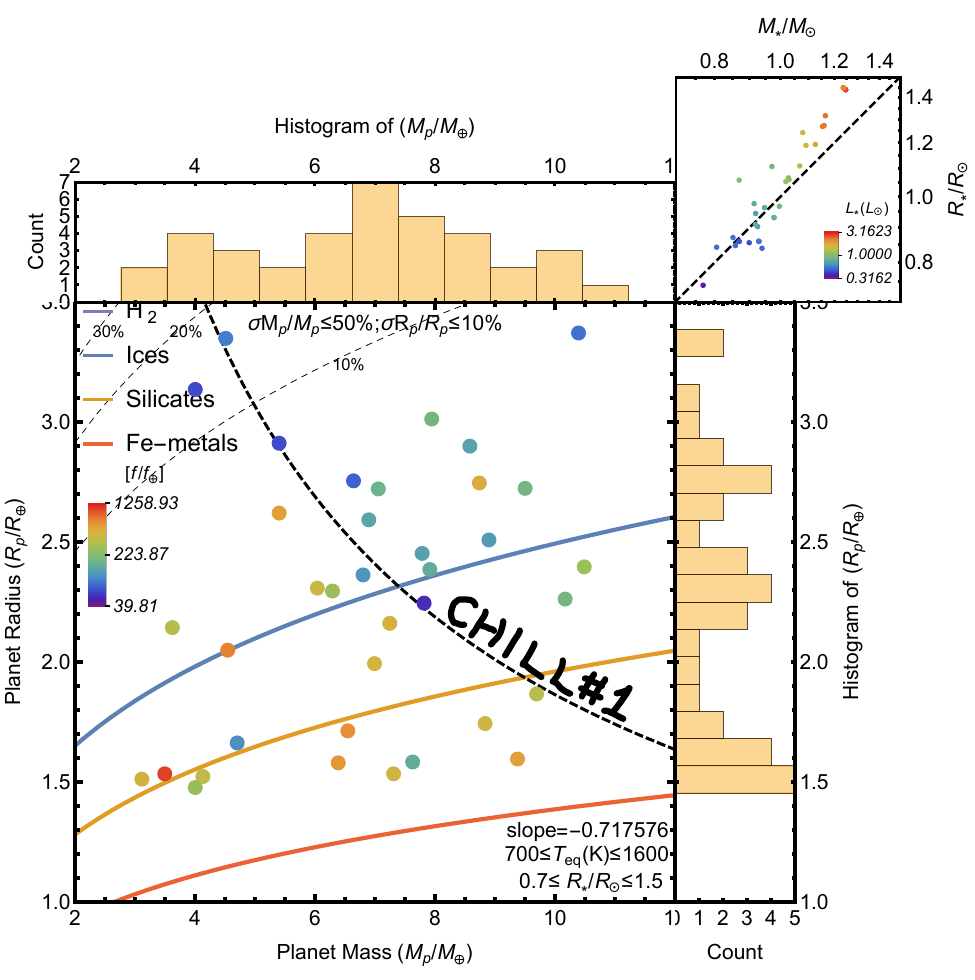}
    \caption{CHILL\#1 (thick dashed line) for Small Exoplanets: Super-Earths/Mini-Neptunes/Cores with H$_2$-Envelopes. Linear-Linear Plot in Planet Masses ($M_p$) and Radii ($R_p$), Zoom-in. Planets are color-coded logarithmically according to their incident bolometric host-stellar flux ($f/f_{\oplus}$) where $f_{\oplus}=1.36$ kW/m$^2$ is the standard flux received at Earth-distance from the Sun (solar constant). According to the planet radius ($R_p$) histogram attached to the right of the main mass-radius diagram, it seems there is a possibility to further sub-divide the planets which are in between 2-3 $R_{\oplus}$ into two sub-populations. Furthermore, the clustering of exoplanets near the intersection of CHILL\#1 and pure-H$_2$O mass-radius curve is interesting.}
    \label{fig:eb1_2}
\end{figure}

First, let us discuss the CHILL\#1 for small exoplanets, as shown in Figure \ref{fig:eb1_1} and Figure \ref{fig:eb1_2}. 

These small exoplanets possess a certain degree of their primordial hydrogen envelope or thick atmosphere, which place them at their corresponding positions on the mass-radius diagram. Their envelope or thick atmosphere is expected to undergo escape process as discusses earlier. They are also expected to possess a core made of either rock-metals or hydrogen-bearing ices. 

Due to the selective escape of lighter species, their planetary atmosphere is expected to be gradually enriched in heavier species over the course of time, which increases the mean molecular weight of the atmosphere and decreases the scale height. Thus, eventually, their planetary radii will become close to that of bare cores, either rock-metal or water-ice, as calculated from theory. 

Therefore, we interpret the observed bi-modal distribution of small exoplanet radii and the small exoplanet radius gap at around 2 R$_{\oplus}$ as caused by the density difference between a rock-metal core and a water-ice-rich core~\citep[]{LPSC2017:Zeng2017PlanetFormation, Zeng2018SurvivalMNRAS, PNAS:Zeng2019, Zeng2021}.

Other possibilities of explaining the small exoplanet radius gap can be found in the following literature~\citep{Berger2018Revised2, Berger2020, Fulton2017ThePlanets, Fulton2018TheGap, Petigura2018ThePlanets, Petigura2022, Owen2013, Owen2017, Wu2018MassSuper-Earths, VanEylen2018, Ginzburg2018, Rogers2021, Rogers2023, Izidoro2022, Lehmer2017RockySaturation}.


Furthermore, we consider the CHILL\#1 as a physical boundary differentiating two group of small exoplanets: one group which still partially retains their primordial hydrogen envelope of near \emph{nebular} composition, and the other group which has completely lost their primordial hydrogen envelope due to higher planet equilibrium temperature, but may possess a highly-evolved secondary envelope/atmosphere of near \emph{meteoritic}/\emph{planetary} composition.

Thus, there is a significant temperature contrast for small exoplanets on either side of this physical boundary. The critical temperature of this physical boundary CHILL\#1 is estimated at T$_{\text{eq}}\approx$1000K, or equivalently, a flux-level of two hundred times the Earth's flux-level ($f/f_{\oplus}\approx200$). 


This flux cut-off is further confirmed and elucidated in additional plots in Section~\ref{sec:PuffyPlanets} and Section~\ref{sec:SmallerHosts}.


{\color{black} This physical boundary of CHILL\#1 in the log-log plot of mass-radius can be linearly-fitted to yield a negative slope of $-\frac{2}{3}$ ($\pm 0.05$). We can approximately express this relation as:}

\begin{equation}
    (M_p/10M_{\oplus})^2 \times (R_p/2R_{\oplus})^3 \approx 1
\end{equation}

where the product of the planet mass squared and the planet radius cubed equals a constant approximately along this physical boundary. Another equivalent expression is:

\begin{equation}
    (M_p/10M_{\oplus})^{2/3} \times (R_p/2R_{\oplus}) \approx 1
\end{equation}

where the product of the planet mass to the two-thirds power and the planet radius equals a constant along this physical boundary.

{\color{black}This physical boundary CHILL\#1 is represented as the thick dashed curve in Figure~\ref{fig:eb1_1} and Figure~\ref{fig:eb1_2}. Note that both figures are rendered linear-linear in mass-radius and that is why this cut-off appears curved instead of straight. There is a special functionality in our \emph{ManipulatePlanet} Mathematica code (\url{https://community.wolfram.com/groups/-/m/t/2445247}) which allows a one button switch between log-log plot versus linear-linear or linear-log or log-linear plot. The \emph{ManipulatePlanet} Mathematica code also provides an interactive fitting tool which determines the slope of an intercept. That is how we were able to identify and carefully examine the boundaries and cut-offs.}

The most interesting feature of CHILL\#1 is its negative slope, as opposed to the positive slope of a typical mass-radius curve of a uniform composition in that regime that we are interested. This negative slope is most likely due to the intrinsic physics and chemistry of the escape process.

\subsection{CHILL\#2 for giant exoplanets}
\begin{figure}[!ht]
    \centering
    \includegraphics[width=0.45\textwidth, angle=0]{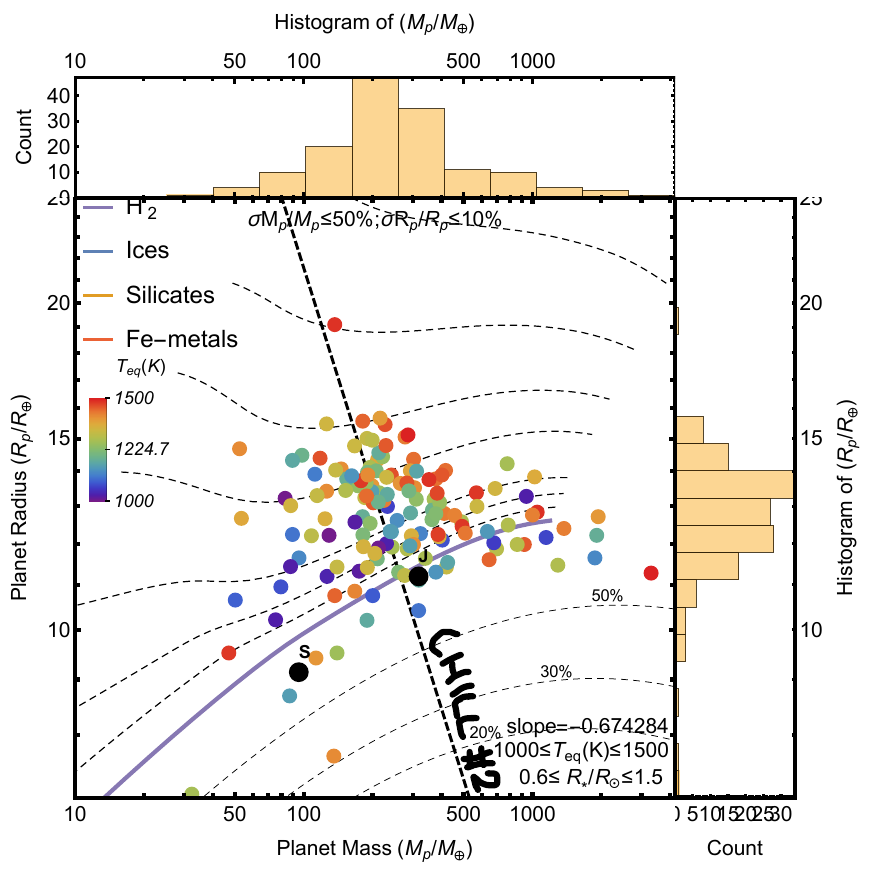}
    \caption{CHILL\#2 for Large Exoplanets: Gas Giants/Hot Jupiters. 1000K-1500K (200$f_{\oplus}$-800$f_{\oplus}$). Significant inflation of planet radii occur within this equilibrium temperature range. Notice the approximate triangular shape of planet distribution region in lgM-lgR, which tapers off towards higher masses. This overall distribution is a consequence of three factors combined: (1) planet mass (M$_p$); (2) insolation received ($f/f_{\oplus}$) or equilibrium temperature (T$_{\text{eq}}$); (3) metallicity [Fe/H]. {\color{black}Theoretical mass-curves are given for pure-hydrogen with its variations in both directions: (1) upward-direction with increasing internal specific entropy (S (eV/1000 K/atom) = 0.3 (purple solid curve which is considered to be cold), 0.4, 0.5, 0.6, 0.7, 0.8, 0.9, and 1.0 correspondingly). The nominal surface of truncation of calculation for these eight mass-radius curves is taken to be at the density of 0.01 g/cc. Hydrogen EOS data from~\citep{Becker2014AbDwarfs}; (2) downward-direction with progressively mixing with heavier elements/species. The thin dashed mass-radius curves in between that of cold H$_2$ fluid and cold H$_2$O fluid are for the 50\%, 30\%, 20\%, and 10\% mass-mixtures of the two end members, i.e., they represent '50\%', '30\%', '20\%', and '10\%' of (cold) hydrogen envelope by mass fractions as labelled. The host stars are mostly Main-Sequence (MS) stars hotter than M-dwarfs.}}
    \label{fig:eb3_1}
\end{figure}

\begin{figure}[!ht]
    \centering
    \includegraphics[width=0.45\textwidth, angle=0]{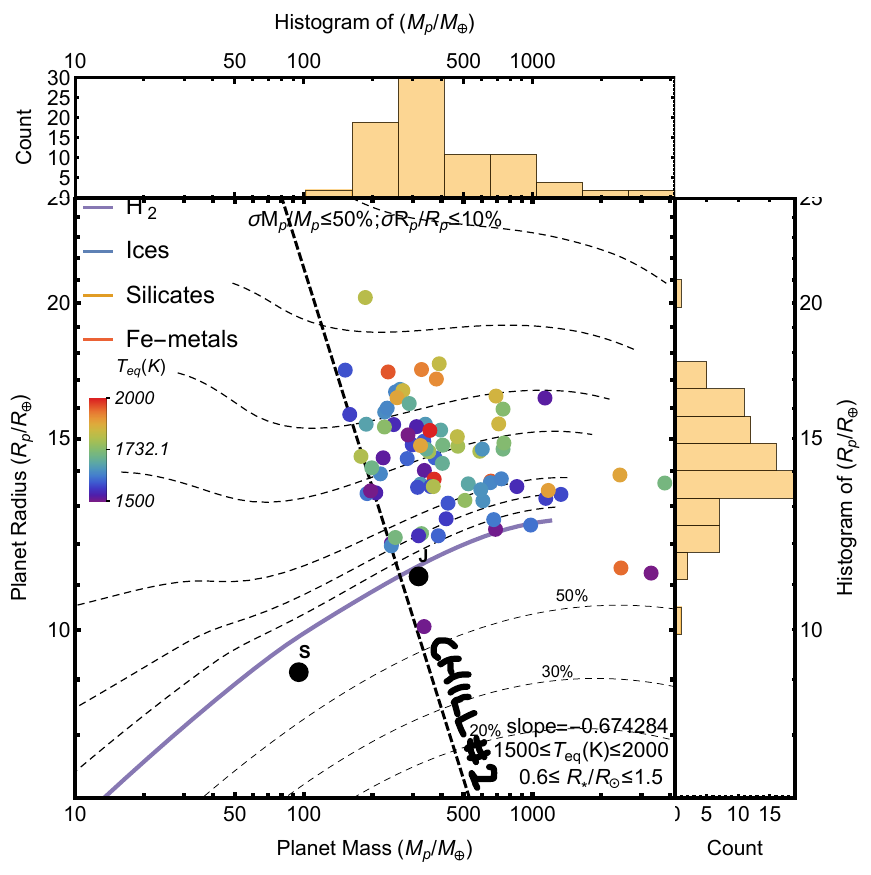}
    \caption{CHILL\#2 for Large Exoplanets: Gas Giants/Hot Jupiters. 1500K-2000K (800$f_{\oplus}$-3000$f_{\oplus}$). Significant evaporative loss and destruction of gas giants occur within this equilibrium temperature range. The cut-off with (-2/3) slope in lgM-lgR is evident. Notice the change of the temperature scale. {\color{black}Theoretical mass-curves are given for pure-hydrogen with its variations in both directions: (1) upward-direction with increasing internal specific entropy; (2) downward-direction with progressively mixing with heavier elements/species. The hosts are mostly Main-Sequence (MS) stars hotter than M-dwarfs.}}
    \label{fig:eb3_2}
\end{figure}

Then, let us consider the CHILL\#2 of gas giant exoplanets, as shown in Figure \ref{fig:eb3_1} and Figure \ref{fig:eb3_2}. 

These gas giant exoplanets are dominated by hydrogen and helium in their composition. Therefore, the CHILL\#2 represents very significant mass loss or even complete destruction of these exoplanets. A simultaneous physical factor to consider is the significant inflation of the planet radius due to its host-stellar insolation, which is also the physical mechanism which drives the escape process. Thus, a strong correlation is expected between the measured planet (inflated) radius, planet mass, and its flux-level. 

Another important physical factor which affects the measured planet radius is the metallicity of the planet hydrogen envelope. Due to internal convective motion and mixing, a significant portion of metals, elements heavier than hydrogen and helium, is likely dispersed throughout the envelope, with a possible gradient concentrating towards the planet core. It has been demonstrated by calculation that what matters is the total metallicity or total mass of metals contained within the planetary body, regardless of its distribution in the envelope or core~\citep[]{Stevenson2017Ge131:PlanetaryEvolution}. Thus, to the leading order, we can express the relationship between planet radius, planet mass, flux-level, and planet metallicity in the following functional form: 

\begin{equation}
R_p = \mathcal{F}(M_p,\text{flux},\text{metallicity})
\end{equation}

where the terms on the right-hand side inside the function $\mathcal{F}$ are: (1) the planet mass (M$_p$); (2) the incident bolometric host-stellar flux; (3) the metallicity. 

Here we choose the host-stellar [Fe/H] to represent the planet metallicity because it is most readily measurable. However, we need to keep in mind that there may be significant difference between the host-stellar metallicity and the planet metallicity as caused by the ongoing escape process. In other words, the planets are expected to be gradually enriched in metals due to the escape process. 

Also, observations of warm Jupiters indicate that most possess significantly higher metallicity than their host stars~\citep[]{Thorngren2016}. This suggests that planetary metallicity can be enriched by processes other than atmospheric escape.

Also, an implicit assumption of the correlation between the iron abundance with the abundance of other important planet-building elements such as characterized by [C/O], [Mg/Si], [Fe/Si], [O/H], etc.) are made. 

\begin{figure}[!ht]
    \centering
    \includegraphics[width=0.45\textwidth, angle=0]{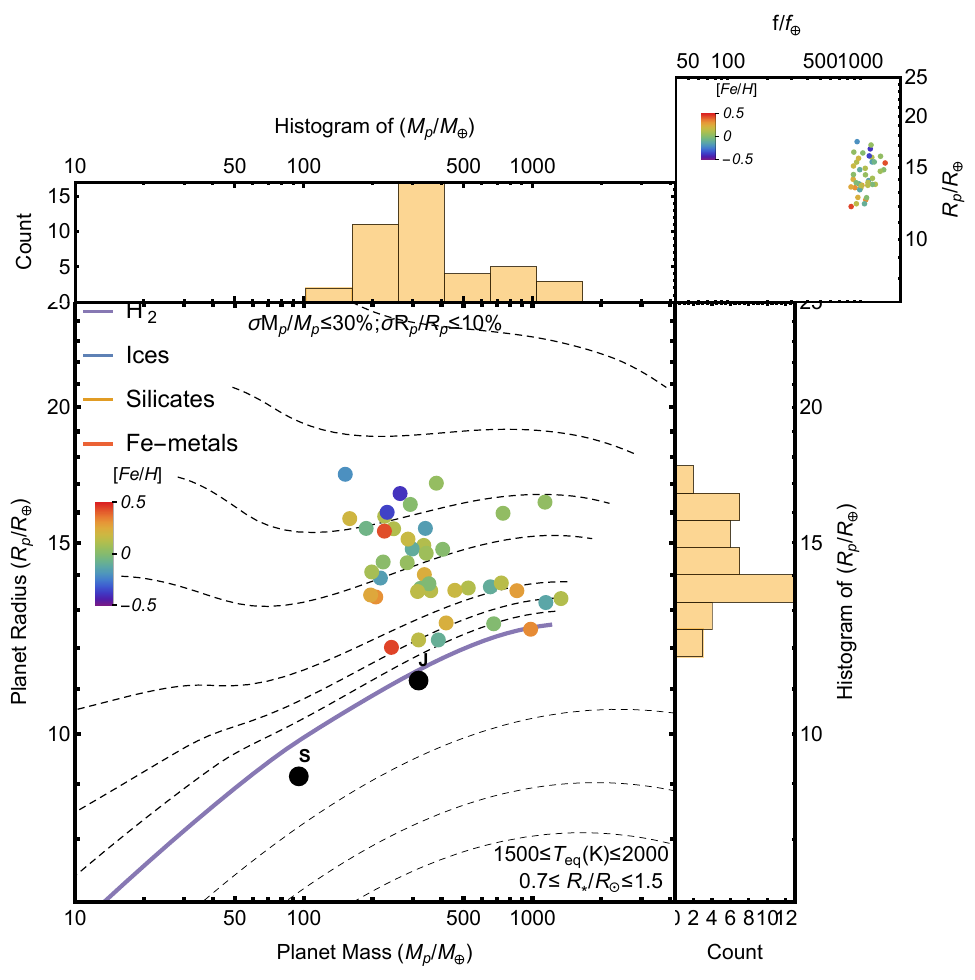}
    \caption{CHILL\#2 for Large Exoplanets: Gas Giants/Hot Jupiters. 1500K-2000K. This is a re-rendering of Figure~\ref{fig:eb3_2}, with a stricter selection in planet errorbars, and a new subplot showing the insolation flux received ($f/f_{\oplus}$) versus planet radii ($R_p/R_{\oplus}$). The flux is measured in the unit of what Earth receives from the Sun, i.e., the solar constant ($f_{\oplus}=1.36$ kW/m$^2$). The cut-off in the lgM-lgR diagram shows up as a cut-off in flux of ($f_{\text{crit}}/f_{\oplus}\approx800$). This is expected because the lower limit of the T$_{\text{eq}}$-range is set at 1500K which exactly corresponds to this $f_{\text{crit}}$. The gas giants which can survive at or above this threshold flux level generally are more massive than Saturn ($\gtrsim$ 100 M$_{\oplus}$). Furthermore, the planets in the main plot as well as the subplot are now color-coded according to their host-stellar metallicity [Fe/H]. Their inflated radii correlate with their metallicity [Fe/H] with the general trend of lower-metallicity ones having larger radii (more-inflated) and higher-metallicity ones having smaller radii (less-inflated), as expected.}
    \label{fig:eb3_3}
\end{figure}

Nonetheless, adding in the host-stellar [Fe/H] as an extra dimension in the mass-radius plot gives us insight into the physics which shapes the planet distribution, as demonstrated in Figure~\ref{fig:eb3_3}.

Detailed examinations of Figure~\ref{fig:eb3_1}, Figure~\ref{fig:eb3_2}, and Figure~\ref{fig:eb3_3} show that significant inflations of giant exoplanet radii begin to occur in the planet equilibrium temperature range of 1000-1500K (T$_{\text{eq}}$).

Moreover, complete destruction of giant exoplanets begins to occur on a grand scale in the planet equilibrium temperature range of 1500K-2000K (T$_{\text{eq}}$). In Figure~\ref{fig:eb3_2}, we find a sharp truncation or physical boundary in the mass-radius diagram which we identify as the CHILL\#2.

{\color{black} This physical boundary of CHILL\#2 in the log-log plot of mass-radius can be linearly-fitted to yield a well-defined negative slope of $-\frac{2}{3}$ ($\pm 0.03$). Again, we can also approximate this relation as:}

\begin{equation}
    (M_p/300M_{\oplus})^2 \times (R_p/10R_{\oplus})^3 \approx 1
\end{equation}

or equivalently, in unit of Jupiter mass (M$_{\text{J}}$) and Jupiter radius (R$_{\text{J}}$), 

\begin{equation}
    (M_p/M_{\text{J}})^2 \times (R_p/0.9R_{\text{J}})^3 \approx 1
\end{equation}

In particular, Figure~\ref{fig:eb3_3} and additional plots in Section~\ref{appendix:GasGiantsSubPlots} show that the CHILL\#2 is located at the planet equilibrium temperature of 1500K (T$_{\text{eq}}$) assuming zero bond albedo, or equivalently, eight hundred times the Earth's flux-level ($f/f_{\oplus}\approx800$). This suggests that the giant exoplanets which lie along the edge of this physical boundary in Figure~\ref{fig:eb3_2} receives about eight hundred times the Earth's flux. When converted into standard units, this quantity is $10^{6}$W/m$^2$, or one million Watts per square meter on a surface perpendicular to the host-stellar irradiation summed over the entire electro-magnetic spectrum. Thus, we believe this strong insolation serves as the major power source which drives the planet envelope and atmosphere escape.

\section{Discussion}

The key feature that we want to solve and explain by theory is the negative slope of the CHILL\#1 and the CHILL\#2.


Contrary to intuition, the negative slope suggests along the physical boundary of escape, the less massive and more puffier exoplanets which have a shallower gravitational potential can somehow last for the same timescale as compared to the more massive and less puffier exoplanets which have a deeper gravitational potential. Moreover, the puffier exoplanets have proportionally larger surface area to intercept the incoming host-stellar radiation which should result in more escape. Therefore, we need to consider and calculate the efficiency of the aforementioned planet-scale rocket engine under different parameters, that the rocket engine on puffier exoplanet may be of poorer efficiency as compared to the rocket engine on a less puffier and more compact exoplanet, due to self-shielding effect and metallicity effect. 

{\color{black}We speculate there is a \emph{shock front} emerging in the critical scenario which accelerates the flow to super-sonic speed (please see discussion in~\ref{tropopause}). Thus, there exists a \emph{phase transition} in the escape scenario from ordinary thermal escape to \emph{shock-wave} accelerated scenario. This might explain the change of the positive-slope boundary delineating the boundary of hot Saturns to the negative-slope boundary cutting off the hot Jupiter population as well as the small exoplanet population in the mass-radius diagram.}







\subsection{On Planet Temperature}\label{appendix:PlanetTemp}
This average equilibrium temperature ($T_{\text{eq}}$) for a planet on an ecliptic orbit can be calculated as follows. 

First, let's consider the total amount of stellar insolation the planet receives averaged over one full orbital period: 

\begin{equation}
    \bar{f} = \frac{1}{P} \cdot \int_{0}^{P} \frac{L_{\star}}{4 \pi r^2} \,dt
\end{equation}

Then, we can substitute the $r^2$ in the denominator with the angular velocity $\dot{\theta}$ using the Kepler's Second Law.

\begin{equation}
    \text{const} = \frac{1}{2} r^2 \dot{\theta} = \frac{1}{2} r^2 \frac{d\theta}{dt} = \frac{\pi a b}{P} = \frac{\pi a^2 \sqrt{1-e^2}}{P}
\end{equation}

Then, recall the Kepler's Third Law we can re-write $P$ as:

\begin{equation}
    P = \frac{2\pi}{\sqrt{G(M_{\star}+M_p)}} \cdot a^{3/2} \approx \frac{2\pi}{\sqrt{GM_{\star}}} \cdot a^{3/2}
\end{equation}

Then, we can figure out the "const" as:

\begin{equation}
    \text{const} = \frac{\pi a^2 \sqrt{1-e^2}}{P} \approx \frac{1}{2} \sqrt{GM_{\star} \cdot a (1-e^2)}
\end{equation}

Thus, the inverse of orbital distance ($r$) squared is directly proportional to the angular velocity ($\dot{\theta}$):

\begin{equation}
    \frac{1}{r^2} = \frac{\dot{\theta}}{\sqrt{GM_{\star} \cdot a (1-e^2)}}
\end{equation}

Then, the integral can be easily carried out because a planet simply revolves about $2\pi$-angle in each orbital period:

\begin{equation}
    \begin{split}
    \bar{f} &= \frac{1}{P} \cdot \int_{0}^{P} \frac{L_{\star}}{4 \pi} \cdot \frac{1}{r^2} \,dt = \frac{1}{P} \cdot \int_{0}^{P} \frac{L_{\star}}{4 \pi} \cdot \frac{\dot{\theta} \,dt}{\sqrt{GM_{\star} \cdot a (1-e^2)}} \\ &= \frac{1}{P} \cdot \frac{L_{\star}}{4 \pi} \cdot \frac{2\pi}{\sqrt{GM_{\star} \cdot a (1-e^2)}} = \frac{L_{\star}}{4\pi a^2 \sqrt{1-e^2}}
    \end{split}
\end{equation}

Notice the extra factor of $1/\sqrt{1-e^2} \approx 1+\frac{1}{2}e^2+\frac{3}{8}e^4+...$, is the fundamental argument behind \emph{Milankovitch cycle}~\citep{Milankovic1941}. The orbit eccentricity change affects the average insolation. In the long-run, it builds up and affects the overall climate of the planet. On the other hand, the change of planet \emph{obliquity} only affects the re-distribution of stellar insolation over different areas of the planet surface without changing the total amount of energy received, assuming albedo does not change.

Thus, then the equilibrium temperature of the planet can be calculated as follows:

First, the \emph{average} equilibrium temperature $T_{\text{eq}}$ of the planet surface can be considered as the temperature if the total stellar insolation received over $\pi R_p^2$ is transported and uniformly distributed over its entire surface area of $4\pi R_p^2$:

\begin{equation}
    \begin{split}
    T_{\text{eq}} &= \left( \frac{\bar{f}}{4\sigma} \right)^{1/4} = \left( \frac{L_{\star}}{4\sigma \cdot 4\pi a^2 \sqrt{1-e^2}} \right)^{1/4} \\ &= \left( \frac{4\pi R_{\star}^2 \sigma T_{\star}^4}{4\sigma \cdot 4\pi a^2 \sqrt{1-e^2}} \right)^{1/4}
    \end{split}
\end{equation}

where $\sigma$ is the \emph{Stefan-Boltzmann constant}, $a$ is orbit semi-major axis, and the effect of orbit eccentricity $e$ is included, assuming bond-albedo = 0 (meaning complete absorption of incident host-stellar bolometric flux) and efficient heat transport and re-distribution over the planet surface. Therefore, 

\begin{equation}
    \begin{split}
    T_{\text{eq}} &= T_{\text{star}}/(2 a/ R_{\text{star}})^{1/2}/(1 - e^2)^{1/8} \\ &= \frac{T_{\star}}{\sqrt{2a/R_{\star}}} \cdot \left( 1+\frac{1}{8}e^2+\frac{9}{128}e^4+... \right)
    \end{split}
\end{equation}

In this paper, when we refer to the equilibrium temperature we always refer to $T_{\text{eq}}$, i.e., the temperature averaged over the entire planet sphere and over one full orbital period, assuming zero-albedo. Locally, certain areas on planet surface can become hotter or colder due to inefficient or incomplete or different mode of heat re-distribution.

Now, if this close-in planet is tidally locked to its host star, and if its eternal day-side never communicates heat with its eternal night-side but averages within its hemisphere, and then re-radiate its received energy as infrared photons into space, then, we can estimate its day-side average hemispherical temperature as:

\begin{equation}
    T_{\text{dayside-hemisphere}} = 2^{1/4} \cdot T_{\text{eq}}
\end{equation}

Furthermore, if the sub-stellar point of this planet, which is the point of the planet under direct vertical insolation from its host star, never communicates heat with other areas, and then re-radiate its received energy as infrared photons back into space, then, we can estimate its sub-stellar point temperature as (the highest possible temperature achievable on such a planet):

\begin{equation}
    T_{\text{sub-stellar}} = 2^{1/2} \cdot T_{\text{eq}}
\end{equation}

\begin{figure}[!ht]
    \centering
    \includegraphics[width=0.45\textwidth, angle=0]{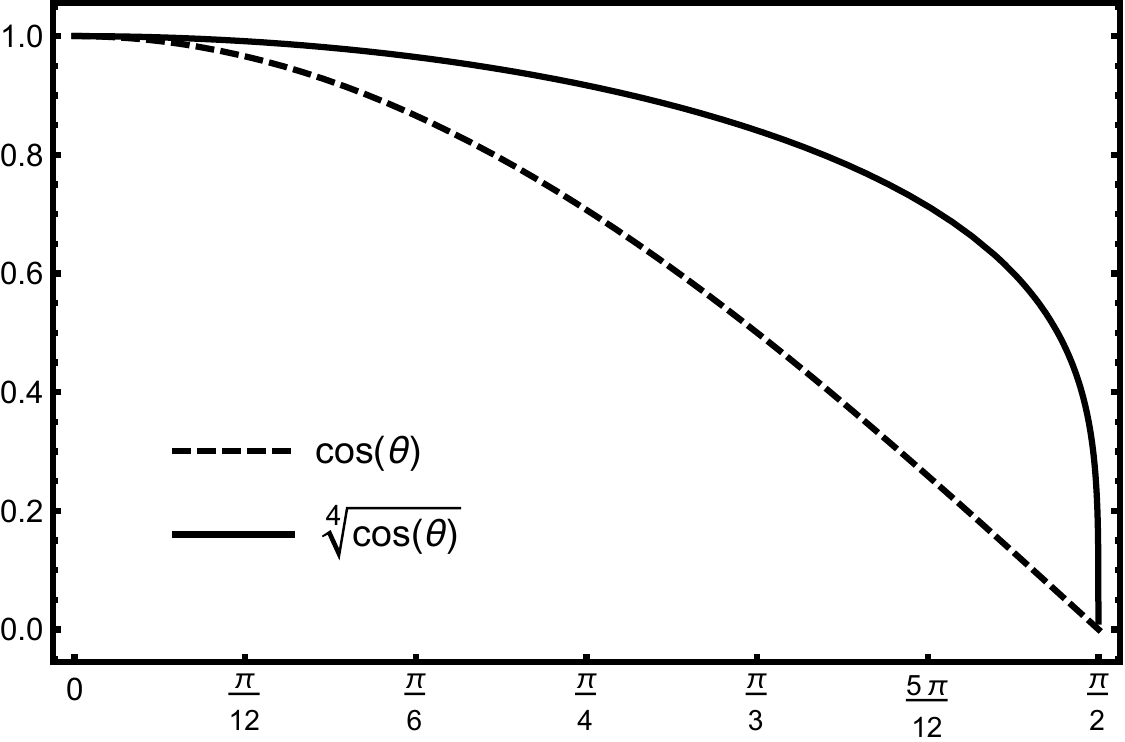}
    \caption{$\left( \cos{\theta} \right)^{1/4}$-dependence. Rapid change occurs near the rim $\frac{5}{12}\pi$-$\frac{1}{2}\pi$.}
    \label{fig:Cos1:4}
\end{figure}

Likewise, in such a scenario of no heat communication, for a spot which subtends angle $\theta$ from the sub-stellar point on the day-side its temperature is estimated to be (please see Fig.~\ref{fig:Cos1:4}):

\begin{equation}
    T_{\text{dayside}}\left( \theta \right) = T_{\text{sub-stellar}} \cdot \left( \cos{\theta} \right)^{1/4} = 2^{1/2} \cdot T_{\text{eq}} \cdot \left( \cos{\theta} \right)^{1/4}
\end{equation}

Its night-side may have some arbitrarily low temperature. 

Thus, for more detailed quantitative calculation, one needs to consider which hemisphere or portion of the planet is heated and is responsible for the main escape flux. This also depends on the geometry of heat re-distribution on such a planet. 





In terms of orbital parameters alone, if not including planet albedo effect, the equilibrium temperature of a planet primarily depends on its orbit semi-major axis $a$, and secondarily on the effect of its orbit eccentricity $e$,

\begin{equation}
    T_{\text{eq}} = T_{\text{star}}/(2 a/ R_{\text{star}})^{1/2}/(1 - e^2)^{1/8}
\end{equation}


Observationally speaking, the $T_{\text{eq}}$-characteristic of 
Cosmic Hydrogen-Ice Loss Line (CHILL\#1) and Hydrogen Loss Line (CHILL\#2) are: T$_{\text{eq}}\sim$1000K (or approximately 200 times the Earth's insolation ($f_{\oplus}$)) for Super-Earths/Mini-Neptunes---planets of masses on the order of a few up to a few tens of Earth masses and T$_{\text{eq}}\sim$1500K (or approximately 800 times the Earth's insolation ($f_{\oplus}$)) for massive gas giants---planets of masses on the order of a few hundred Earth masses or one Jupiter mass. This strong T$_{\text{eq}}$-dependence of the escape process can be thought of as an On-Off switch which determines whether a planet with significant volatile envelope can survive over the planetary system age at a specified host-stellar irradiation level.

The difference in this critical value between the small and large exoplanets (200 $f_{\oplus}$ versus 800 $f_{\oplus}$) may be explained by different escaping geometry required for substantial depletion of the hydrogen envelope for a small planet versus for a massive gas giant. Assuming inefficient heat circulation (or zero heat re-distribution), the sub-stellar hot spot of a highly-irradiated exoplanet can get as high as $\sqrt{2}$ of its globally-averaged equilibrium temperature T$_{\text{eq}}$, 

\begin{equation}
    \text{T}_{\text{hotspot}} \approx \sqrt{2}\cdot \text{T}_{\text{eq}} \sim \begin{cases}
    1500\text{~K}, & \text{$f/f_{\oplus} \approx 200$ for small planet}.\\
    2000\text{~K}, & \text{$f/f_{\oplus} \approx 800$ for large planet}.
  \end{cases}
\end{equation}


The depth of penetration of dissociation fronts is likely different for small planet versus large planet, and thus, results in different loss rates. 


\subsection{On Puffy and TTV Planets}
\label{sec:PuffyPlanets}

\begin{figure}[!ht]
    \centering
    \includegraphics[width=0.45\textwidth, angle=0]{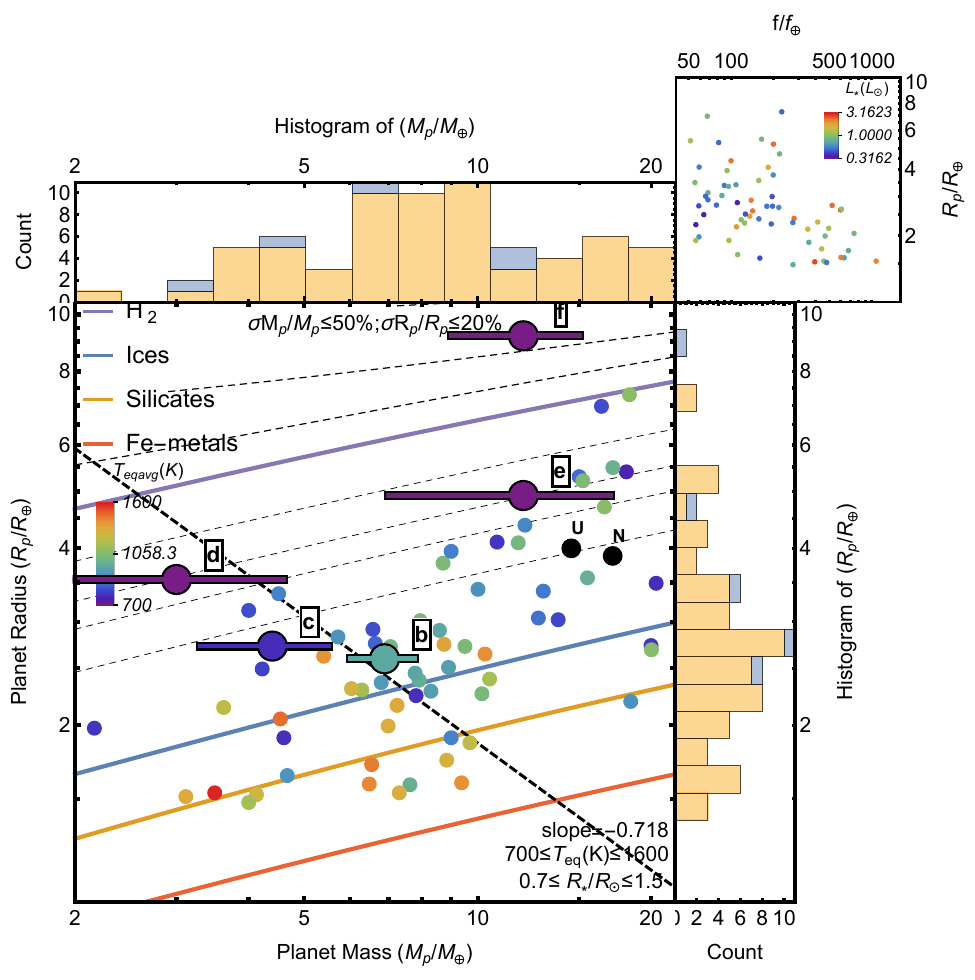}
    \caption{CHILL\#1 for Small Exoplanets: Super-Earths/Mini-Neptunes/Cores with H$_2$-Envelopes. Log-Log Plot in Planet Masses ($M_p$) and Radii ($R_p$), Zoom-out. Here we focus on the higher-flux portion of the planet radius versus flux subplot. A cut-off flux of ($f_{\text{crit}}/f_{\oplus}\approx200$) is identified in the $(f/f_{\oplus})-R_p$ subplot, as measured in the unit of solar constant ($f_{\oplus}=1.36$ kW/m$^2$). Below this cut-off flux ($f_{\text{crit}}$), some planets can retain a low-mean-molecular-weight H$_2$/He-Envelope, which are consistent with their locations in the planet mass-radius main diagram. Above this cut-off flux level ($f_{\text{crit}}$), no planets bigger than $\sim 3R_{\oplus}$ exist. This suggests a likely loss of such low-mean-molecular-weight H$_2$/He-Envelope. An example of a multi-planet system: HIP41378 Planetary System is shown here for comparison~\citep{Santerne2019}. Several planets in this system have received very low flux compared to many other exoplanets shown in this figure, and thus, are expected to retain their envelopes. The equilibrium temperatures of Planet d, e, and f actually lie below the lower limit of the temperature-scale, and thus, they show up as purple altogether. Planet b, c, and d seem to lie along the CHILL\#1 critical for envelope loss.}
    \label{fig:eb1_7}
\end{figure}

Figure~\ref{fig:eb1_7} shows an additional subplot of planet radius versus stellar insolation flux. A critical cut-off flux of about 200 times the equivalent of Earth's insolation flux ($f_{\text{crit}}/f_{\oplus}\approx200$) is clearly identified in this subplot. Beyond this cut-off flux the planets bear no H$_2$-envelope. To make a comparison, this cut-off flux level is equivalent to moving current Earth 15 times closer to the Sun so there is $15^2=225$ times increase in flux. The overall picture of the flux-dependent mass-radius diagram is consistent with a flux-driven H$_2$-envelope escaping scenario. 

\subsubsection{HIP41378 Planetary System}

One multi-planet system (HIP41378) of relatively low insolation flux is shown for comparison. Many planets there bear significant H$_2$-envelope and are consistent with our hypothesis~\citep{Santerne2019}~\citep{Vanderburg2016HIP41378}.

This system is of particular interest.  Its host star is a slightly metal-poor late F-type star. It has been pointed out that metal-poor stars, if their chemistry only reflects the nucleo-synthetic contributions of Big Bang and perhaps the pp chain and carbon cycle, would mostly include the elements representing the early steps of the \emph{alpha-process}: Carbon, Nitrogen, Oxygenn, but not much the next two \emph{alpha-process} elements beyond Neon: Magnesium and Silicon which are important rock-building elements. Thus, the planetary systems formed around such metal-poor stars may only contain gas-giant and icy planets~\citep[p.~105]{Lewis2004PhysicsSystem}.

This idea seems to be applicable to many other Transiting-Timing-Variation (TTV) planetary systems as well. These systems overlap greatly with the so-called compact multi-planet systems which often feature adjacent planets near 2:1 mean motion resonance or other small integer mean motion resonances, in order to allow significant TTV effect to be detectable. It has been noted that there is a positive correlation between the occurrence of such TTV systems (sometimes also called compact multi-planet systems) with lower host-stellar metallicity ([Fe/H])~\citep{Brewer2018CompactHosts}. 

If the planets in such a compact multi-planet system are H$_2$-gas-rich or used to be H$_2$-gas-rich, and puffy, and are or have been experiencing significant mass loss through H$_2$-envelope escape, then we may explain their compact orbital configuration as orbital evolution due to mass loss. In such models, significant mass loss often leads to orbital decay/drift, or the shrinkage of orbital size~\citep{Thom2012}~\citep{Kurokawa2014}, that is, the decrease of orbital semi-major axis ($a$). If this decay happens to all planets in the same system, then as planets' orbits evolve and become more compact together, it would be natural for adjacent ones to be caught in mean motion resonance.

\subsubsection{Kepler-29 Planetary System}

Another prominent example which happens to be shown in Figure~\ref{fig:eb1_7} is the Kepler-29 system. This system features two adjacent planets in 7:9 mean motion resonance (very adjacent indeed, with one planet orbiting the host star at 0.09 AU and the other orbiting at 0.11 AU!)~\citep{Borucki2011}~\citep{Fabrycky2012}. They (K-29b and K-29c) are the two un-labelled dots in between planet d \& planet c of HIP41378 in Figure~\ref{fig:eb1_7} with slightly different colors. They are of similar masses and radii and densities, and from the mass-radius relations they possess a small but non-trivial (10\%$\sim$20\% by mass fraction) amount of H$_2$-envelope.

It seems that they (HIP41378 b,c,\&d, K-29b\&c), along with some other exoplanets, all lie roughly along this mass-radius line with negative slope in log-log plot. It has a slope of $\sim$ (-2/3) in lgM-lgR space. It indicates that M$_p^2 \times$ R$_p^3$ $\sim$ const, or equivalently, M$_p^{2/3} \times$ R$_p$ $\sim$ const. Moreover, this line serves as a dividing line between two groups of exoplanets on the mass-radius diagram: Group A to the left which are hotter and poorer in H$_2$ envelope; and Group B to the right which are colder and richer in H$_2$ envelope. This dividing line is also evident in Figure~\ref{fig:eb1_1} and Figure~\ref{fig:eb1_2}, and there it shows up as a curve. Therefore, we hypothesize it as a boundary separating the planets with H$_2$-envelope-largely-retained versus H$_2$-envelope-lost over billion-year timescale orbiting sun-like Main-Sequence star. Since we observe mostly mature planet systems, their ages are generally similar to that of our own solar system in terms of the order-of-magnitude.

Then, a natural explanation for the observed bi-modal exoplanet radius is that there exist generally two classes of planet cores: rocky versus icy, and the evaporation of the primordial H$_2$ help to reveal their true nature to us the observers. This returns to the argument that metal-poor stars which lack the elements beyond Neon would only be capable of forming icy cores (made primarily of elements Carbon, Oxygen, Nitrogen, plus some Hydrogen) and gas giants, while the stars which have enough Magnesium, Silicon, and/or iron, would be capable of forming rocky cores~\citep[p.~105]{Lewis2004PhysicsSystem}, and on top of that forming icy cores and gas giants also. This idea can be tested in future when more precise measurements of host-stellar elemental abundances become available. 

\subsection{On Smaller Host-Stars and Younger Age}\label{sec:SmallerHosts}

\subsubsection{TOI-560 (HD 73583) Planetary System}

It is worthwhile to extend this discussion to younger planetary systems ($\lesssim 1$Gyr), as well as systems with colder-and-smaller host-stars ($\lesssim$0.7R$_{\odot}$). We expect that the factors of (1) younger age and (2) smaller host stellar size act in a similar way on planets: (1) Younger planetary systems would tend to harbor more volatile-rich and puffier planets at a given host-stellar flux level since there is not enough time yet for the volatiles to completely escape from the planets; Similarly, (2) Planets around smaller host stars tend to retain more volatile envelopes at a given age since the total luminosity of the host star is much less, and the energy of each individual photon emitted by the host star is less.

Recent observations reveal that a young ($\sim500-600$ Myr) mini-Neptune HD 73583 b (TOI 560.01)~\citep{Barragan2021, Zhang2022, Zhang2023} orbiting a K-dwarf star is currently losing its gas in a form of an outflow at velocity comparable in scale to its escape velocity with a mass-loss rate estimated to be $\sim0.2$M$_{\oplus}$ Gyr$^{-1}$~\citep{Zhang2022}. If such outflow is confirmed as the main physical mechanism of atmospheric loss on other exoplanets, then it can be thought of as a nozzle which drains the H$_2$-envelope over age.

\begin{figure}[!ht]
    \centering
    \includegraphics[width=0.45\textwidth, angle=0]{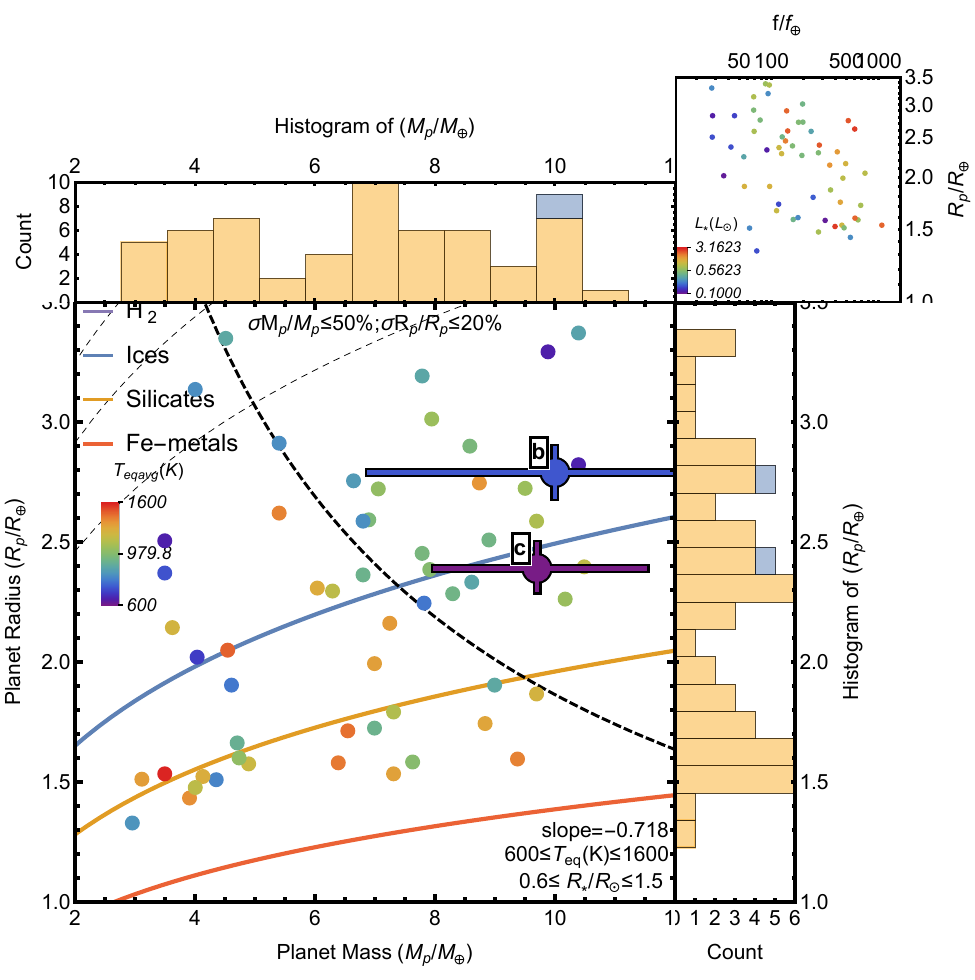}
    \caption{CHILL\#1 for Small Exoplanets: Super-Earths/Mini-Neptunes/Cores with H$_2$-Envelopes. Linear-Linear Plot in Planet Masses ($M_p$) and Radii ($R_p$), Zoom-in. Similar to Figure~\ref{fig:eb1_2} but with slightly different and wider parameter' selection reaching down to lower mass, lower radius (lower limit from 0.7 R$_{\odot}$ in Figure~\ref{fig:eb1_2} down to 0.6 R$_{\odot}$ in this plot), and less luminous host stars; and also colder and less-irradiated exoplanets; and also allowing bigger errorbars in planet radius. One prominent example of a twin-planet system: HD73583 b and HD73583 c are shown~\citep{Barragan2021}. HD73583 b has been studied in detail from ground-based observation to show an atmospheric outflow directed its host-star~\citep{Zhang2022} at high velocity. A subplot of planet-radius-versus-flux is shown on the upper right corner, with host stellar luminosity as colorcoding, whereas planet HD73583 c is too cold and receives too little flux ($f_{\text{HD73583c}}/f_{\oplus}\approx10$) so it lies outside the range of this subplot. Planet HD73583 b receives about 50 times Earth's insolation ($f_{\text{HD73583b}}/f_{\oplus}\approx43$) and lies in the upper left region of the subplot. Again, according to the $R_p$ histogram on the right-hand-side, there is a possibility to further sub-divide the planets which are in between 2-3 $R_{\oplus}$ into two sub-populations. Also, smaller host-stars (0.6-0.7 R$_{\odot}$) seem to host more volatile-rich planets, because the host stellar bolometric luminosity decreases rapidly with its mass according to the mass-luminosity relationship ($L_{\star} \propto M_{\star}^{3.5}$) and also each photon emitted is less energetic on average. Although, there might be observational biases involved. Thus, the CHILL\#1 may be blurred somewhat due to less-irradiated planets around smaller colder host-stars.}
    \label{fig:eb1_3}
\end{figure}

\begin{figure}[!ht]
    \centering
    \includegraphics[width=0.45\textwidth, angle=0]{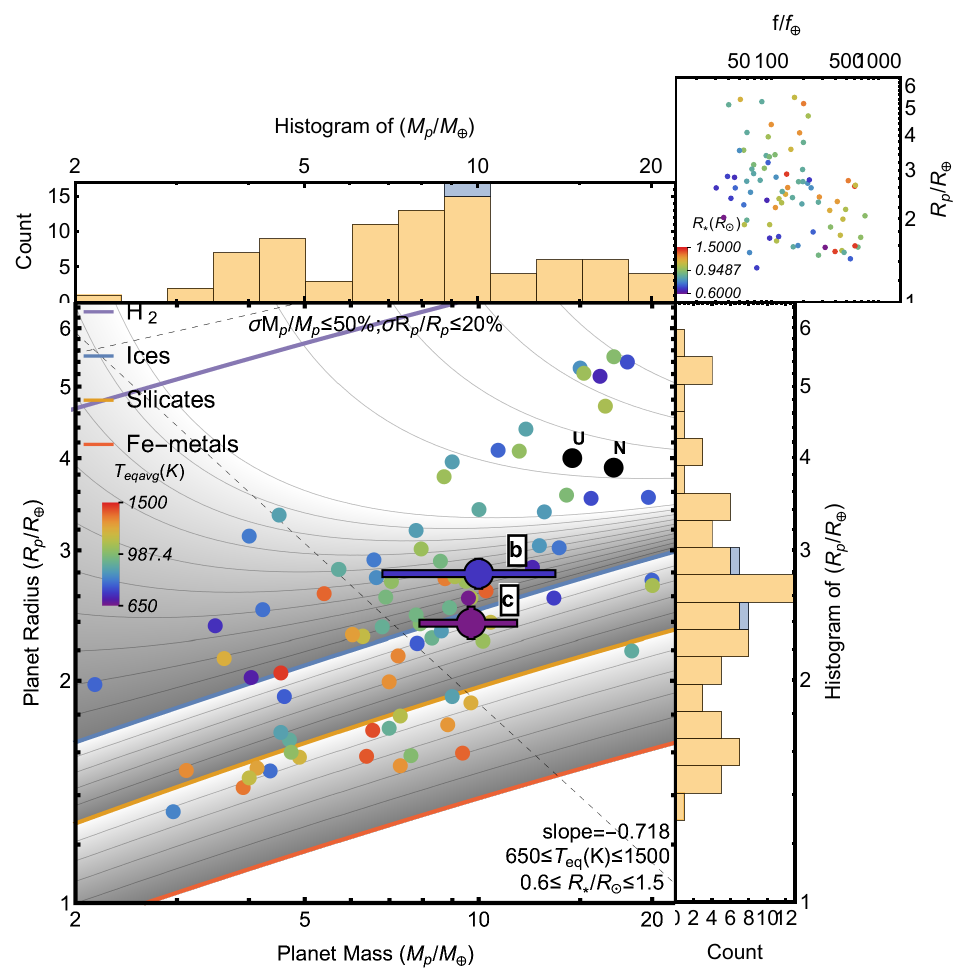}
    \caption{CHILL\#1 for Small Exoplanets: Super-Earths/Mini-Neptunes/Cores with Envelopes. Log-Log Plot in Planet Masses ($M_p$) and Radii ($R_p$), Zoom-out. Similar to Figure~\ref{fig:eb3_1} but zoom out in planet mass and radius. Again, HD73583 b and HD73583 c are shown~\citep{Barragan2021}. Again, a cut-off flux of ($f_{\text{crit}}/f_{\oplus}\approx200$) is identified in the $(f/f_{\oplus})-R_p$ subplot on the upper-right corner. Above this cut-off flux level, no planets bigger than $\sim 3R_{\oplus}$ exist. Dashed line in M-R plot corresponds to  $(M_p/M_{\oplus})^2 \times (R_p/R_{\oplus})^3 \sim 800$, or alternatively, $(M_p/10M_{\oplus})^2 \times (R_p/2R_{\oplus})^3 \sim 1$. It can be viewed as the boundary dividing two groups of exoplanets: Group A to the left, which are hotter and poorer in H$_2$ envelope, but with some diversity; and Group B to the right which are uniformly colder and richer in volatile including H$_2$. Shaded region above pure ices mass-radius curve shows contours of $z$ from 0 to 1 in step of 0.1.}
    \label{fig:eb1_4}
\end{figure}

Thus, we decide to make additional plots (see Figure~\ref{fig:eb1_3} and Figure~\ref{fig:eb1_4}) to include this particular system, and also, include other systems around slightly smaller host stars by lowering radius threshold from earlier plots (lower limit from 0.7 R$_{\odot}$ in Figure~\ref{fig:eb1_2} down to 0.6 R$_{\odot}$ in this plot), to see the effect. As expected, the CHILL\#1 is somewhat blurred but is still existent, if one counts the ratio of hot (orange) planets versus cold (blue-green) planets on either side of this boundary--CHILL\#1. Moreover, there seems to be a pile-up of planets along this boundary. Thus, we may think of planets not as fixed points on the mass-radius diagram, but rather, as slowly (or occasionally fast) moving dots tracking along each of their evolutionary trajectory. One possibility is that the CHILL\#1 is not impenetrable, but can be regarded as a \emph{stagnation line} where the evolving planet stagnate there for a while before moving further along its evolutionary trajectory to lose envelope entirely. This feature needs to be explored in detail in future.

To further investigate the nature of these planets, we invoke the contours of dimensionless parameter $z$ in mass-radius plot (Figure.~\ref{fig:eb1_4}) as introduced in \citep{Zeng2021}, and calculate them \emph{only} for an atmosphere instead of an envelope:

\begin{equation}
    z \equiv  \frac{ \left[ \int \frac{dP}{\rho} \right] }{\left( \frac{G \cdot M_{\oplus}}{R_{\oplus}} \right)}
\end{equation}

The numerator corresponds to an integral which runs from the bottom of the atmosphere to the top. In practicality, without loss of generality, we can invoke the ideal gas law for atmosphere and show that: 

\begin{equation}
    \begin{split}
    \left[ \int \frac{dP}{\rho} \right] &=  \frac{\text{R}T}{\overline{\mu}} \cdot \ln{ \frac{\overbrace{P_{\text{bottom}}}^{\text{$\sim10^8$Pa }}}{\underbrace{P_{\text{top}}}_{\text{$\sim10^2$Pa}}} } \\ &= \frac{\text{R}T}{\overline{\mu}} \cdot \ln{ \frac{\overbrace{n_{\text{bottom}}}^{\text{defined by fluid density:$\sim 10^{22}$cm$^{-3}$}}}{\underbrace{n_{\text{top}}}_{\text{defined by depth transit obs. probes:$\sim 10^{16}$cm$^{-3}$}}} }
    \end{split}
\end{equation}

$n_{\text{bottom}}$ is defined by the fluid density where the inter-molecular distance becomes comparable to the dimension (van der Waal radius) of the molecule itself so inter-molecular interaction becomes strong. Any number density above this threshold level should be considered as belonging to an envelope instead of an atmosphere. This typically corresponds to a pressure level of 0.1$\sim$1 GPa ($10^8 \sim 10^9$ Pa). This level corresponds to a number density of dense super-critical fluid ($n\sim 10^{22}$cm$^{-3}$). Any number density above this threshold level H$_2$ molecules are squeezed strongly against each other in the dense fluid and is considered an envelope instead of an atmosphere. 

$n_{\text{top}}$ is defined by the number density level where transit observation probes by a grazing ray passing through the planetary atmosphere. Depending on the composition and scale height of the atmosphere, it is typically of the pressure level of 1$\sim$10 millibar ($10^2 \sim 10^3$ Pa). This level corresponds to the number density of von Karman line previously defined ($n\sim 10^{16}$cm$^{-3}$). Thus, the von Karman line can be viewed as the edge of the atmosphere of the planet or the boundary between the planet atmosphere and the outer space. Any number density below this threshold level the stellar photon can mostly pass right through without much absorption. 

Thus, from the bottom of the atmosphere to the top, the number density is rarefied by roughly one-million-fold. That is, the linear separation between molecules is increased by roughly one-hundred-fold. So,

\begin{equation}
    \begin{split}
    \left[ \int \frac{dP}{\rho} \right] &= \frac{\text{R}T}{\overline{\mu}} \cdot \ln{\left(10^6\right)} = 13.8 \cdot \frac{\text{R}T}{\overline{\mu}} \\ &= 115 \cdot \frac{T}{\overline{\mu}} = 1.15\times10^5 \cdot \frac{T/10^3\text{K}}{\overline{\mu}}
    \end{split}
\end{equation}

The mean molecular weight $\overline{\mu}$ of a cosmic H$_2$-He mixture with heavy component mass fraction $Z$ is:

\begin{equation}
    \overline{\mu} \approx \frac{\mu_{\text{H$_2$-He}}}{(1-Z)} = \frac{2.3 \text{g/mol}}{(1-Z)} = \frac{2.3\times10^{-3} \text{kg/mol}}{(1-Z)}
\end{equation}

Therefore, in S.I. unit:

\begin{equation}
    \begin{split}
    \left[ \int \frac{dP}{\rho} \right] &= \left( \frac{1.15\times10^5}{2.3\times10^{-3}} \right) \cdot \left(\frac{T}{10^3\text{K}}\right) \cdot (1-Z) \\ &= \overbrace{\left( \frac{1}{2}\cdot 10^8 \right)}^{\text{S.I. unit}} \cdot T_3 \cdot (1-Z)
    \end{split}
\end{equation}

On the other hand, the normalization factor in the denominator expressed in S.I. unit is:

\begin{equation}
    \left(\frac{G \cdot M_{\oplus}}{R_{\oplus}} \right) = \underbrace{\left(0.625\times10^8 \right)}_{\text{S.I. unit}}
\end{equation}

Thus, the dimensionless parameter $z$ which characterizes the thickness of planetary atmosphere or envelope is:

\begin{equation}
    \begin{split}
    z \equiv  \frac{ \left[ \int \frac{dP}{\rho} \right]}{\left( \frac{G \cdot M_{\oplus}}{R_{\oplus}} \right)} &= \left( \frac{0.5\times10^8}{0.625\times10^8} \right) \cdot T_3 \cdot (1-Z) \\ &= \underbrace{\left( \frac{1}{1.25} \right)}_{\approx 1} \cdot T_3 \cdot (1-Z)
    \end{split}
\end{equation}

Higher metallicity ($Z$) or Lower Temperature ($T_3$) of the atmosphere will decrease the value of $z$, and thus, will decrease the thickness of atmosphere.

Lower metallicity ($Z$) or Higher Temperature ($T_3$) of the atmosphere will increase the value of $z$, and thus, will increase the thickness of atmosphere.

Therefore, for the range of physical parameters that we are interested in ($T_3\sim1$ and $Z \geqslant 0$), the planets which \emph{only} have an atmosphere but not an envelope should have $z\lesssim1$, because the bottom pressure is limited by $\sim0.1$ GPa. In Figure~\ref{fig:eb1_4}, the gray shaded region immediately above the pure H$_2$O mass-radius curve corresponds to $z$ contours of values from 0 to 1 in increment of 0.1. So the planets lying in this gray shaded region of the mass-radius diagram can in principle be explained by \emph{only} having an extended/puffed atmosphere but not a H$_2$-envelope. 

On the other hand, planets with $z\gtrsim1$ must have a H$_2$-envelope which has a bottom pressure higher than $\sim0.1$ GPa. This includes many planets in the neighborhood of Uranus and Neptune in Figure~\ref{fig:eb1_4}. We can take this argument one-step further and argue that if we take the CHILL\#1 of $(M_p/10M_{\oplus})^2 \times (R_p/2R_{\oplus})^3 \sim 1$ as boundary dividing two groups of exoplanets: Group A to the left, and Group B to the right. Then, all planets of Group A can be explained by possessing only an atmosphere (with various amount of metallicity) but not an envelope. Meanwhile, many planets of Group B certainly or likely possess a H$_2$-envelope. 

We expect that as the escape process occurs on these planets, their atmosphere-envelope will become gradually enriched in heavier components (most likely H$_2$O vapor and H$_2$O super-critical fluid) than H$_2$ and He, and the escape process of hydrogen on them will eventually become \emph{diffusion-limited}. They are like cosmic refinery which selectively keep element Oxygen and its subsidiary compound: water (H$_2$O), while allowing hydrogen to be lost to space.

Therefore, a future theoretical problem to tackle is to understand the escape process of hydrogen in an H$_2$O-dominated atmosphere-envelope system~\citep{Goldblatt2015,Pierrehumbert2023}.

\begin{figure*}
\centering
\begin{subfigure}{1\columnwidth}
    \includegraphics[width=\textwidth]{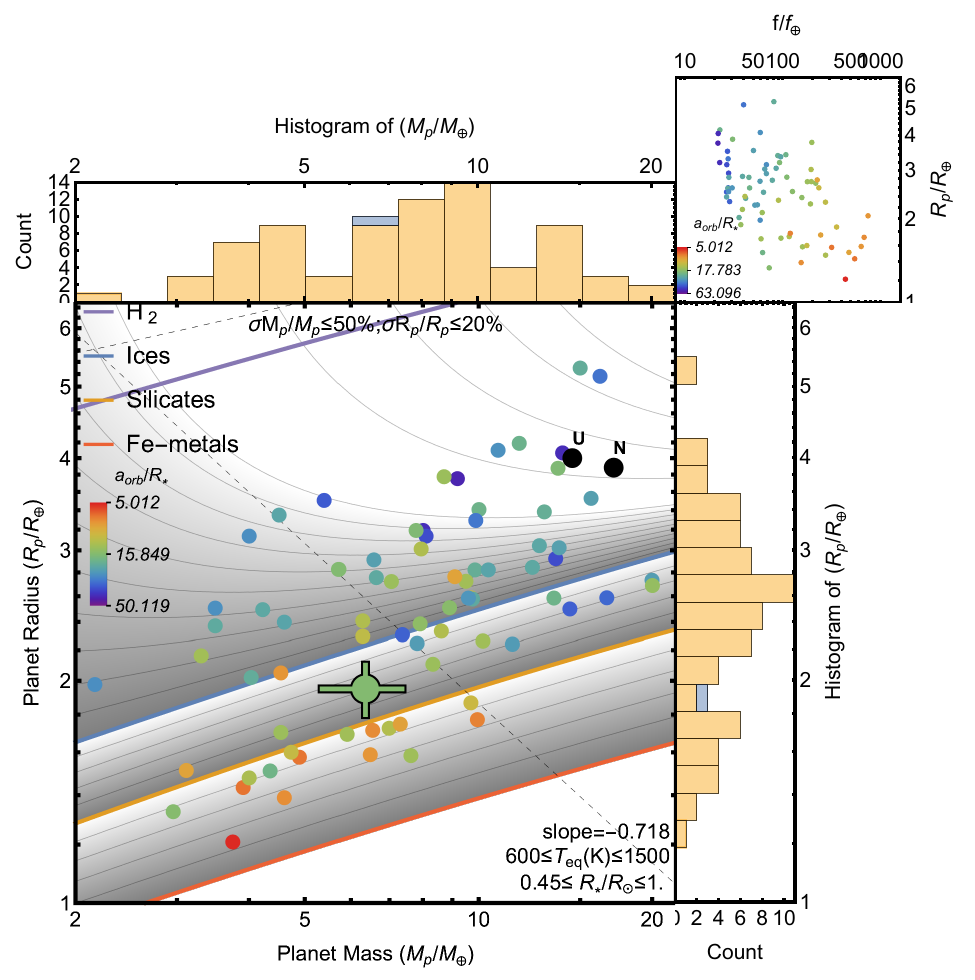}
    \caption{}
    \label{fig2:first}
\end{subfigure}
\hfill
\begin{subfigure}{1\columnwidth}
    \includegraphics[width=\textwidth]{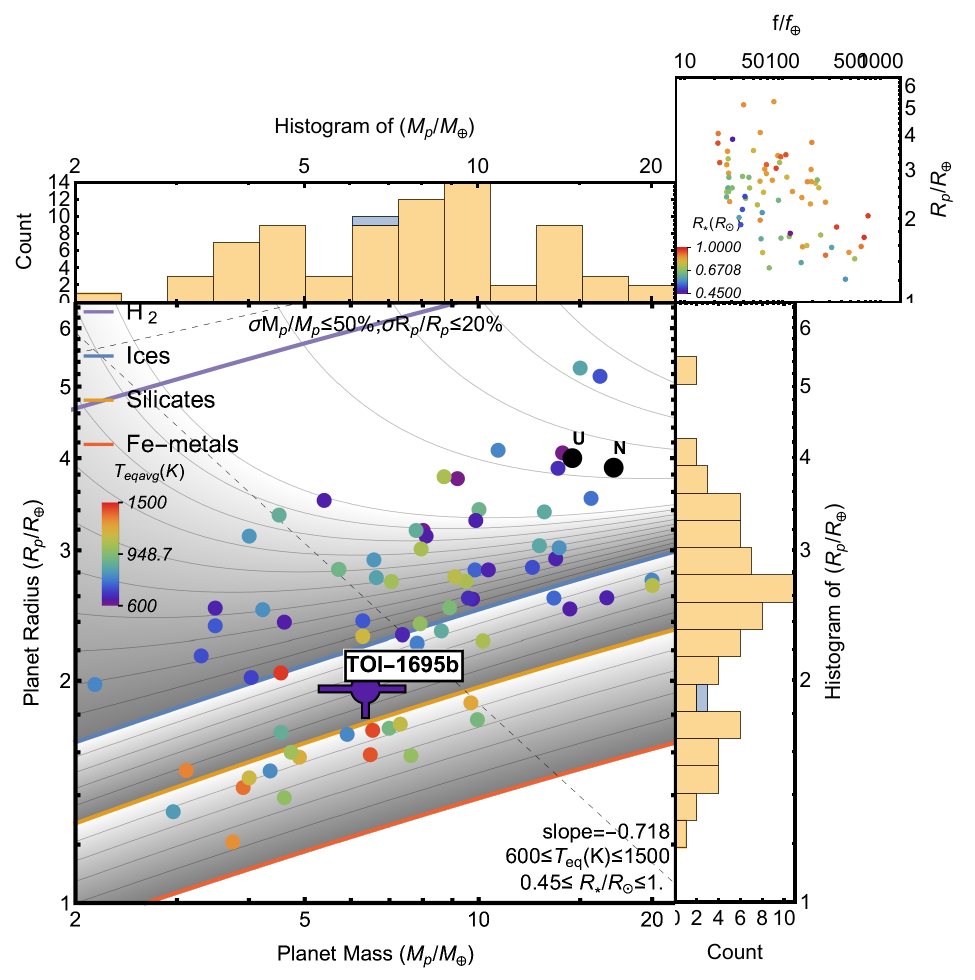}
    \caption{}
    \label{fig2:second}
\end{subfigure}
\hfill
\begin{subfigure}{1\columnwidth}
    \includegraphics[width=\textwidth]{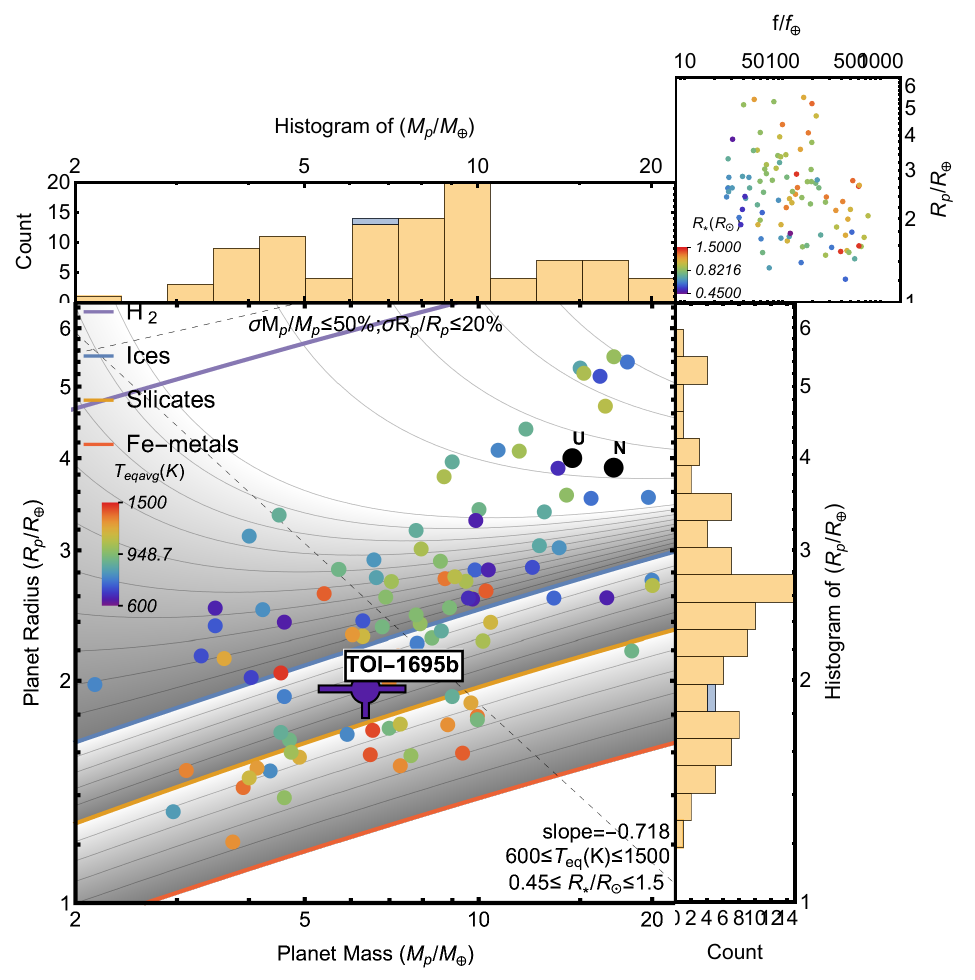}
    \caption{}
    \label{fig2:third}
\end{subfigure}
\hfill
\begin{subfigure}{1\columnwidth}
    \includegraphics[width=\textwidth]{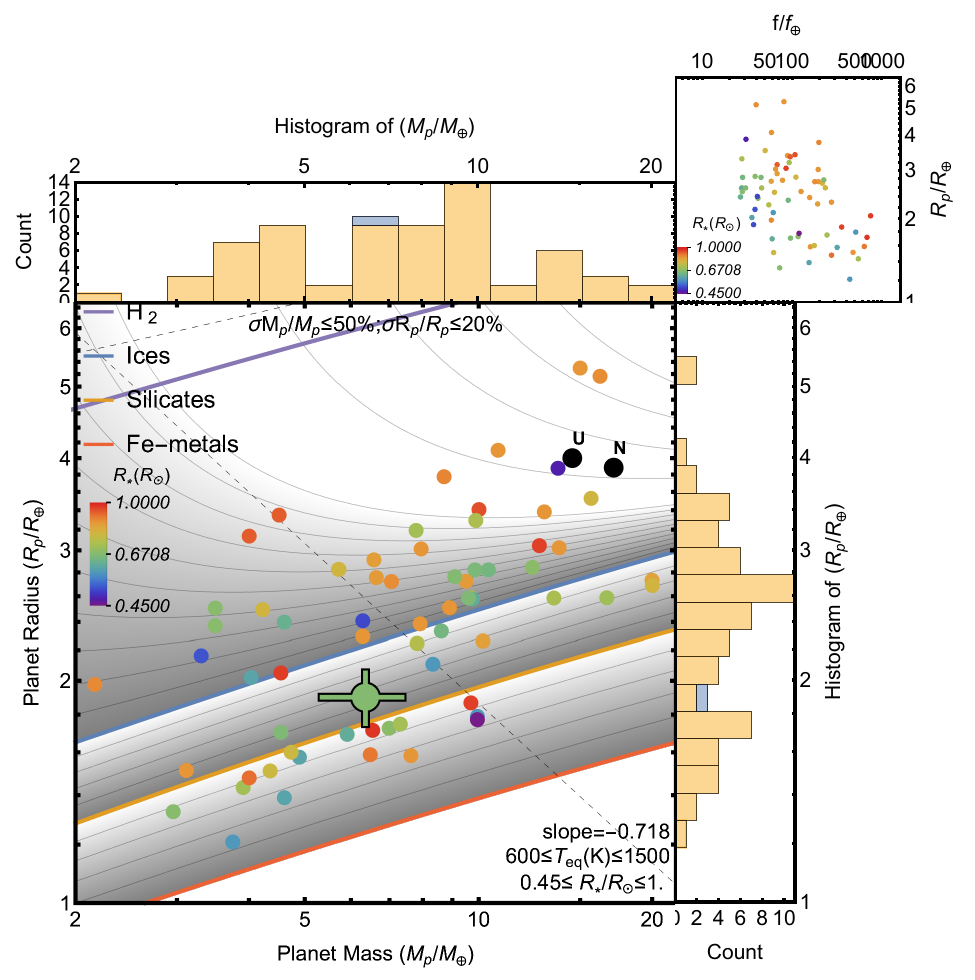}
    \caption{}
    \label{fig2:fourth}
\end{subfigure}

\caption{{\color{black}Mass-radius plots and flux-planet radius subplots including TOI-1695 as an example, reaching down to smaller host stars (0.45 R$_{\oplus}$). First important feature is the contrast between the sequence of rocky planets versus the volatile-rich planets. This contrast is most evident in (a) color-coding using $a_{\text{orb}}/R_{\star}$ and (b)  color-coding using planet globally-average equilibrium temperature T$_{\text{eqavg}}$. Secondly, (c) reveals that CHILL is a boundary of a group of planets of almost uniform temperature (green-colored), while blending with another group of planets which are either hotter or colder in temperature but lie closely with the water-EOS mass-radius curve.  Finally, (d) suggests the abundance of water worlds with almost no envelopes around smaller host stars.}}
\label{fig:TOI1695}
\end{figure*}

{\color{black}
\subsubsection{TOI-1695 Planetary System}
Another prominent example of exoplanet around a cooler and less luminous M-dwarf host-star is the TOI-1695 system~\citep{Cherubim2022, Kiefer2022}. In this particular case, the host star is very slightly larger than 0.5 solar radius ($R_{\odot}$).
Here we provide four plots and sub-plots of the TOI-1695 system illustrating different coloring schemes (Fig.~\ref{fig:TOI1695}). Small host stars such as M-Dwarfs are expected to host a lot of water worlds, because their bolometric luminosity is weak so unlikely to perturb the states of planets from after their formation.} 

\subsection{On More Systematic Comparison of Gas Giant Exoplanets}\label{appendix:GasGiantsSubPlots}

\begin{figure*}
\centering
\begin{subfigure}{0.66\columnwidth}
    \includegraphics[width=\textwidth]{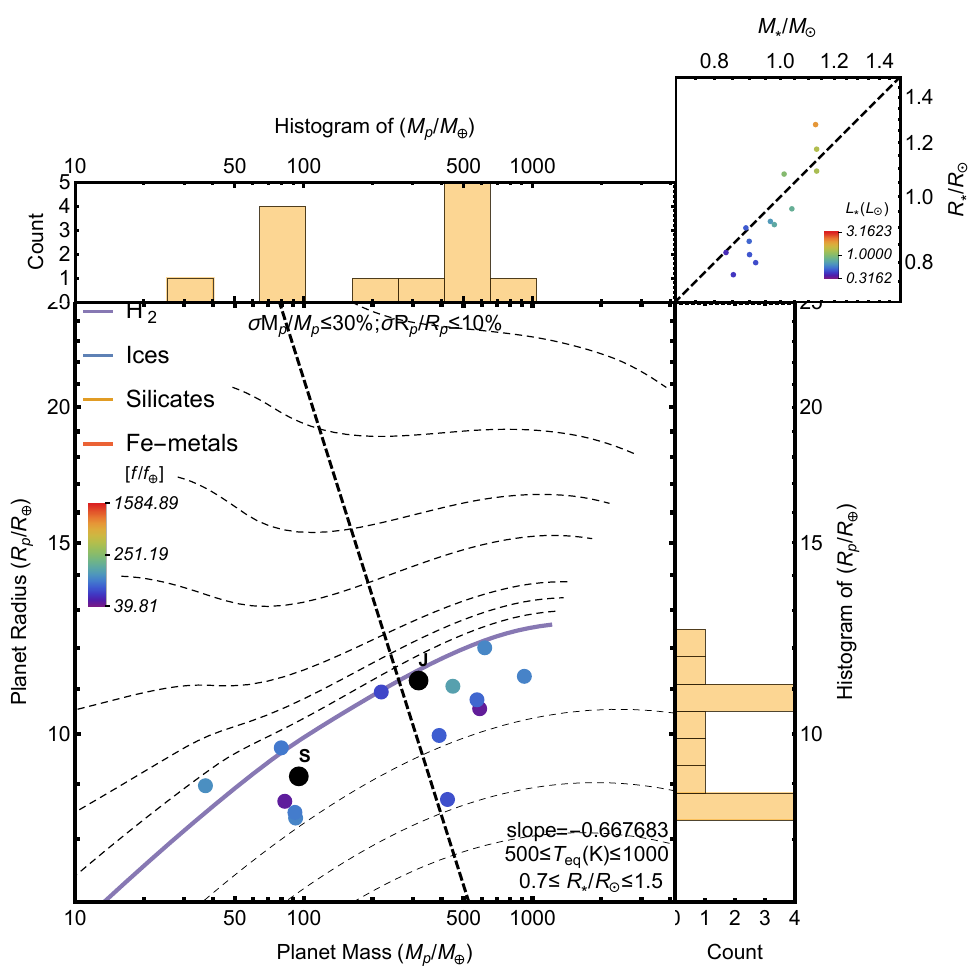}
    \caption{}
\end{subfigure}
\hfill
\begin{subfigure}{0.66\columnwidth}
    \includegraphics[width=\textwidth]{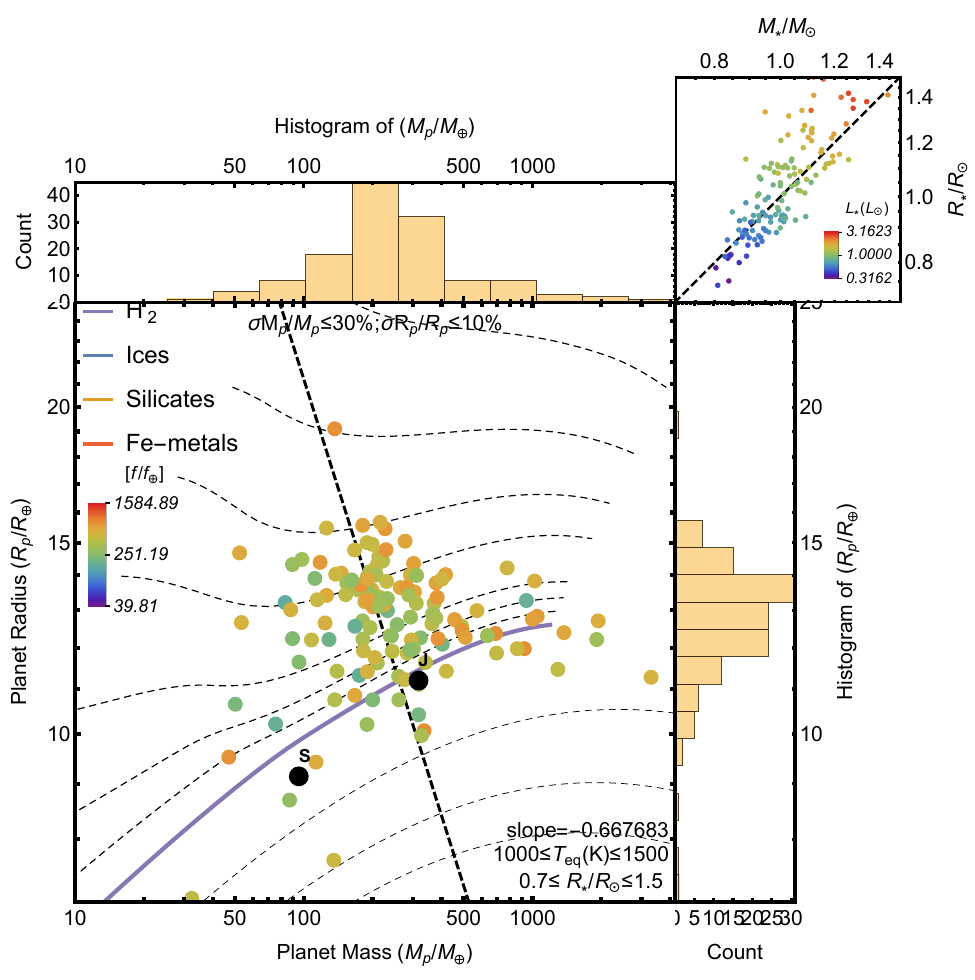}
    \caption{}
\end{subfigure}
\hfill
\begin{subfigure}{0.66\columnwidth}
    \includegraphics[width=\textwidth]{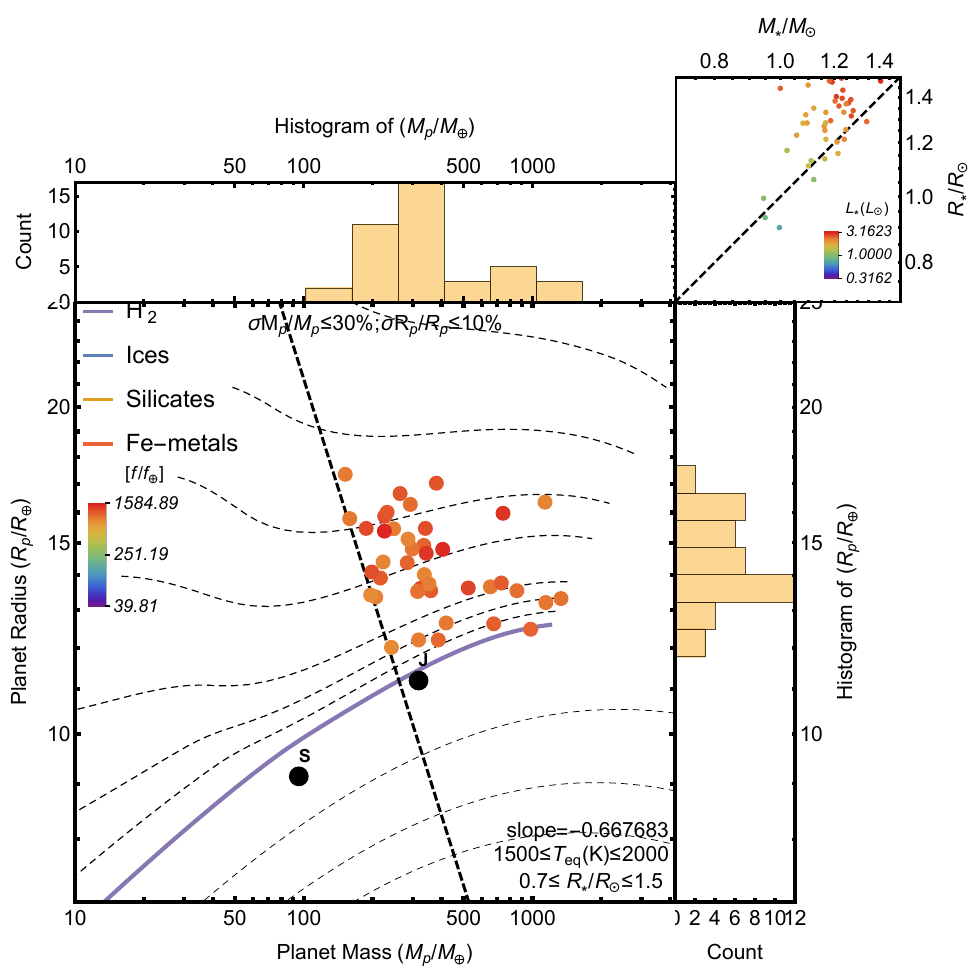}
    \caption{}
\end{subfigure}
\hfill
\begin{subfigure}{0.66\columnwidth}
    \includegraphics[width=\textwidth]{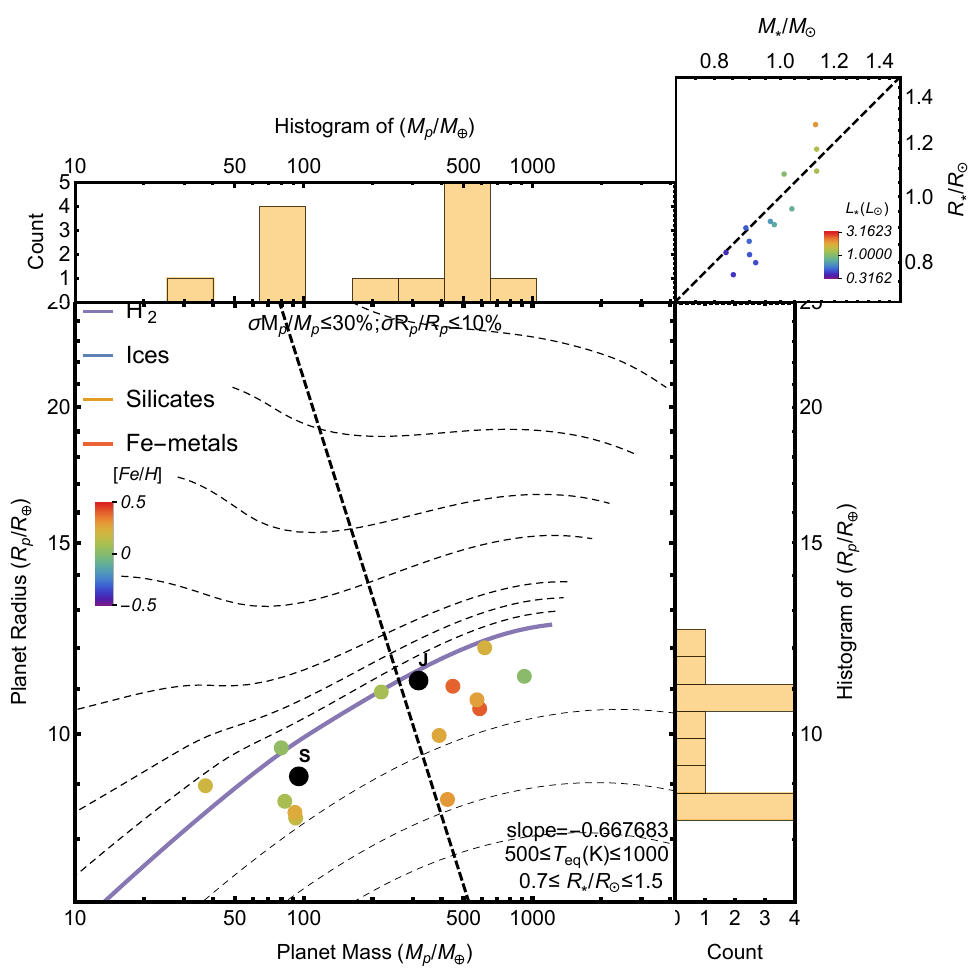}
    \caption{}
\end{subfigure}
\hfill
\begin{subfigure}{0.66\columnwidth}
    \includegraphics[width=\textwidth]{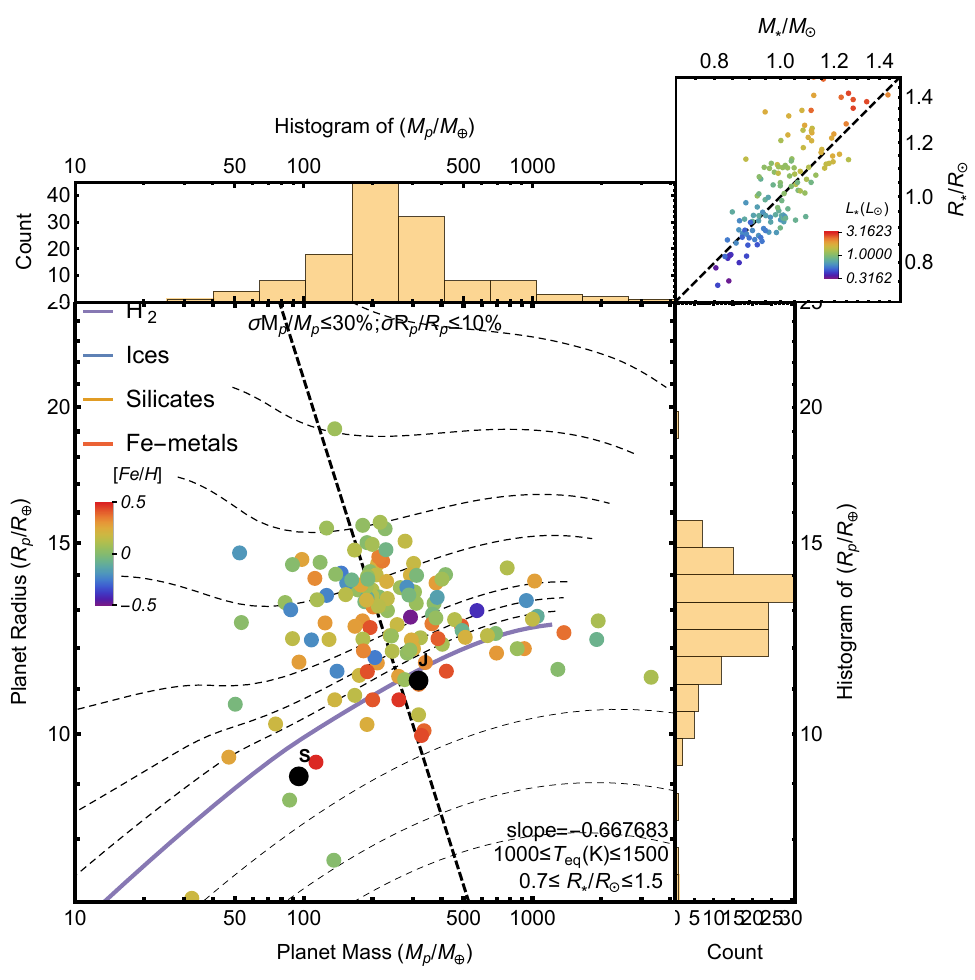}
    \caption{}
\end{subfigure}
\hfill
\begin{subfigure}{0.66\columnwidth}
    \includegraphics[width=\textwidth]{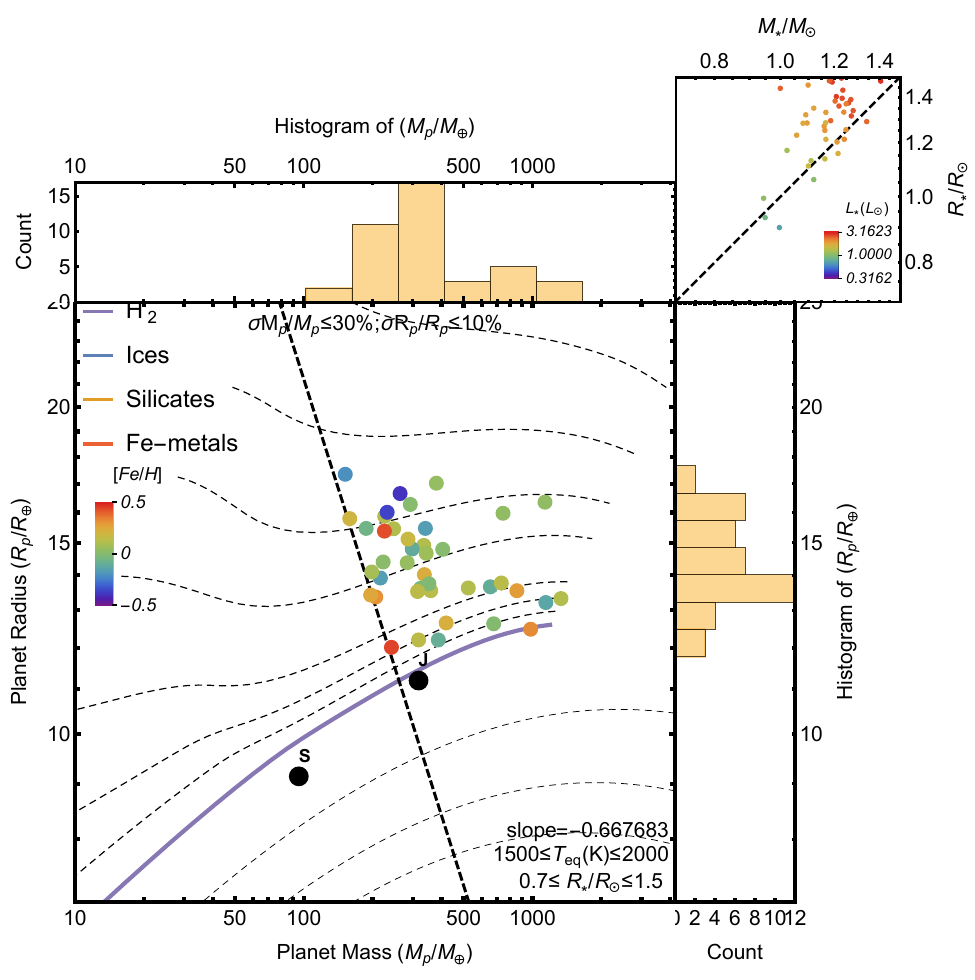}
    \caption{}
\end{subfigure}
\hfill
\begin{subfigure}{0.66\columnwidth}
    \includegraphics[width=\textwidth]{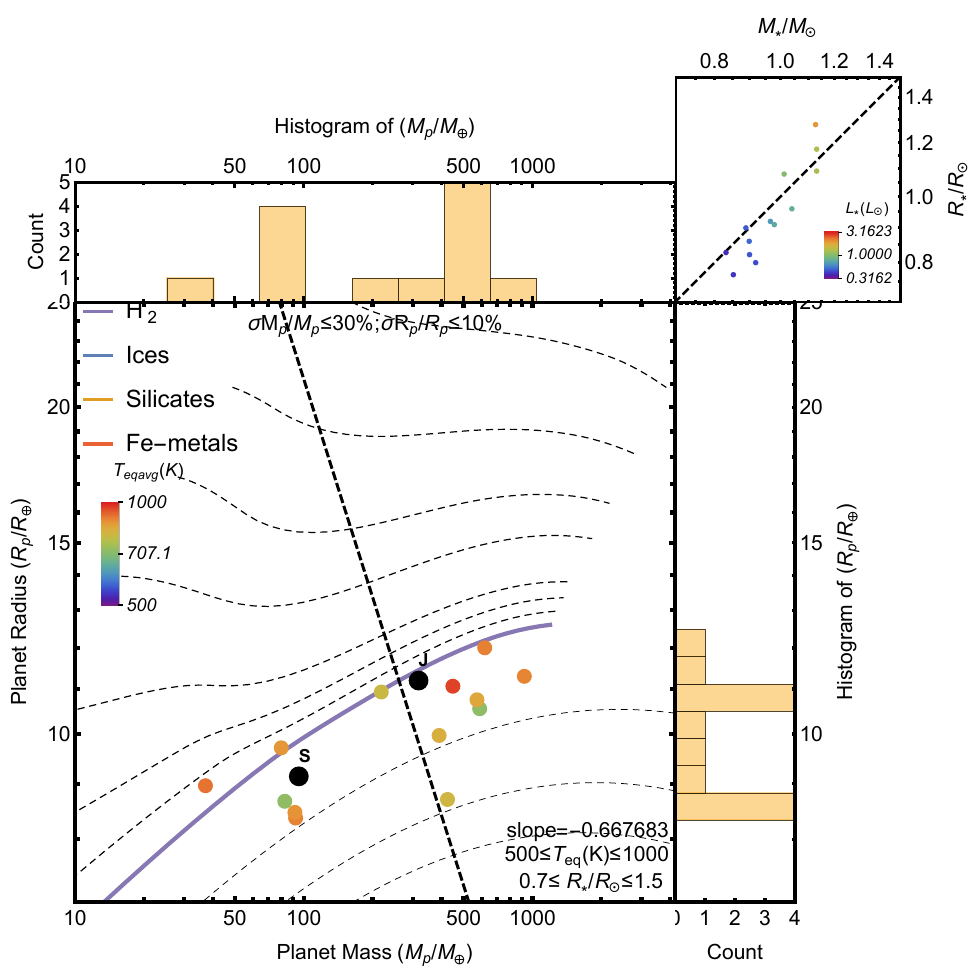}
    \caption{}
\end{subfigure}
\hfill
\begin{subfigure}{0.66\columnwidth}
    \includegraphics[width=\textwidth]{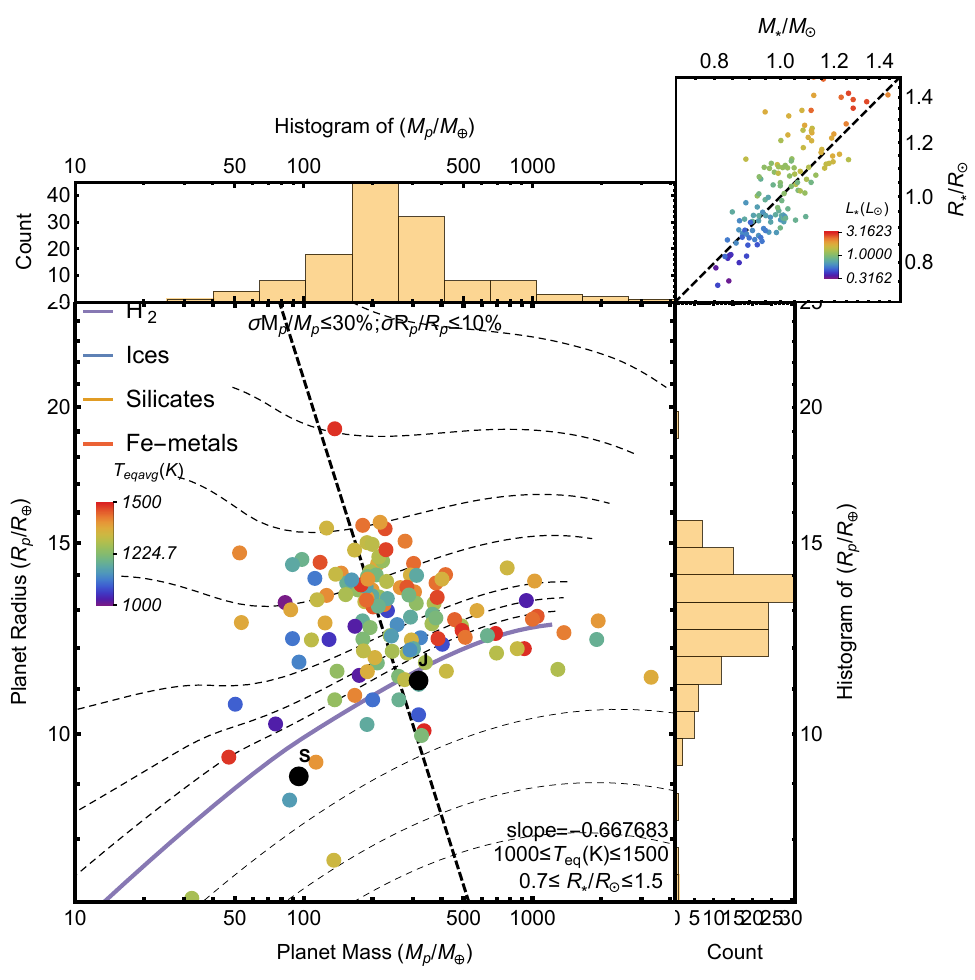}
    \caption{}
\end{subfigure}
\hfill
\begin{subfigure}{0.66\columnwidth}
    \includegraphics[width=\textwidth]{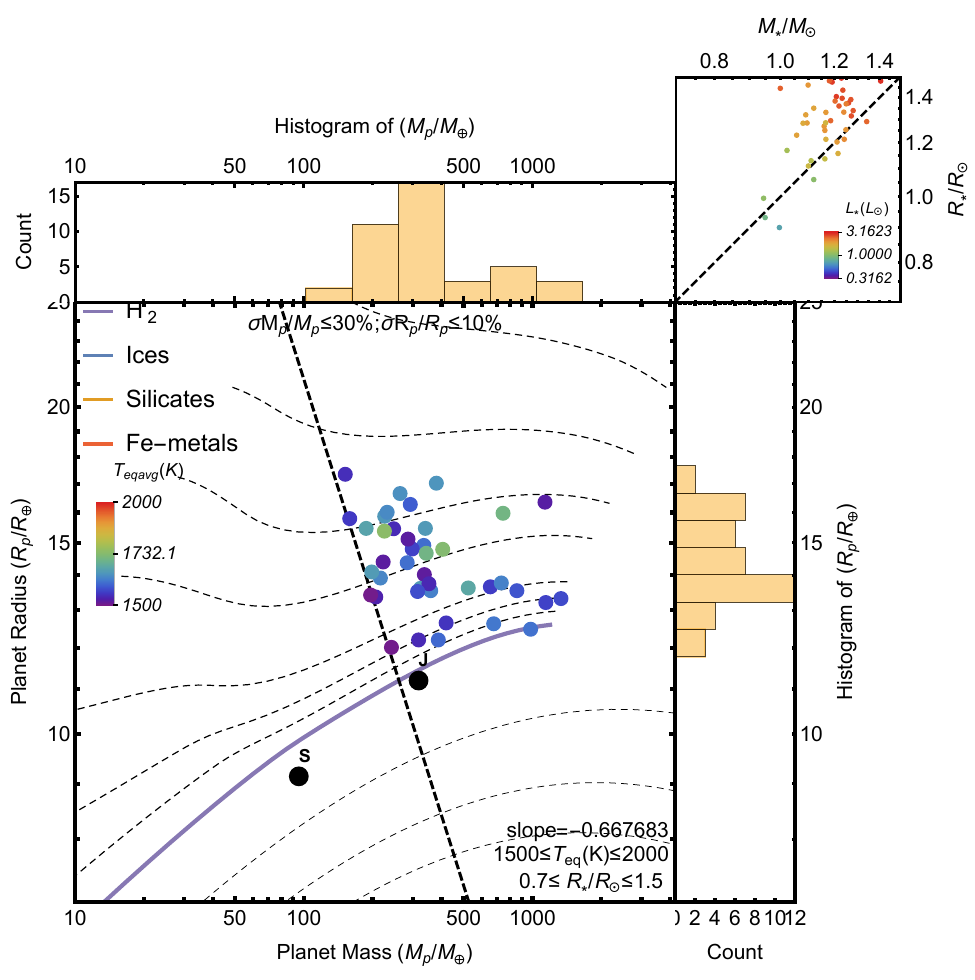}
    \caption{}
\end{subfigure}
\caption{CHILL\#2 for Large Exoplanets: Gas Giants/Hot Jupiters. Each column corresponds to a selected equilibrium temperature range: 500K-1000K, 1000K-1500K, 1500K-2000K. Each row corresponds to a different color-coding scheme: $f/f_{\oplus}$, [Fe/H], T$_{\text{eq}}$. Theoretical mass-curves are given for pure-hydrogen with its variations in both directions: (1) upward-direction with increasing internal specific entropy; (2) downward-direction with progressively mixing with heavier elements/species. Each subplot on the upper-right corner shows the mass-radius diagram of host stars ($M_{\star}$-$R_{\star}$) color-coded by their luminosities ($L_{\star}$) in logarithmic-scale. The hosts are mostly Main-Sequence (MS) stars hotter than M-dwarfs.}
\label{fig:eb2}
\end{figure*}

{\color{black}
The exact physical mechanism of gas giant radius inflation needs to be explored further. However, there is ample observational evidence that this radius inflation correlates directly with the \emph{current} level of bolometric incident flux from the host star. Please see Figure~\ref{fig:eb2} and the captions therein. The observation of the re-inflation of gas giants surrounding evolved red giant host stars supports this point~\citep{Grunblatt2017SeeingStars}.

We speculate that the \emph{hot spot} on such a gas giant planet may also be responsible for its radius inflation, in addition to being the main engine driving the escape processes. The \emph{hot spot} itself may produce an anti-jet (thermal jet) injecting and depositing stellar heat into the deep interior of these planets, and by doing so, increasing their interior specific entropy. The measurement of planetary radius is described in~\ref{sec:OnTransitDepth}.}

\section{Conclusion}

Two Cosmic Hydrogen and Ice Loss Lines (CHILLs \#1 \& \#2) have been identified for small exoplanets and gas giant exoplanets in the mass-radius diagram. The CHILL\#1 corresponds to a critical boundary of planetary equilibrium temperature (T$_{\text{eq}}$) of 1000 K. The CHILL\#2 corresponds to a critical boundary of planetary equilibrium temperature (T$_{\text{eq}}$) of 1500 K.


This negative slope requires further investigation, and calls for the attention of the study of the \emph{planetary-scale rocket engine}. In conclusion, we hypothesize that many planets ($\gtrsim 1$M$_{\oplus}$) have initially acquired various degrees of thick primordial H$_2$-He envelopes from being embedded in gas-rich disks. This primordial H$_2$-He envelope on any close-in planet is then gradually lost over time which is powered by the host-stellar irradiation. {\color{black}A viable} escape mechanism is the {\color{black}\emph{shock-front-driven}} hydrodynamic outflow of neutral hydrogen atoms which are dissociated from hydrogen molecules (H-H) and other hydrogen-bearing molecules (X-H) such as the gaseous form of cosmic ices including water, ammonia, and methane, {\color{black}due to the absorption of host-stellar photons and subsequent heating, molecular dissociation, and acceleration}. 

{\color{black}Another interesting topic to explore is the possible formation scenarios of these ice-rich/water-rich super-Earths around their host stars. This must suggest different disk properties and volatile distributions within those disks, as compared to the pre-solar disk. The pre-solar disk may be very special due to the formation and presence of the gas giant proto-Jupiter at around 5 AU, which had depleted the inner disk of materials and volatiles, in particular, water ices significantly. }

{\color{black}\emph{Marvelous is our Universe, where photons generated by hydrogen fusion inside stars, ultimately drive the hydrogen dissociation and escape on the planets which orbit them!}}




\section{Acknowledgement}
The author LZ would like to thank Robin Wordsworth, Junjie Dong, Collin Cherubim, Kaitlyn Loftus, Michail I. Petaev, Joe Gonzeles, and James G. Anderson of Harvard University Department of Earth and Planetary Sciences for fruitful discussions. {\color{black}The authors acknowledge the support from the DOE-NNSA grant DE-NA0004084 awarded to Harvard University: Z Fundamental Science Program (PI: SBJ).}


\clearpage
\appendix

\section{Atmospheric Layering}
\label{layering}

{\color{black}
Under the influence of planet gravity field, a planetary atmosphere typically exhibits layering structure. This atmospheric layering induced by gravitational potential as well as material property of gas will have strong influence when we consider the physics of the escape processes occurring on these planets. For simplicity, let us first consider the thermal escape process (see~\ref{appendix:JeansEscape} and~\ref{appendix:HydrodynamicEscape}) {\color{black}under the relatively low host-stellar photon flux level}. 

The thermal escape process is typically modelled by two limiting scenarios, (i) Jeans escape (see~\ref{appendix:JeansEscape}), and (ii) Hydrodynamic escape (see~\ref{appendix:HydrodynamicEscape})~\citep{Jeans1925, Chapman1939, Chapman2000, Cercignani2000, Parker1958, Hunten1974, Hunten1971, Hunten1973, Hunten1993, Hunten2002, Chamberlain1989, Walker1977, Goody1972, Goody1995, Anderson1975, Pudovkin1983, Zahnle1986, Zahnle2017TheB, Catling2017EscapeSpace, Lehmer2017ThePlanets, Owen2018AtmosphericExoplanets, Gronoff2020, Mierkiewicz2021, Kang2021}. For Scenario (i) Jeans escape, the atom or molecule escapes mostly by its own (random) thermal velocity ($v_{\text{random}}$) from the most tenuous part of the upper atmosphere (virtually collisionless), with typically a near-Maxwellian velocity distribution assumed for the purpose of calculation. For Scenario (ii) Hydrodynamic escape, the atoms or molecules escape collectively as a bulk outflow characterized by an altitude-dependent bulk outflow (mean) velocity ($\overline{v}$) sourced from a certain region in the atmosphere, with the flow itself driven by an \emph{effective pressure gradient}, powered by certain physical mechanism such as the host-stellar irradiation.

In reality, both Scenario (i) escape due to the random thermal velocity ($v_{\text{random}}$) and Scenario (ii) escape due to the average group flow velocity ($\overline{v}$), should occur simultaneously. Thus, they are merely two limiting cases of the same thermally-driven escape process. Collectively speaking, they are both considered thermal escape processes. Therefore, it is important to identify the \emph{source region} and the \emph{heating mechanism} of this thermal escape process. 

As mentioned before, due to the planet surface gravity and the compressibility of the atmospheric gas species, the planet atmosphere typically exhibits a multi-layered structure with different critical number densities at different altitude levels in the atmosphere, which correspond to different physics operating at each level. This number density can vary across many orders of magnitude across the thickness of the planetary atmosphere. It is also important to view the escape process in the context of this multi-layered atmosphere structure. According to the literature and previous works by other researchers, we tentatively identify four important layers in the planet atmosphere as follows: (1) the von Karman line, (2) the exobase, (3) the transonic front, and (4) the H-H dissociation front.

(1) The first critical level in the atmosphere is the von Karman line~\citep{VonKarman1967,McDowell2018}, or the physical edge of the planetary atmosphere. It is also a critical level below which the atmosphere is well-mixed~\citep{Wallace2006}. At this altitude level the molecular and eddy diffusion coefficients are approximately equal. Below this level the mean free path of molecular collisions becomes small enough ($\lambda\lesssim 1$ cm) to allow effective mixing of different gas species so that they all have the same scale height, which is calculated by the average molecular mass of the mixture at a given temperature. Above this level, each gas species then assumes its own scale height according to its own molecular weight. For example, this level is about $\sim$80 km height above sea-level in the Earth's atmosphere. It is also closely related with the \emph{homopause}, \emph{turbopause}, or \emph{mesopause} (boundary between mesosphere and thermosphere). It is considered as the physical edge of the atmosphere and defines its boundary with the outer space.

Below the von Karman line, $\lambda$ is short enough so that each atmospheric particle may experience many collisions in its neighborhood before transporting to other levels. Thus, Local Thermal Equilibrium (LTE) can be safely assumed~\citep[]{Goody1995}. Here $n$ is the number density (cm$^{-3}$), $\sigma$ is the molecular collisional cross-section (cm$^2$), and $\lambda$ is the mean free path for molecular collision (cm), 

\begin{equation}
1\text{~cm} \sim \lambda_{\text{von Karman line}} \sim \frac{1}{n_{\text{von Karman line}} \cdot \sigma}
\end{equation}\label{eq1:1}

The collisional cross-section $\sigma$ is typically the atomic or molecular dimension ($\sim 1~\AA$) squared:

\begin{equation}
\sigma \sim (1\AA)^2 \sim 10^{-16} \text{~cm}, \text{~so that~} n_{\text{von Karman line}}\sim10^{16}\text{~cm}^{-3}.
\end{equation}\label{eq1:2}

{\color{black} Generally speaking, there are two types of collisional cross-sections to consider: low-energy elastic collisions versus energetic inelastic collisions. So that the collisional cross-section is energy-dependent. Thus, the cross-section for energetic in-elastic collision required for molecular dissociation may be one order or even two orders of magnitude smaller than the low-energy elastic collisional cross-section~\citep{LandauLifshitz1981}.


In Section~\ref{sec:H2EOS}, we explain that the H$_2$ molecule are "squeezy", in fact, any molecules are "squeezy". They would appear to have different collisional cross-sections under different energy transfer situations~\citep{Landau1932a,Landau1932b}.

The typical van der waal radius for the low-energy "elastic" collisional cross-sections (two molecules just gently touch each other and bounce away) is usually a few times larger, giving rise to the so-called \emph{kinetic cross-section}, being about 10-100 times larger, than the \emph{reaction cross-section} for inelastic collisions when the energy exchange is so great (on the order of a few electron-volts in order to rip a chemical bond).

For our purpose, we need high-energy "inelastic collisions" that can deposit the kinetic energy into the vibrational mode, that is the stretching of the H-H (hydrogen) molecule or the H-O-H (water) molecule and eventually break the X-H covalent bond. We need "head-on" collisions, not the "sideway" collisions. High-energy "head-on" "inelastic" collisional cross-section is smaller compared to that calculated from the gentler low-energy "sideway" collisional cross-section from the van der waal radius. On the other hand, sideway collisions are capable of maintaining the rotational mode of (H-H) molecule in equilibrium with its translational mode down to much lower number density.}


(2) The second critical layer in the upper atmosphere is the exobase (p$_{\text{exob}}$) where the mean free path (MFP or $\lambda$) of atom/molecule becomes so large as comparable to the local atmospheric scale height ($\mathcal{H}\approx10^{2}\sim10^{3}$km) of the upper atmosphere, which is of course temperature-dependent and species-dependent. There the atmosphere becomes extremely tenuous and virtually collisionless~\citep[]{Goody1995}, 

\begin{equation}
\mathcal{H} \sim \lambda_{\text{exobase}} \sim \frac{1}{n_{\text{exobase}} \cdot \sigma}
\end{equation}\label{eq1:3}

Recall the definition of the molecular or atomic collisional cross-section $\sigma$:

\begin{equation}
\sigma \sim (1\AA)^2 \sim 10^{-16} \text{~cm}, \text{~so that~} n_{\text{exobase}}\sim10^{8}-10^{9}\text{~cm}^{-3}.
\end{equation}\label{eq1:4}

(3) The third critical layer in the upper atmosphere is the sonic level (r$_c$) or transonic front where the outward escaping flow transitions from \emph{sub-sonic} to \emph{super-sonic}, which is a characteristic of the flow enforced by the continuity equation. Let's define $u_0 \equiv \sqrt{k_B T /\mu_{\text{H}}}$ as the local isothermal sound speed for the gas species (neutral atomic hydrogen) considered~\citep{Lehmer2017ThePlanets, Walker1977}, and let $T_3 \equiv T/10^{3}$K, we have,

\begin{equation}
    \begin{split}
    r_c &= \frac{GM_p}{2 \cdot u_0^2} = \frac{GM_p \cdot \mu_{\text{H}}}{2 k_B T} = \frac{R_p^2}{2\cdot \mathcal{H}} \\ &= \bigg( \frac{R_p^2}{2\cdot (k_B T)/(\mu_{\text{H}} \cdot g)} \bigg) =  3.75 R_{\oplus} \cdot \bigg( \frac{M_p/M_{\oplus}}{T/10^3\text{K}} \bigg) \propto \frac{M_p}{T_3}
    \end{split}
\end{equation}

Due to its inverse-proportionality to Temperature T, the sonic level (r$_c$) is lowest and closest to the planet surface where the temperature is highest, which occurs typically near the sub-stellar region of the planet---hot spot.

(4) The fourth critical layer to consider is the H-H bond \emph{dissociation front} (see the following discussion as well as~\ref{appendix:H2Dissociation} and~\ref{appendix:H2PhotoDissociation}). It is considered as the source region of the neutral atomic hydrogen outflow (Figure~\ref{fig:draw}). Because a hydrogen atom is the lightest species among all atmospheric species, which is only half the molecular mass of an H$_2$-molecule, and thus, a hydrogen atom is expected to escape into outer space more easily than an H$_2$-molecule under similar conditions such as temperature and planetary gravitational field (see~\ref{appendix:JeansEscape} and~\ref{appendix:HydrodynamicEscape}). 

{\color{black}Under higher host-stellar photon flux level, the atmosphere will become much more dynamic as compared to the static case. Layering is still expected to exist due to planet gravity, however, it geometry may change, especially within the hot spot region where the interaction between the host-stellar photon and planetary matter is the most intense. Some of the layers may become less important or even irrelevant, while some of the layers, in particular, the \emph{sonic front} and the \emph{dissociation front} may even merge to form a \emph{shock front}, so that the escape process becomes much more energetic and is no-longer diffusion-limited. We will discuss this photon-driven expansive shock escape mechanism in Section~\ref{phasetranstion}.} 

}


\section{Jeans Escape}\label{appendix:JeansEscape}

We can express the pressure at the \emph{exobase} (p$_{\text{exob}}$) as a function of collisional cross-section ($\sigma$), the mean-molecular weight of the escaping species ($\mu$), and the surface gravity of the planet at the level of exobase ($g_{\text{exob}}$): 

\begin{equation}
    \begin{split}
    \bigg( p_{\text{exob}} = n_{\text{exob}} \cdot k_B \cdot T \bigg) &\sim \frac{k_B \cdot T}{\mathcal{H} \cdot \sigma} \\ &\sim \frac{\mu \cdot g_{\text{exob}}}{\sigma} = \bigg(\frac{\mu \cdot G M_p/r_{\text{exob}}^2}{\sigma} \bigg)
    \end{split}
\end{equation}\label{eq2}

Collisional cross-sections $\sigma$ for atoms (e.g. H, O, ...) or small molecules (e.g. H$_2$, OH, O$_2$, etc.) are similar and reflect their characteristic dimensions of $\sim10^{-8}$cm and are commonly near $\sim10^{-16}$cm$^2$. This is the same cross-section for photon interaction (scattering and absorption) with the electron cloud in atoms and small molecules~\citep[p.79]{Lewis2004PhysicsSystem}~\citep[Chap.~1]{Tsien1965}. 

Therefore, we can plug in this number of $\sigma\approx10^{-16}$cm$^2$ and obtain: 

\begin{equation}
    \begin{split}
    p_{\text{exob}} &\sim \frac{(1\text{g}/N_A) \cdot 10^3\text{cm}/\text{s}^2}{10^{-16} \text{cm}^2} \cdot \frac{M_p/M_{\oplus}}{(r_{\text{exob}}/R_{\oplus})^2} \\ &\approx 10^{-6} \text{Pa} \cdot \frac{M_p/M_{\oplus}}{(r_{\text{exob}}/R_{\oplus})^2} \approx 10^{-11} \text{atm} \cdot \frac{M_p/M_{\oplus}}{(r_{\text{exob}}/R_{\oplus})^2}
    \end{split}
\end{equation}

Therefore, the pressure (p$_{\text{exob}}$) which defines the \emph{exobase} is a low number ($10^{-11}$ atm), modulated by a factor proportional to surface gravity. 

The escape flux ($\Phi_{\text{Jeans}}$ in cm$^{-2}$ s$^{-1}$): number of atoms/molecules escaping into outer space per unit surface area at the \emph{exobase}-level per unit time) can be found by integrating over the Maxwellian velocity distribution of the escaping species at the \emph{exobase}~\citep{Walker1977, Goody1995, Catling2017EscapeSpace}:

\begin{equation}
    \Phi_{\text{Jeans}} = \frac{n_{\text{exob}}\cdot u}{2 \cdot \pi^{1/2}} \cdot e^{-v_{\text{esc}}^2/u^2} \cdot \left(\frac{v_{\text{esc}}^2}{u^2}+1\right)
\end{equation}

Here $n_{\text{exob}}$ is the number density of escaping species evaluated at the escape level, that is, the exobase-level. 

$v_{\text{esc}} \equiv \sqrt{2GM_p/r_{\text{exob}}} = \sqrt{2} \cdot \sqrt{GM_p/r_{\text{exob}}}$ is the escape velocity and $u \equiv \sqrt{2k_B T/\mu} = \sqrt{2}\cdot u_0$ is the most probable velocity of the Maxwellian velocity distribution under thermal equilibrium\footnote{where $u_0 \equiv \sqrt{k_B T /\mu}$ is the isothermal sound speed, where $\mu$ is the molecular weight for the gas species considered.}, both evaluated at the escape level.

It is convenient to define a dimensionless \emph{Jeans parameter} $\lambda$ as~\citep{Goody1995, Catling2017EscapeSpace, Owen2018AtmosphericExoplanets}:

\begin{equation}
    \begin{split}
    \lambda &\equiv \frac{v_{\text{esc}}^2}{u^2} = \frac{GM_p}{r_{\text{exob}}\cdot u_0^2} = \frac{GM_p/r_{\text{exob}}}{k_B T/\mu} \\ &=  \frac{g_{\text{exob}} \cdot r_{\text{exob}}}{k_B T/\mu} =  \left( \frac{r_{\text{exob}}}{\mathcal{H}_{\text{exob}}} \right) \propto \mu
    \end{split}
\end{equation}

\emph{Jeans parameter} $\lambda$ measures how difficult it is for a gas species to escape, the higher the value of $\lambda$, the more difficult it is to escape. Since $\lambda \propto \mu$, it is harder for heavier species to escape compared to lighter species, under the same temperature T condition.

For example, because $\mu_{\text{H}_2} = 2 \cdot \mu_{\text{H}}$, the \emph{Jeans parameter} $\lambda$ doubles for a H$_2$-molecule when it is compared to that of an H-atom: $\lambda_{\text{H}_2} = 2 \cdot \lambda_{\text{H}}$. Thus, theoretically it is much easier for an H-atom to thermally escape than a H$_2$-molecule.

The last expression of $\lambda = \left( r_{\text{exob}}/\mathcal{H}_{\text{exob}} \right) =  \left(\mathcal{H}_{\text{exob}}/r_{\text{exob}} \right)^{-1}$ suggests that $\lambda$ can be viewed over the planetary scale, as the ratio of planet radius at the exobase-level over the scale height of the escaping species evaluated at the exobase-level. 

In other words, $\lambda$ is the reciprocal of the scale height $\mathcal{H}$ measured in the unit of planet radius $\left( R_p \sim r_{\text{exob}} \right)$. The smaller the value of $\lambda$, the higher the ratio of $\left(\mathcal{H}_{\text{exob}}/r_{\text{exob}} \right)$. 

The escape flux ($\Phi_{\text{Jeans}}$) in terms of $\lambda$ is:

\begin{equation}
    \Phi_{\text{Jeans}} = \bigg(\frac{n_{\text{exob}}\cdot u}{2 \cdot \pi^{1/2}} \bigg) \cdot \bigg(\left(\lambda+1\right) \cdot e^{-\lambda} \bigg)
\end{equation}

Ignoring all the factors of the order of unity including the geometric factor, the escape flux $\Phi_{\text{Jeans}} \propto \exp{(-\lambda})$.

\begin{equation}
    \Phi_{\text{Jeans}} \sim n_{\text{exob}} \cdot u \cdot e^{-\lambda}
\end{equation}

The smaller the value of $\lambda$, the more intense the escape. In fact, this formula is not sensitive to the exact level (as long as it is below the sonic level and above the von Karman line, in the upper atmosphere) at which both $n$ and $\lambda$ are evaluated because the product $n \cdot e^{-\lambda}$ is largely conserved due to the \emph{barometric law}~\citep{Catling2017EscapeSpace}. Later we will show the advantage of evaluating it right at the von Karman line. 

On the other hand, the ratio $\left( \Phi_{\text{Jeans}}/n_{\text{exob}} \right)$ is often referred to as the effusion velocity~\citep{Goody1995}:

\begin{equation}
    u \cdot e^{-\lambda} \sim u_0 \cdot e^{-\lambda}
\end{equation}

Therefore, the effusion velocity is of the same order of magnitude as the sound speed $u_0 \equiv \sqrt{k_B T /\mu}$ attenuated by the exponential factor $\exp{(-\lambda})$.

With the definition of $\lambda$, the critical number density of H-atom at the exobase can be expressed as:

\begin{equation}
    \begin{split}
    n_{\text{H,exob}} = \frac{p_{\text{H,exob}}}{k_B T} \sim \frac{p_{\text{exob}}}{k_B T} &= \frac{1}{k_B T} \cdot \frac{\mu_{\text{H}}}{\sigma} \cdot \frac{G M_p}{r_{\text{exob}}^2} \\ &= \frac{1}{u_0^2} \cdot \frac{1}{\sigma} \cdot \frac{G M_p}{r_{\text{exob}}^2} = \frac{\lambda_{\text{H}}}{\sigma \cdot r_{\text{exob}}}
    \end{split}
\end{equation}

In all cases, it is much easier for an H-atom to escape compared to a H$_2$-molecule by a factor of $\sim e^{\left(-\lambda_{\text{H}}\right)}/e^{\left(-\lambda_{\text{H}_2}\right)} \sim e^{\left(\lambda_{\text{H}_2}-\lambda_{\text{H}}\right)} \sim e^{\lambda_{\text{H}}}$ evaluated at the \emph{exobase}, which amounts to a factor of at least a few hundred $\sim$ a few thousand times and sometimes much more. Therefore, the escaping flux of H$_2$ is often negligible compared to that of the H-atom. Meanwhile, H$_2$-molecule converts into H-atom by molecular dissociation occurring at high Temperature and low Pressure regions in the upper atmosphere.

\section{Hydrodynamic Escape}\label{appendix:HydrodynamicEscape}

Applying \emph{Bernoulli's Principle} (Conservation of Energy) to a compressible flow, we have \emph{(along a streamline)}:

\begin{equation}
    \Delta \left[ \underbrace{\frac{1}{2}\cdot \overline{v}^2}_{\text{kinetic energy}} + \underbrace{\int \frac{dP}{\rho(P)}}_{\text{internal energy}} + \underbrace{\Psi}_{\text{gravitational potential}} \right] = 0
\end{equation}

There is exchange in between the three types of energy, and therefore, the total energy is always conserved in the flow. Here $\overline{v}$ refers to the group (average) velocity of the flow, not of its individual constituent atom or molecule. The random (thermal) velocity distribution is included in the internal energy part.

The escape flux ($\Phi_{\text{hydrodynamic}}$) due to hydrodynamic escape is of similar functional form as the Jeans flux attenuated by the same exponential factor $\exp{(-\lambda})$, to within a factor of unity including geometric factor:

\begin{equation}
    \Phi_{\text{hydrodynamic}} \sim n_{\text{source}} \cdot u_0 \cdot e^{-\lambda}
\end{equation}

This is easy to understand because both the hydrodynamic escape and the Jeans escape are end-member cases of thermal escape, of which the \emph{Boltzmann} factor $\exp{(-\lambda})$ is characteristic. 

The only important difference is that the number density in $\Phi_{\text{hydrodynamic}}$ is no longer the number density of the escaping species evaluated at the exobase, but substituted by the number density evaluated at the origin of the hydrodynamic escape flow: $n_{\text{source}}$, is the number density (cm$^{-3}$) of the escaping species in the source region where the hydrodynamic outflow originates. In our consideration, we equate $n_{\text{source}}$ as the the number density of H-atom at the von Karman line. So, 

\begin{equation}
    \Phi_{\text{hydrodynamic}} \sim n_{\text{H,von Karman line}} \cdot u_0 \cdot e^{-\lambda}
\end{equation}

Then, this $\lambda$ is evaluated at the von Karman line as well. This $n_{\text{H,von Karman line}}$ can be viewed as sourcing an effective pressure which drives the outflow into space. 

Due to the Continuity Equation or (Conservation of Mass Flux), this outflow must be initially sub-sonic when the number density is high, but eventually transforms into super-sonic when the number density is low. Thus, a transonic front is expected to occur somewhere along the streamline. 

\section{Collisional Dissociation}\label{appendix:H2Dissociation}

{\color{black}
\subsection{Covalent bond}

One H-H molecule can be visualized as a dumbbell consisting of two weights (protons) held together by a spring (a covalent bond): Fig.~\ref{fig:h2}. The elastic modulus and tensile strength of this spring are two physical parameters which characterize this molecule in study. 

\begin{figure}[!ht]
    \centering
    \includegraphics[width=0.45\textwidth, angle=0]{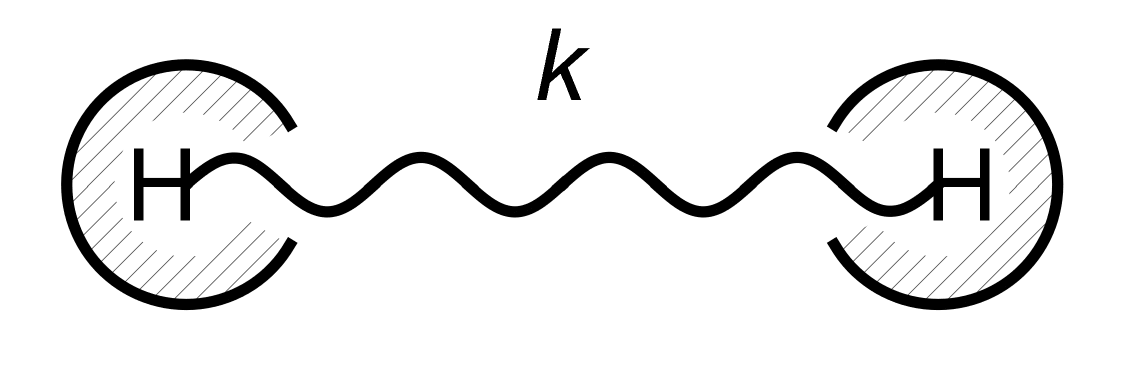}
    \caption{Schematic diagram of a hydrogen molecule, showing two hydrogen atoms connected by a spring (covalent bond).}
    \label{fig:h2}
\end{figure}

On one hand, the elastic modulus or spring constant ($k$) of this spring determines the fundamental vibrational frequency of this molecule. For small displacement $r$ from equilibrium, the potential energy $V$ can be approximated as~\citep{Tsien1965,Lewis2004PhysicsSystem,Draine2011}: $V(r) \approx V(0) + \frac{1}{2} \times k \cdot r^2$, with fundamental frequency $\nu_0 = \frac{1}{2\pi}\cdot\sqrt{k/m_r}$, with $m_r$ being the \emph{reduced mass} of this molecule $m_r = m_1 \cdot m_2 /(m_1+m_2)$. Quantum-mechanically, this molecule would have vibrational energy mode spacing as follows: 

\begin{equation}
    E_n = \left( n+\frac{1}{2} \right) \cdot h \cdot \nu_0, \text{~where $n = 0, 1, 2, 3,...$}
\end{equation}

On the other hand, its tensile strength determines the \emph{maximum strength} ($\sim \frac{1}{2} k \cdot {r_{\text{max}}}^2$) of this spring beyond which it is broken, where $r_{\text{max}}$ being the maximum displacement. Then, the molecule become dissociated into its constituent atoms and fly apart. Interestingly enough, this bond strength of a H-H bond is very similar to that of an O-H bond, or even a N-H bond or a C-H bond~\citep{Kim2020,Karapetyants1978,Tsien1965}, all being at about 4.5 eV. 

Both $k$ and $r_{\text{max}}$ stay roughly the same for any X-H single covalent bond. The main difference comes about from the \emph{reduced mass} $m_r$. Hydrides (X-H) have the smallest reduced masses, with H$_2$ molecule (H-H) being the extreme. Thus, H$_2$ has an un-usually high $\nu_0 \approx 1.5\times10^{14}$ Hz corresponding to a wavelength  $\lambda \approx 2 \mu$m, or an energy $h\nu_0 \approx 0.6$ eV, which is equivalent to the energy of a near-infrared photon~\citep{Lewis2004PhysicsSystem,Draine2011}.  

Thus, for H$_2$ molecule, there exist about $10-15$ vibrational energy levels (counting from ground state up and taking into account \emph{anharmonicity} which diminishes the spacing of higher-lying energy levels), before it is fully dissociated into hydrogen atoms. Therefore, H$_2$ is the most quantum-mechanical and least classical of all molecules. For other molecules, the spacing between energy levels becomes tighter due to its larger \emph{reduced mass} $m_r$, and thus, smaller $\nu_0$, so that more vibrational energy levels can fit in before molecular dissociation.}

{\color{black}
\subsection{Thermal Dissociation Line of H$_2$ Molecule}
{\color{black}So far we have discussed the dissociation of a H$_2$ molecule from the point-of-view of a single molecule. Now, we will place the molecule into a thermodynamic setting as surrounded by numerous other neighboring molecules constantly interacting and exchanging energy with each other, and discuss the physical and chemical process from the point-of-view of an assembly, where both temperature and pressure can be well-defined.}

Assuming Local Thermodynamic Equilibrium (LTE) and Local Chemical Equilibrium (LCE), the relative proportion between the number of hydrogen atoms and the number of hydrogen molecules can be estimated from the partial pressure equilibrium constant ($K_p$) according to the \emph{law of mass action} as:


\begin{equation}
 K_p(T) = \frac{(p_{\text{H}})^2}{p_{\text{H}_2}}
\end{equation}

$p$'s are partial pressures of each species are in the unit of atm ($\sim$bar=$10^5$Pa). 

Equivalently, $K_p$ can be expressed in terms of the degree of dissociation $\alpha$, and the total pressure $P$ as:

\begin{equation}
    \begin{split}
    K_p(T) &= \frac{((2\alpha)\cdot n R T)^2}{(1-\alpha)\cdot n R T} =  \frac{(2\alpha)^2}{(1-\alpha)} \cdot \bigg( n R T \bigg) \\ &=\frac{(2\alpha)^2}{(1-\alpha)} \cdot \bigg( \frac{P} {1+\alpha} \bigg) = \frac{4\alpha^2}{(1-\alpha^2)} \cdot P
    \end{split}
\end{equation}

where $n$ is the number density of total H$_2$ \emph{originally} present, $P$ is the total pressure of the mixture under \emph{current} equilibrium condition: $P = p_{\text{H}} +p_{\text{H}_2}$, and $\alpha$ is the degree of dissociation defined as:


\begin{equation}
 \alpha = \frac{p_{\text{H}_2\text{(decomposed)}}}{p_{\text{H}_2\text{(total H$_2$ originally present)}}}
\end{equation}

For the temperature range from 1000 K to 4000 K, the value of $K_p$ is approximately~\citep{Lewis2004PhysicsSystem}:

\begin{equation}
    \log_{10} K_p(T) = 6.16 -\frac{23,500}{T(\text{K})}=6.16-\frac{23.5}{T_3}
\end{equation}

The first term (6.16) reflects the entropy difference between the products and the reactants measured in the appropriate unit, 
whereas the numerator above the temperature in the second term basically reflects the bond strength of an H-H bond measured in a temperature scale, which is about 4.5 eV, {\color{black}or 2.25 eV per H-atom liberated from this H-H chemical bond.} This energy (enthalpy) is much higher compared to the typical thermal energy of an atom or molecule in the temperature range ($\sim10^{3}$ K) that we are interested. Therefore, only the most energetic molecules in the tail-end of the Maxwellian velocity distribution can participate in this dissociation reaction.




Below $\sim$2000 K, hydrogen is present almost exclusively as molecular H$_2$~\citep{Lewis2004PhysicsSystem}. The \emph{transition} from hydrogen-molecule-dominated regime to hydrogen-atom-dominated regime occurs along a narrow zone in the P-T diagram. This narrow zone can be approximated as a narrow curve in the lgP-lgT diagram. Let's now find such a curve.

\begin{equation}
    \log_{10} \left[\frac{4\alpha^2}{(1-\alpha^2)} \cdot P_{\text{atm}} \right] = 6.16-\frac{23.5}{T_3}
\end{equation}

For small $\alpha$, we have:

\begin{equation}
    \log_{10} (4\alpha^2) + \log_{10} P_{\text{atm}}  \approx 6.16-\frac{23.5}{T_3}
\end{equation}

so,

\begin{equation}
    \log_{10} (\alpha^2) \approx 6.16 - \log_{10}{4} - \log_{10} P_{\text{atm}} - \frac{23.5}{T_3}
\end{equation}

or,

\begin{equation}
    \log_{10} \alpha \approx 2.78 - \frac{1}{2} \cdot\log_{10} P_{\text{atm}} -\frac{11.75}{T_3}
\end{equation}

Consider the temperature-dependent critical pressure ($P_{\text{diss1\%H$_2$}}$(T)) at which dissociation reaches one-percent level ($\alpha=1$\%). It traces out a curve in P-T space (Fig.~\ref{fig:thermaldissociation}). Along this curve approximately 1\% hydrogen molecules are dissociated into hydrogen atoms under equilibrium condition. Equivalently, $P_{\text{diss1\%H$_2$}}$ can be expressed as: 

\begin{equation}
    \log_{10}P_{\text{diss1\%H$_2$}}(\text{atm}) \approx 9.56 -\frac{23.5}{T_3}
\end{equation}

Alternative pathways of forming atomic hydrogen through photo-dissociation: the direct dissociation of hydrogen-bearing molecule species by the incoming energetic UV photons from the host-star are further explained in~\ref{appendix:H2PhotoDissociation}. If those hydrogen atoms formed from photo-dissociate make frequent enough collisions with other atmospheric species before their escape, as they have cascaded their energy to other species we can consider them being \emph{thermalized} and include them in our discussion of thermal escape here.}


{\color{black}In the lower atmosphere}, collisions are so frequent as to ensure the local thermodynamic equilibrium (LTE), which then allows the exchange of energy across different energy modes of a molecule---translational, vibrational, rotational, through frequent collisions. This is indeed needed to convert the translational kinetic energy and deposit it through collisions into the molecular vibrational energy in order to break up the H-H bond. Due to the symmetry of a diatomic H$_2$-molecule, it has zero dipole-moment, and thus, is almost transparent to the electromagnetic radiation field by itself in the visible or infrared wavelengths of either stellar or planetary origin (see~\ref{appendix:H2Opacity}). However, the presence of trace amount of other polar molecules such as H$_2$O can have strong absorptions and transport the energy to H$_2$ through collision cascades.



{\color{black}In the upper atmosphere}, collisions become rarer so that a liberated hydrogen atom can deviate from its collisional/chemical equilibrium with H$_2$ and other hydrogen-bearing species, and {\color{black}easily} make escape. Due to the reduced collisional frequency, the atmospheric temperature above the von Karman line also starts to deviate from the planet equilibrium temperature, and becomes highly sensitive and variable as influenced by the host-stellar electro-magnetic activities and also space environments. Thus, this is another reason for us to evaluate the thermal escape process right at the von Karman line, which is the physical boundary between the atmospheric body and outer space, where we can still rely upon the temperature calculated from the planet equilibrium temperature calculation (see Section~\ref{appendix:PlanetTemp}). 

{\color{black}Under the more energetic and dynamic settings with higher incident stellar photon flux, the planetary atmospheric layering geometry considered under the more static scenario in~\ref{layering} may be \emph{warped} significantly within the hot spot region. The actual escape geometry and mechanism may then involve a local \emph{shock front} formed from the merging of two or more critical layers in the atmosphere driven by the strong interaction between the incoming stellar photons and the planetary surface matter, which we will discuss in detail in the following Section~\ref{phasetranstion}.}

{\color{black}However, the discussions in~\ref{layering} and the calculations in~\ref{appendix:HydrodynamicEscape} and~\ref{appendix:EnergeticView} are still valuable and meaningful because those arguments boil down to the general principle of energy conservation (\emph{Bernoulli's Principle} for flow and the exploitation of the degree-of-freedom of molecular dissociation), such that one can evaluate the temperature-dependence of the escape rate near the source.}



{\color{black}
\subsection{Thermal Dissociation Line of Water Molecule}

Likewise, the dissociation of a water molecule into hydrogen and oxygen (or OH radical) {\color{black}proceeds} at increasing temperature or decreasing pressure. This chemical equilibrium has been experimentally investigated by \emph{Walther Nernst} and his colleagues~\citep{Jellinek1986,Karapetyants1978}. For not-too-high-temperatures where further dissociation into H atoms and OH radicals does not play important role, we have:

\begin{equation}
2~\text{H}_2\text{O} (\text{g}) \rightleftharpoons 2~\text{H}_2 (\text{g}) + \text{O}_2 (\text{g}) + 57~\text{kcal/(mol of H$_2$O)}
\end{equation}

This enthalpy of 57 kcal/(mol of H$_2$O) translates to 114 kcal/(mol of H$_2$O-pairs), which is $\sim$4.9 eV per pair of H$_2$O from the stoichiometry of the formula above. The partial pressure (in unit of atm) equilibrium constant $K_p$ is:

\begin{equation}
 K_p(T) = \frac{(p_{\text{H$_2$}})^2 \cdot (p_{\text{O$_2$}})}{(p_{\text{H$_2$O}})^2}
\end{equation}

Equivalently, $K_p$ can be expressed in terms of the degree of dissociation $\alpha$, and the total pressure $P$ as:

\begin{equation}
    \begin{split}
    K_p(T) &= \frac{(\alpha \cdot n R T)^2 \cdot (\frac{1}{2} \alpha \cdot n R T)}{((1-\alpha)\cdot n R T)^2} =  \frac{\alpha^3}{2 \cdot (1-\alpha)^2} \cdot \bigg( n R T \bigg) \\ &=\frac{\alpha^3}{2 \cdot (1-\alpha)^2} \cdot \bigg( \frac{P} {1+\frac{1}{2} \alpha} \bigg) = \frac{\alpha^3 \cdot P}{(2+\alpha) \cdot (1-\alpha)^2}
    \end{split}
\end{equation}

where $n$ is the number density of total H$_2$O \emph{originally} present, $P$ is the total pressure of the mixture under \emph{current} equilibrium condition: $P = p_{\text{H}_2\text{O}}  + p_{\text{H}_2} + p_{\text{O}_2}$, and $\alpha$ is the degree of dissociation defined as:


\begin{equation}
 \alpha = \frac{p_{\text{H}_2\text{O}\text{(decomposed)}}}{p_{\text{H}_2\text{O}\text{(total H$_2$O originally present)}}}
\end{equation}

For the temperature range from 1000 K to 3000 K, the value of $K_p$ is approximately (which agree with the experimental data within $\sim1$\% of $\log_{10} K_p(T)$)~\citep{Karapetyants1978}:

\begin{equation}
    \log_{10} K_p(T) = \bigg(-0.24+ 1.75 \cdot \log_{10}\left[T(\text{K})\right] \bigg) -\frac{25,266}{T(\text{K})}
\end{equation}

Let's again introduce $T_3 \equiv T/10^3$K to simplify the above expression: 

\begin{equation}
    \begin{split}
    &\log_{10} K_p(T_3) =  5.01 + \\ &\bigg(  \underbrace{1.75 \cdot \log_{10}T_3}_{\text{due to the entropy difference between products and reactants: }\int_{0}^{T} \frac{\Delta C_p \cdot dT}{R \cdot T}} \bigg) \\ &-\frac{25.266}{T_3}
    \end{split}
\end{equation}

For 1000K-3000K, $K_p$ can be further simplified to:

\begin{equation}
    \log_{10} K_p(T_3) \approx 5.5 -\frac{25.0}{T_3}
\end{equation}


again, 

\begin{equation}
    \log_{10} \left[\frac{\alpha^3}{(2+\alpha)(1-\alpha)^2} \cdot P_{\text{atm}} \right] = 5.5-\frac{25.0}{T_3}
\end{equation}

For small $\alpha$, we have:

\begin{equation}
    \log_{10} \left( \frac{\alpha^3}{2} \right) + \log_{10} P_{\text{atm}}  \approx 5.5-\frac{25.0}{T_3}
\end{equation}

so,

\begin{equation}
    \log_{10} \left( \alpha^3 \right) \approx 5.5 + \log_{10}{2} - \log_{10} P_{\text{atm}} -\frac{25.0}{T_3}
\end{equation}

or,

\begin{equation}
    \log_{10} \alpha \approx 1.93 - \frac{1}{3} \cdot \log_{10} P_{\text{atm}} -\frac{8.3}{T_3}
\end{equation}

{\color{black}Therefore, on the surface of some water worlds, the strong interaction with host-stellar photons would lead to gradual dissociation of water molecules and the escaping of liberated hydrogen over time.

\begin{figure}[!ht]
    \centering
    \includegraphics[width=0.45\textwidth, angle=0]{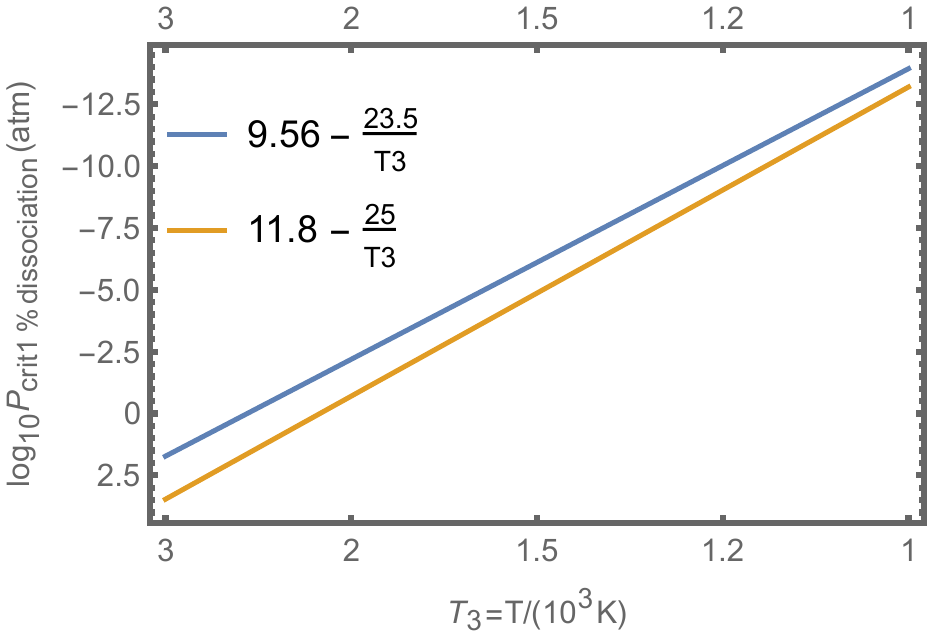}
    \caption{1\%-dissociation front of H$_2$ (blue) and H$_2$O (orange) in the P-T diagram under equilibrium consideration.}
    \label{fig:thermaldissociation}
\end{figure}

The critical pressure $P_{\text{diss1\%H$_2$O}}$ at which dissociation reaches one-percent level ($\alpha=1$\%) in P-T space (Fig.~\ref{fig:thermaldissociation}) is:}

\begin{equation}
    \log_{10}P_{\text{diss1\%H$_2$O}}(\text{atm}) \approx  11.8 -\frac{25.0}{T_3}
\end{equation}


1-bar or 1-atm level is approximately the photosphere of these planets we consider, where collisionally-induced opacity becomes significant to absorb most incoming stellar photons. 

Notice that although the H$_2$O dissociation front penetrates deeper to reach higher pressure-levels under the same temperature condition, it is more difficult to \emph{completely} dissociate a neutral hydrogen atom out of a water molecule, because the dissociation of water generally proceeds in two steps: first one would have to liberate the hydrogen from the strong chemical affinity of oxygen, then one would have to break the H-H covalent bond.}


\subsection{Detailed Calculations of the H$_2$ Collisional Dissociation}

Here we consider the direct dissociation of H$_2$ in a hydrogen-dominated atmosphere:

\begin{equation}
\text{H}_2 \rightleftharpoons \text{H} + \text{H}
\end{equation}

The pressure equilibrium constant $K_p$:

\begin{equation}
    \lg K_p = \lg \frac{p_{\text{H}}^2\text{(atm.)}}{p_{\text{H}_2}\text{(atm.)}} = \left[ 6.16 - \frac{23,500}{T(\text{K})} \right]
\end{equation}

where

\begin{align}
p_{\text{H}} &= n_{\text{H}} \cdot k_B T \\ 
p_{\text{H}_2} &= n_{\text{H}_2} \cdot k_B T
\end{align}

and

\begin{align}
n_{\text{H}} + 2 \cdot n_{\text{H}_2} &= \text{total number of H-atoms in both species, conserved} \\ 
n_{\text{H}_2} + n_{\text{H}_2} &= n \text{,~total number of particles, not conserved}
\end{align}

Let

\begin{align}
n_{\text{H}} &\equiv \mathcal{X} \cdot n \\ 
n_{\text{H}_2} &\equiv (1-\mathcal{X}) \cdot n
\end{align}

So that, 

\begin{equation}
    \lg K_p = \lg \left( \frac{n_{\text{H}}^2 \cdot (k_B T)^2}{n_{\text{H}_2} \cdot (k_B T)} \right) = \lg \left( \frac{n_{\text{H}}^2}{n_{\text{H}_2}} \cdot k_B T \right) = \lg \left( \frac{\mathcal{X}^2}{(1-\mathcal{X})} \cdot \frac{n\cdot k_B T}{\text{1 atm}} \right)
\end{equation}

Thus,

\begin{equation}
    n \approx \left( \frac{(1-\mathcal{X})}{\mathcal{X}^2} \right) \cdot \left( \frac{\text{1 atm}}{k_B T} \right) \cdot 10~\widehat{} \left[ 6.16 - \frac{23,500}{T(\text{K})} \right]
\end{equation}

and,

\begin{align}
n_{\text{H}}\equiv \mathcal{X} \cdot n &\approx \left( \frac{1-\mathcal{X}}{\mathcal{X}} \right) \cdot \left( \frac{\text{1 atm}}{k_B T} \right) \cdot 10~\widehat{} \left[ 6.16 - \frac{23,500}{T(\text{K})} \right] \\ 
n_{\text{H}_2}\equiv (1-\mathcal{X}) \cdot n &\approx \left( \frac{1-\mathcal{X}}{\mathcal{X}} \right)^2 \cdot \left( \frac{\text{1 atm}}{k_B T} \right) \cdot 10~\widehat{} \left[ 6.16 - \frac{23,500}{T(\text{K})} \right]
\end{align}

At the H$_2$ dissociation front, $\left( \frac{1-\mathcal{X}}{\mathcal{X}} \right)$ is of the order of unity, and thus,

\begin{equation}
    n_{\text{H,dissociation-front}} \sim \left( \frac{\text{1 atm}}{k_B T} \right) \cdot 10~\widehat{} \left[ 6.16 - \frac{23,500}{T(\text{K})} \right]
\end{equation}

For T$\sim 10^3$K, 

\begin{equation}
    \left( \frac{\text{1 atm}}{k_B T} \right) \sim \left( \frac{10^5 \text{Pa}}{k_B \cdot 10^3\text{K}} \right) \approx 7.25 \cdot 10^{18} \text{cm}^{-3} \approx 10^{18.86} \text{cm}^{-3}
\end{equation}

Thus, to within a factor of unity, for T$\sim 10^3$K-$4 \times 10^3$K, 

\begin{equation}
    n_{\text{H,dissociation-front}} \sim \left( \frac{1}{T/(10^3\text{K})} \right) \cdot
    10~\widehat{} \left[ 25.0 - \frac{23,500}{T(\text{K})} \right] \text{cm}^{-3}
\end{equation}

Notice the strong temperature-dependence of $n_{\text{H}}$, assuming chemical equilibrium, we have the following table:

\begin{table}[ht!]
\caption{$n_{\text{H}}$ dependence on temperature T} 
\begin{center}
\begin{tabular}{ |c|c|}
 \hline
 T (K) & $n_{\text{H}}$(cm$^{-3}$)  \\
 \hline
 1000 K & $3.\times10^1$  \\ 
 1200 K & $2.\times10^5$  \\
 1500 K & $1.5\times10^9$ \\ 
 2000 K & $1.\times10^{13}$  \\ 
 3000 K & $5.\times10^{16}$  \\ 
 \hline
\end{tabular}
\end{center}
\end{table}

Therefore, starting around $\sim1500$K, the H$_2$ dissociation front penetrates into the exobase ($n\sim10^{(8\sim9)}$cm$^{-3}$). Starting around $\sim2000$K, the H$_2$ dissociation front penetrates deeper into the atmosphere and approaches the level of the von Karman line ($n\sim10^{16}$cm$^{-3}$) below which different gas species are well-mixed with a common scale-height. For comparison, the mass density of super-critical fluid H$_2$ is approximately $\rho_{\text{H}_2\text{~fluid}} \sim0.07$g cm$^{-3}$ or $\sim \frac{1}{16}$g cm$^{-3}$. The corresponding number density is then $n_{\text{H}_2\text{~fluid}} \sim2\times10^{22}$cm$^{-3}$. Approaching this number density, EOS needs to take into account the strong inter-molecular repulsion.


However, a major flaw in this calculation is whether the local chemical equilibrium can be achieved when the number densities of H$_2$ and thus the collisional rates fall below a certain value in the upper atmosphere. One needs to carefully consider the reaction kinetics of a uni-molecular decomposition of H$_2$ through molecular collisions~\citep{Jordan1979}:

\begin{equation}
    \begin{split}
        \text{H}_2 + \text{H}_2 &\underset{k_2}{\stackrel{k_1}{\rightleftharpoons}}  \text{H}_2^{*} + \text{H}_2 \\ 
        \text{H}_2^{*} &\stackrel{k_d}{\rightarrow} \text{H} + \text{H} \\
        \hline
        \text{\emph{net}~~~} \text{H}_2 &\rightarrow \text{H} + \text{H}
    \end{split}
\end{equation}

A H$_2$ molecule is activated by collision with another H$_2$ molecule. In the collisional process the relative kinetic energy of the two colliding molecules is transformed, in part, to (vibrational) internal energy of one of the molecules, and thus, activating it. Such inelastic collisions provide the basis for understanding the activation process. The excited molecule H$_2^{*}$ either de-excites through another collision, or, decomposes into the products---atomic hydrogen.

Above the von Karman line ($n\sim10^{16}$cm$^{-3}$) the gas becomes highly rarefied so that vibrational energy of a molecule decouples from the thermal equilibrium, because a large number of collisions is necessary to excite the vibrational degree of freedom. On the other hand, rotational degree of freedom remains in equilibrium to much lower density because it is more easily perturbed by collisions. Therefore, it is more reasonable to evaluate the number density of H-atom ($n_{\text{H}}$) at the von Karman line where the H$_2$-molecule ($n_{\text{H}_2}$) is still the dominating species and the collisions among them are frequent enough. Recall that, 

\begin{equation}
    \lg K_p = \lg \left( \frac{n_{\text{H}}^2 \cdot (k_B T)^2}{n_{\text{H}_2} \cdot (k_B T)} \right) = \lg \left( \frac{n_{\text{H}}^2}{n_{\text{H}_2}} \cdot \frac{k_B T}{\text{1 atm}} \right) = \left[ 6.16 - \frac{23,500}{T(\text{K})} \right]
\end{equation}

Thus,

\begin{equation}
    n_{\text{H}}^2 \sim \left(n_{\text{H}_2} \cdot \frac{\text{1 atm}}{k_B T} \right) \cdot 10~\widehat{} \left[ 6.16 - \frac{23,500}{T(\text{K})} \right]
\end{equation}

Thus, for T$\sim 10^3$K-$4 \times 10^3$K, 

\begin{equation}
    n_{\text{H}}^2 \sim \left( \frac{n_{\text{H}_2}}{T/(10^3\text{K})} \right) \cdot
    10~\widehat{} \left[ 25.0 - \frac{23,500}{T(\text{K})} \right]
\end{equation}

Now, for convenience of calculation, let:

\begin{equation}
    T_3 \equiv T/(10^3\text{K})
\end{equation}

Then,

\begin{equation}
    n_{\text{H}}^2 \sim \left( \frac{n_{\text{H}_2}}{T_3} \right) \cdot
    10~\widehat{} \left[ 25.0 - \frac{23.5}{T_3} \right]
\end{equation}

So, 

\begin{equation}
    n_{\text{H}} \sim \sqrt{\left( \frac{n_{\text{H}_2}}{T_3} \right) \cdot
    10~\widehat{} \left[ 25.0 - \frac{23.5}{T_3} \right]} \text{cm}^{-3}
\end{equation}

Equivalently,

\begin{equation}
    n_{\text{H}} \sim \sqrt{ \frac{n_{\text{H}_2}}{T_3}} \cdot
    10~\widehat{} \left[ 12.5 - \frac{11.75}{T_3} \right] \text{cm}^{-3}
\end{equation}

Finally, 

\begin{equation}
    n_{\text{H,von Karman line}} \sim \sqrt{ \frac{n_{\text{H}_{2\text{,von Karman line}}}}{T_3}} \cdot
    10~\widehat{} \left[ 12.5 - \frac{11.75}{T_3} \right] \text{cm}^{-3}
\end{equation}

Plug in the number density estimated at the von Karman line $n_{\text{H}_{2\text{,von Karman line}}} \sim 10^{16} \text{~cm}^{-3}$, 

\begin{equation}
    n_{\text{H,von Karman line}} \sim \frac{1}{\sqrt{T_3}} \cdot
    10~\widehat{} \left[ 20.5 - \frac{11.75}{T_3} \right] \text{cm}^{-3}
\end{equation}

Note that this formula will inevitably fail at some high temperature when the H$_2$ dissociation front penetrates deep enough so that the number density of the escape flux is no longer limited by the temperature itself but instead limited by the energy input. Because now the escape flux produces a non-negligible effect of \emph{cooling}, this \emph{cooling} has to be compensated by a significant portion of the energy input of host-stellar incident radiation energy. In other words, breaking H-H bond provides effective cooling. At this point, the escape flow saturates and transitions from an \emph{entropy-limited} to an  \emph{energy-limited} regime. {\color{black}We will re-visit this point when we discuss the energetics, and show that this is a \emph{phase transition} from ordinary thermal escape scenario to \emph{shock-front}-driven hydrodynamic high-speed outflow.}


\section{H$_2$ Photo-Dissociation}
\label{appendix:H2PhotoDissociation}

The other possibility is the ability of energetic host-stellar photons with wavelengths less than 277 nm (corresponding to H-H bond strength of 4.5 eV) to rupture the strong H-H bond in molecular hydrogen. The sequence of events is:

\begin{equation}
    \begin{split}
        \text{H}_2 + \gamma &\rightarrow  \text{H}_2^{*} \\ 
        \text{H}_2^{*} &\rightarrow \text{H} + \text{H} \\
        \hline
        \text{\emph{net}~~~} \text{H}_2 + \gamma &\rightarrow \text{H} + \text{H}
    \end{split}
\end{equation}

$\gamma$ is a UV photon with wavelength shorter than 277 nm. The decay of the excited state H$_2^{*}$ is rapid, and the reaction that will be observed is the \emph{net} reaction. Thus, the rate equation becomes, 

\begin{equation}
    \frac{d[\gamma]}{dt} = \frac{d[\text{H}_2]}{dt} = -\frac{1}{2} \cdot \frac{d[\text{H}]}{dt} = -J_{\text{H}_2} \cdot [\text{H}_2]
\end{equation}

where the square brackets indicate concentrations. $J_{\text{H}_2}$ is the \emph{photochemical loss rate} of molecular hydrogen per hydrogen molecule or the \emph{inverse lifetime} of a hydrogen molecule for photo-dissociation for our purpose. 

The rate of change of photon density is the rate of change of radiant energy density divided by $(h\cdot\nu)$. Thus,

\begin{strip}
\begin{align}
	\frac{d[\gamma]}{dt} & = \int_{\nu_0}^{\infty} \frac{d\nu}{h\nu} \,\Big( \mathbf{\nabla} \cdot \mathbf{F}_{\nu} \Big) &&\text{($\mathbf{F}_{\nu}$ is the monochromatic flux)}\nonumber\\
    & = \int_{\nu_0}^{\infty} \frac{d\nu}{h\nu} \, \bigg( \int_{4\pi} d{\omega}_{\vec{l}} \cdot [\text{H}_2] \cdot \sigma_{\nu} \cdot (-I_{\nu}) \bigg) &&\text{(integrate over all solid angle of direction $\vec{l}$)}\\
    & = -[\text{H}_2] \cdot \underbrace{\int_{\nu_0}^{\infty} \frac{d\nu}{h\nu} \,\bigg( \int_{4\pi} d{\omega}_{\vec{l}} \cdot \sigma_{\nu} \cdot I_{\nu} \bigg)}_{\text{photochemical loss rate constant $J_{\text{H}_2}$}} &&\text{($[\text{H}_2]$ is the number density of H$_2$)}\\
     & = -[\text{H}_2] \cdot \underbrace{\int_{\nu_0}^{\infty} \frac{d\nu}{h\nu} \,\bigg( \sigma_{\nu} \cdot f_{\nu,0} \cdot \exp{\bigg(-\int_{z}^{\infty} \sigma_{\nu}(z') \cdot n(z') \cdot \frac{dz'}{\xi_{\star}} \bigg)} \bigg)}_{\text{photochemical loss rate constant $J_{\text{H}_2,d} \Big( z \Big) $ evaluated at height $z$}} &&\text{($f_{\nu,0}$ is host-stellar irradiance before attenuation)}
\end{align}
\end{strip}

$\nu_0$ is the minimum frequency that can give rise to a dissociation, corresponding to 277 nm for molecular hydrogen. $\sigma_{\nu}$ is the monochromatic \emph{absorption coefficient per molecule} which by definition is again a collisional cross-section, but this time for a photon of negligible size to interact with a molecule with an effective frequency-dependent collisional cross-section. Thus, $\sigma_{\nu}$ is the cross-section for light-matter interaction compared to $\sigma$ for molecular collision. Usually,

\begin{equation}
    \sigma_{\nu} < \sigma
\end{equation}

$I_{\nu}$ is the monochromatic radiance which originates from the direct host-stellar beam within the small solid angle ($\Delta\omega_{\star}$) of host-stellar disc but has already been attenuated by the planetary atmosphere from $\infty$ until height $z$. Thus, before any planet atmospheric attenuation, we have, 

\begin{equation}
    \begin{split}
     f_{\nu,0} = I_{\nu,0} \cdot \Delta\omega_{\star},~\text{where $\Delta\omega_{\star}$ is the small solid angle} \\ \text{of host-stellar disc viewed from planet.}
    \end{split}
\end{equation}

\begin{equation}
    \begin{split}
    \xi_{\star} = \cos{\theta_{\star}},~
    \text{where $\theta_{\star}$ is the zenith angle subtended} \\ \text{by the host-star as viewed from a location on planet.}
    \end{split}
\end{equation}

The frequency-dependent absorption cross-section $\sigma_{\nu}$ for H$_2$ and He from 0-300 nm wavelength is needed to model the photo-dissociation process---the absorption bands/spectrum of H$_2$ and He in the UV. Fortunately, H$_2$ ceases to absorb at 100 nm, and He, 50 nm~\citep{Lewis2004PhysicsSystem, Draine2011}. If the product from photo-dissociation collides with many other atmospheric species and shares its energy widely, then its energy enters the thermal reservoir and is considered thermalized.


\section{H$_2$ Opacity}\label{appendix:H2Opacity}


In order for the dissociation to occur, the energy must be deposited into the vibrational mode (degree-of-freedom) of H$_2$-molecule, through collisions, but NOT through absorption of visible or (thermal) infrared photons. A symmetric diatomic molecule such as H$_2$ has zero electric-dipole moment. Its dimension ($\sim 1 \AA$) is much smaller compared to the wavelength of an infrared photon ($\sim 10^4 \AA$), so the effect of the changing E-field is seen as the same by both H-atoms in a H$_2$-molecule. So it cannot absorb an infrared photon even if the photon is precisely at the natural vibration frequency of the molecule. Thus, the direct interaction of H$_2$-molecule with an infrared photon is \emph{forbidden}.

On the other hand, electronic transitions of H$_2$ lie in the ultra-violet part of the spectrum~\citep{Draine2011}. Thus, H$_2$ is mostly transparent to the planetary radiation field which emits infrared photons, as well as the stellar radiation field except at discrete lines due to the electronic transitions in the ultra-violet, which can only provide very limited opacity. Thus, the stellar bolometric radiation can penetrate all the way down to great depth to reach $\gtrsim1$ bar level ($n \sim 10^{19}$cm$^{-3}$) where the pressure-induced (i.e. collisionally-induced) absorption cross-section and the Rayleigh scattering cross-section for H$_2$ become significant~\citep{Hayashi1979, Goody1995, Lewis2004PhysicsSystem}. There H$_2$-fluid becomes dense enough to absorb all incoming photons as a continuum. This level would define the photo-sphere of such a planet. 

Now, if there is trace amount of polar molecules with non-zero dipole moments such as H$_2$O and CO mixed in with the H$_2$-gas,  then, H$_2$O and CO molecules can absorb the infrared photons through their vibrational mode, and collisionally cascade this energy to the vibrational and rotational mode of H$_2$-molecule. Depending on the mixing ratio, the presence of these polar molecules can contribute significant opacity and bring the photo-sphere to a shallower depth and lower pressure level. However, all this can only occur in the lower atmosphere (homosphere) where the mean free path ($\lambda$) is short enough ($\lambda \lesssim 1$cm) to allow effective mixing of gas species of different molecular weights. This $\lambda \sim 1$cm roughly corresponds to the pressure level of $\gtrsim1$ Pascal to the order-of-magnitude. This opacity ($\kappa_{\text{polar-molecules}}$) contributed by H$_2$O and CO molecules is important for the temperature range of 1500K-3000K, and can be crudely given by:

\begin{equation}
\kappa_{\text{polar-molecules}} \approx 0.1 \cdot Z,~~~[\text{cm}^2 \text{g}^{-1}] 
\end{equation}

The strongest absorber, the one with the greatest dipole moment and also the most abundant among the available polar molecules, is always H$_2$O gas (vapor). On the other hand, the opacity ($\kappa_{\text{H}^{-}}$) contributed by the presence of negative hydrogen ion, H$^{-}$, only becomes important at higher temperatures of 4000K-8000K, such as in stellar atmospheres of Main-Sequence (MS) stars~\citep{Schwarzschild1958}:

\begin{equation}
\kappa_{\text{H}^{-}} \approx 1.1 \times 10^{-25} \cdot Z^{0.5} \cdot \rho^{0.5} \cdot T^{7.7},~~~[\text{cm}^2 \text{g}^{-1}] 
\end{equation}

where $X$, $Y$, and $Z$ denote the hydrogen, helium, and heavy element fractions by mass, respectively.

\section{Energetic View}\label{appendix:EnergeticView}

Generally, for thermal escape, it is particle which lies in the energetic tail of the \emph{Boltzmann} distribution that can overcome the energy barriers and escape. For our purpose, the energy barriers refer to (1) the barrier to break the H-H bond, and (2) the barrier to escape the gravitational potential of the planet. 

These particles which lie in the energetic tail of the \emph{Boltzmann} distribution have much higher energy compared to the average thermal energy ($k_B T$) of the ensemble. Thus, when they escape, they also carry away their high energy and result in \emph{evaporative cooling} of the planet surface and planet atmosphere. Ultimately, this energy loss due to escape is supplied by a fraction of incoming stellar radiation. This process can be compared to the kinetics and energetics of evaporation~\citep{Lu2019}. Therefore, it is worthwhile to look at the energetics of the escape process. A simple back-of-envelope calculation would illuminate this point. First, let's consider the bolometric solar luminosity:

\begin{equation}
    L_{\odot} = 3.828 \times 10^{26} \text{J/s}
\end{equation}

Earth only intercepts a small fraction of this energy flow:

\begin{equation}
    L_{\oplus} = L_{\odot} \cdot \frac{\pi R_{\oplus}^2}{4\pi a^2} =  \frac{L_{\odot}}{4 \cdot (a/R_{\oplus})^2} = 1.75 \times 10^{17} \text{J/s}
\end{equation}

If all this energy goes into dissociate H$_2$ into H-atoms, we can estimate how long it takes to dissociate 1 Earth mass ($M_\oplus$) equivalent of H$_2$: 

\begin{equation}
    \begin{split}
   \tau &= \frac{\Delta H \cdot M_\oplus /\mu_{\text{H}_2}}{L_\oplus} = \frac{440\text{kJ/mol} \times 6\cdot 10^{24}\text{kg}/(2\text{g/mol})}{1.75 \times 10^{17} \text{J/s}} \\ &= \frac{1.32 \times 10^{33}\text{J}}{1.75 \times 10^{17} \text{J/s}} = 0.75 \times 10^{16} \text{s} \approx 0.2 \times 10^{9} \text{years}
   \end{split}
\end{equation}

Therefore, at Earth's distance, it would take $\sim$0.2 billion years (our estimate) to dissociate one Earth-mass equivalent of H$_2$ into H-atoms, if all the energy that Earth receives from the Sun over this duration goes into this process. Marvelously, this timescale is shorter but not too different from the age of our solar system! 

In reality, only a fraction of the incoming stellar radiation can contribute to the H-H bond dissociation. Thus, it would be extremely important to determine this fraction. Now, many exoplanets we consider are significantly H$_2$-rich compared to our own Earth, that some of them do have $\sim1 M_\oplus$ or more equivalent of H$_2$ mass stored in their envelopes. On the other hand, they are also significantly closer to their host stars, so they typically receive a hundred-fold ($f/f_\oplus \sim$100) or more of incident stellar radiation compared to our own Sun. Thus, we consider this fraction to be of the order of percent-level, in order for these planets to last over billion year timescale. 

Therefore, we need to consider the following energy balance equation, for instance, for the region near the sub-stellar hot spot where the escape is expected to strongest, and ignore any albedo effect for now:

\begin{equation}
    \begin{split}
   \underbrace{f_0}_{\text{absorbed stellar flux}} &= \underbrace{\left( \frac{\Delta H}{\mu_{\text{H}_2}} + \frac{GM_p}{R_p}  \right) \cdot \mu_{\text{H}} \cdot \Phi_{\text{H,escape}}}_{\text{(A): evaporative cooling by escaping H}} \\ &+ \underbrace{\sigma \cdot T^4}_{\text{(B): radiative cooling by escaping photons}}
   \end{split}
\end{equation}

Therefore, the \emph{cooling} in order to balance the incoming stellar radiation can be accomplished in two ways: (A) through escaping particles (H-atoms considered here) which carry away energy, (B) through escaping photons (thermal infrared radiation considered here) which carry away energy. The key difference between (A) and (B) is that for (B) photons can readily escape from planet surface, while for (A) particles need to overcome significant energy barriers so only a very small fraction of them with high enough energy can escape. 

(A) in turn contains two parts: (A1) energy to break up H-H bond, (A2) energy to overcome gravitational potential.

There is a very intricate and subtle relation between term (A) and term (B), because the escaping flux $\Phi_{\text{H,escape}}$ in (A) depends (strongly) on T (thermally driven in our assumption) in (B). 

Thus, there must exist a feedback mechanism between (A) and (B). This feedback mechanism can be understood as follows: if temperature T increases a little bit, then $\Phi_{\text{H,escape}}$ would increase rapidly (because it is strongly temperature-dependent). This rapid increase in $\Phi_{\text{H,escape}}$ would result in even more cooling which would help bring down T, and vice versa. We consider this feedback mechanism acts as a \emph{peak-holding valve} which controls the escape rate.

In other words, we need to figure out the functional dependence of escape flux $\Phi_{\text{H,escape}}$ on temperature T: 

\begin{equation}
    \Phi_{\text{H,escape}} = \Phi_{\text{H,escape}}(T)
\end{equation}

Let's turn back to the equation of thermal (hydrodynamic) escape. Recall the earlier formula:

\begin{equation}
    n_{\text{H,dissociation-front}} \sim \left( \frac{10^{25}\text{cm}^{-3}}{T/(10^3\text{K})} \right) \cdot
    10~\widehat{} \left[-\frac{23,500}{T(\text{K})} \right]
\end{equation}

equivalently, converting to natural-based logarithm by multiplying a factor of $\ln{10} \approx 2.303$ inside the power-index,

\begin{equation}
    n_{\text{H,dissociation-front}} \sim \left( \frac{10^{25}\text{cm}^{-3}}{T/(10^3\text{K})} \right) \cdot
    \exp{ \left[ 2.303 \cdot \bigg( - \frac{23,500}{T(\text{K})} \bigg) \right] }
\end{equation}

\begin{equation}
    n_{\text{H,dissociation-front}} \sim \left( \frac{10^{25}\text{cm}^{-3}}{T/(10^3\text{K})} \right) \cdot
    \exp{ \left[ - \frac{5.4\times10^4}{T(\text{K})} \right] }
\end{equation}

On the other hand,

\begin{equation}
    n_{\text{H,von Karman line}} \sim \left( \frac{10^{20.5}\text{cm}^{-3}}{\sqrt{T/(10^3\text{K})}} \right) \cdot
    10~\widehat{} \left[ -\frac{11,750}{T(\text{K})}  \right]
\end{equation}

equivalently, converting to natural-based logarithm by multiplying a factor of $\ln{10} \approx 2.303$ inside the power-index,

\begin{equation}
    n_{\text{H,von Karman line}} \sim \left( \frac{10^{20.5}\text{cm}^{-3}}{\sqrt{T/(10^3\text{K})}} \right) \cdot
    \exp{ \left[ 2.303 \cdot \bigg( - \frac{11,750}{T(\text{K})} \bigg) \right] }
\end{equation}

\begin{equation}
    n_{\text{H,von Karman line}} \sim \left( \frac{10^{20.5}\text{cm}^{-3}}{\sqrt{T/(10^3\text{K})}} \right) \cdot
    \exp{ \left[ - \frac{2.7\times10^4}{T(\text{K})} \right]
    }
\end{equation}

The reason for this conversion is show that the numerator inside the power-index is just a measure of the bond strength of H-H covalent bond, that is, the enthalpy of H-H covalent bond dissociation $\Delta H \approx 440$kJ/mol $\approx100$kcal/mol divided by the gas constant R$=8.314\text{J}\text{mol}^{-1}\text{K}^{-1}$:

\begin{equation}
    \frac{\Delta H}{\text{R}} = \frac{440\times10^3\text{J}\text{mol}^{-1}}{8.314\text{J}\text{mol}^{-1}\text{K}^{-1}} \approx 5.4\times10^4\text{K}
\end{equation}

(1) Thus, per bond $\Delta E_{\text{H-H bond}} \approx 4.5$eV:

\begin{equation}
    \begin{split}
    \Delta E_{\text{H-H bond}}\text{(eV)} &= \frac{\Delta H}{N_A\times e} \\ &= \frac{440\times10^3\text{J}\text{mol}^{-1}}{6.022\cdot10^{23}\times1.602\times10^{-19}\text{coulombs}} \approx 4.5\text{eV}
    \end{split}
\end{equation}

This is typical value for a covalent bond~\citep{Langes2016}. This value is almost exactly one-third (1/3) of the ionization energy of an H-atom: $\Delta E_{\text{H-ionization}} \approx 13.5$eV.

(2) The energy required for one H-atom to overcome the gravitational potential of the planet is:

\begin{equation}
    \begin{split}
    \Delta E_{\text{grav}}\text{(eV)} &= \frac{GM_p\cdot\mu_{\text{H}}}{R_p\cdot e} \\ &= \frac{M_p/M_{\oplus}}{R_p/R_{\oplus}} \cdot \frac{(11.2\cdot10^{3}\text{m/s})^2/2\times1.6735\cdot10^{-27}\text{kg}}{1.602\times10^{-19}\text{coulombs}} \\ &\approx \frac{M_p/M_{\oplus}}{R_p/R_{\oplus}} \cdot 0.65\text{eV} \sim 1.\text{eV}
    \end{split}
\end{equation}

(3) In contrast, the thermal energy $k_B T=k_B\cdot (10^3\text{K})\approx0.086$eV$\sim0.1$eV, 

\begin{equation}
    \begin{split}
    \Delta E_{\text{thermal}}\text{(eV)} &= \frac{k_B T}{e} = \frac{T}{10^3\text{K}} \cdot \frac{1.38\cdot10^{-23}\text{J}\text{K}^{-1}\times10^3\text{K}}{1.602\times10^{-19}\text{coulombs}} \\ &\approx \frac{T}{10^3\text{K}} \cdot 0.086\text{eV} \sim 0.1\text{eV}
    \end{split}
\end{equation}

which is about two orders-of-magnitude smaller compared to the energy needed to break the H-H bond, and about one order-of-magnitude smaller compared to the energy needed to escape the gravitational potential well of the planet.

Thus, we perceive the \emph{hierarchy} of energy here, and re-confirms our earlier point that only the energetic tail of the thermal distribution can escape. Recall that,

\begin{equation}
    \Phi_{\text{H-atom, hydrodynamic escape}} \sim n_{\text{H,von Karman line}} \cdot u_0 \cdot e^{-\lambda}
\end{equation}

Therefore, by collecting the terms for the two energy barriers together, we can write the escape flux generally as:

\begin{equation}
    \Phi_{\text{H,escape}} \sim \left( \frac{10^{20.5}\text{cm}^{-3}}{\sqrt{T/(10^3\text{K})}} \right) \cdot u_0 \cdot \exp{\left(-\frac{\Delta E_{\text{H-H bond}}/2+\Delta E_{\text{grav}}}{\Delta E_{\text{thermal}}}\right)}
\end{equation}

plug-in the isothermal sound speed expression $u_0 \equiv \sqrt{k_B T /\mu_{\text{H}}} \approx \sqrt{T/(10^3\text{K})}\cdot3$km/s, we have

\begin{equation}
    \begin{split}
    \Phi_{\text{H,escape}} \sim &\left( \frac{10^{20.5}\text{cm}^{-3}}{\sqrt{T/(10^3\text{K})}} \right) \cdot \sqrt{T/(10^3\text{K})} \cdot \left(3\text{km/s}\right) \times \\ &\exp{\left( -\frac{\Delta E_{\text{H-H bond}}/2+\Delta E_{\text{grav}}}{k_B T} \right)}
    \end{split}
\end{equation}

or, equivalently, 

\begin{equation}
    \Phi_{\text{H,escape}} \sim \bigg( 1.0\times10^{26}\text{cm}^{-2}\text{s}^{-1} \bigg) \cdot \exp{\left( -\frac{\Delta E_{\text{H-H bond}}/2+\Delta E_{\text{grav}}}{k_B T} \right)}
\end{equation}

To simplify the expression, we can write H-atom number escape flux $\Phi_{\text{H,escape}}$ as:

\begin{equation}
    \Phi_{\text{H,escape}} \sim C \cdot \exp{\left( -\frac{a/2 +b \cdot (m/r)}{T} \right)}
\end{equation}

where,

\begin{align}
a &\equiv \frac{E_{\text{H-H bond}}}{k_B} \approx 5.4 \times 10^{4} \text{K} \\ 
b &\equiv \frac{E_{\text{grav}}}{k_B} \approx 7.6 \times 10^{3} \text{K} \\
m &\equiv M_p/M_\oplus \\
r &\equiv R_p/R_\oplus \\
C &\sim \bigg( 1.0\times10^{26}\text{~cm}^{-2}\text{s}^{-1} \bigg)
\end{align}

Now, we can convert this expression to the mass flux of H-atom escape as:

\begin{equation}
    \mu_{\text{H}} \cdot \Phi_{\text{H,escape}} \sim C' \cdot \exp{\left( -\frac{a/2 +b \cdot (m/r)}{T} \right)}
\end{equation}

where $\mu_{\text{H}}=1$ g/mol, so that, 

\begin{equation}
    C' \sim \bigg( 1.66\times10^{2}\text{~g} \cdot \text{cm}^{-2}\text{s}^{-1} \bigg)
\end{equation}

Now, the mass loss rate can be estimated by multiplying the mass flux with an effective surface area of escape ($\eta \cdot R_p^2$) where $\eta$ is a geometric factor of the order of unity depending on the geometry of the escape flow: 

\begin{equation}
    \dot{M_p} = \mu_{\text{H}} \cdot \Phi_{\text{H,escape}} \cdot \eta \cdot R_p^2
\end{equation}

We can non-dimensionalize both sides of the equation so that,

\begin{equation}
    \dot{m} = \mu_{\text{H}} \cdot \Phi_{\text{H,escape}} \cdot \frac{R_\oplus^2}{M_\oplus} \cdot \eta \cdot r^2
\end{equation}

Now, 

Recall the energy balance equation of the absorbed stellar flux ($f_0$). In term (A), we have:

\begin{equation}
    \begin{split}
    \frac{\Delta H}{\mu_{\text{H}_2}} &= \frac{a \cdot k_B \cdot N_A}{\mu_{\text{H}_2}} = \frac{a \cdot \text{R}}{\mu_{\text{H}_2}} \\ &= \frac{a \cdot \text{R}}{2 \cdot \mu_{\text{H}}} = \frac{a}{2} \cdot \frac{\text{R}}{\mu_{\text{H}}} = \frac{a}{2} \cdot \left( 8.314\text{~J}\text{~g}^{-1}\text{K}^{-1} \right) 
    \end{split}
\end{equation}

\begin{equation}
    \begin{split}
    \frac{G M_p}{R_p} &= \frac{b \cdot k_B \cdot N_A}{\mu_{\text{H}}} \cdot \frac{m}{r} = \frac{b \cdot \text{R}}{\mu_{\text{H}}} \cdot \frac{m}{r} = b \cdot \frac{m}{r} \cdot \frac{\text{R}}{\mu_{\text{H}}} \\ &= b \cdot \frac{m}{r} \cdot \left( 8.314\text{~J}\text{~g}^{-1}\text{K}^{-1} \right)
    \end{split}
\end{equation}

So that,

\begin{equation}
    \begin{split}
   \underbrace{f_0}_{\text{absorbed stellar flux}} &= \underbrace{ \left( 8.314\text{~J}\text{~g}^{-1}\text{K}^{-1} \right) \cdot \left( \frac{a}{2} + b \cdot \frac{m}{r} \right) \cdot \mu_{\text{H}} \cdot \Phi_{\text{H,escape}}}_{\text{(A): evaporative cooling by escaping H}} \\ &+\underbrace{\sigma \cdot T^4}_{\text{(B): radiative cooling by escaping photons}}
   \end{split}
\end{equation}

\begin{equation}
    \begin{split}
   &\underbrace{f_0}_{\text{absorbed stellar flux}} = \\ &\underbrace{ \left( 8.314\text{~J}\text{~g}^{-1}\text{K}^{-1} \right) \cdot \left( \frac{a}{2} + b \cdot \frac{m}{r} \right) \cdot C' \cdot \exp{\left( -\frac{a/2 +b \cdot (m/r)}{T} \right)} }_{\text{(A): evaporative cooling by escaping H}} + \\ &\underbrace{\sigma \cdot T^4}_{\text{(B): radiative cooling by escaping photons}}
   \end{split}
\end{equation}

\begin{equation}
    \begin{split}
   &\underbrace{f_0}_{\text{absorbed stellar flux}} = \\ &\underbrace{ \bigg( 1.38\times 10^{7}\text{~W}\text{~m}^{-2}\text{K}^{-1} \bigg) \cdot \left( \frac{a}{2} + b \cdot \frac{m}{r} \right) \cdot \exp{\left( -\frac{a/2+b \cdot (m/r)}{T} \right)} }_{\text{(A): evaporative cooling by escaping H}} + \\ &\underbrace{\left( 5.67\times10^{-8} \text{~W}\text{~m}^{-2}\text{K}^{-4} \right) \cdot T^4}_{\text{(B): radiative cooling by escaping photons}}
   \end{split}
\end{equation}

Now, for convenience of calculation, let's define:

\begin{equation}
    T_3 \equiv T/(10^3\text{K})
\end{equation}

and,

\begin{align}
a_3 &\equiv a/(10^3\text{K}) = 54. \\ 
b_3 &\equiv b/(10^3\text{K}) =7.6 
\end{align}

so that,

\begin{equation}
    \begin{split}
   &\underbrace{f_0}_{\text{absorbed stellar flux}} = \\ &\underbrace{ \left( 1.38\times 10^{10}\text{~W}\text{~m}^{-2} \right)  \cdot \left( \frac{a_3}{2} + b_3 \cdot \frac{m}{r} \right) \cdot \exp{\left( -\frac{a_3/2 +b_3 \cdot (m/r)}{T_3} \right)} }_{\text{(A): evaporative cooling by escaping H-atoms}} + \\ &\underbrace{\left( 5.67\times10^{4} \text{~W}\text{~m}^{-2} \right) \cdot T_3^4}_{\text{(B): radiative cooling by escaping photons}}
   \end{split}
\end{equation}

This is a \emph{transcendental} equation if one wants to exactly solve for $T_3$ for a given $f_0$. It can be solved by recursive method, similar to the way to solve the \emph{Kepler equation}: $E-e\sin{E}=M$.

However, one can quickly estimate at what $T_{3\text{,crit}}$ term (A) and term (B) become comparable in magnitude:

\begin{equation}
    \begin{split}
    &\left( 1.38\times 10^{10}\text{~W}\text{~m}^{-2} \right) \cdot \left( \frac{a_3}{2} + b_3 \cdot \frac{m}{r} \right) \cdot \exp{\left( -\frac{a_3/2+b_3 \cdot (m/r)}{T_{3\text{,crit}}} \right) }
    \\ &\sim 
    \left( 5.67\times10^{4} \text{~W}\text{~m}^{-2} \right) \cdot T^4_{3\text{,crit}}
    \end{split}
\end{equation}

so,

\begin{equation}
     \left( \frac{a_3}{2} + b_3 \cdot \frac{m}{r} \right) \cdot \exp{\left( -\frac{a_3/2+b_3 \cdot (m/r)}{T_{3\text{,crit}}} \right) \sim \left( 4.1\times10^{-6} \right) \cdot T^{4}_{3\text{,crit}}}
\end{equation}

or, equivalently,

\begin{equation}
     \left( \frac{a_3/2+b_3 \cdot (m/r)}{T_{3\text{,crit}}} \right) \cdot \exp{\left( -\frac{a_3/2+b_3 \cdot (m/r)}{T_{3\text{,crit}}} \right) \sim \left( 4.1\times10^{-6} \right) \cdot T^{3}_{3\text{,crit}}}
\end{equation}

if we take natural logarithm on both sides:

\begin{equation}
     \ln{\left( \frac{a_3}{2} + b_3 \cdot \frac{m}{r} \right)}  -\frac{a_3/2 +b_3 \cdot (m/r)}{T_{3\text{,crit}}} \sim -12.4 + 4 \cdot \ln{T_{3\text{,crit}}}
\end{equation}

re-arranging the terms, we have:

\begin{equation}
      \frac{a_3/2+b_3 \cdot (m/r)}{T_{3\text{,crit}}} \sim \ln{\left( \frac{a_3}{2} + b_3 \cdot \frac{m}{r} \right)}  + 12.4 - 4 \cdot \ln{T_{3\text{,crit}}}
\end{equation}

finally,

\begin{equation}
    T_{3\text{,crit}} \sim \frac{a_3/2+b_3 \cdot (m/r)}{\ln{\left( \frac{a_3}{2} + b_3 \cdot \frac{m}{r} \right)}  + 12.4 - 4 \cdot \ln{T_{3\text{,crit}}}}
\end{equation}

by simple inspection, one can see that $T_{3\text{,crit}} \sim 2$ to 3 because the dominating term in the numerator is $a_3/2 \approx 27.$, and the dominating term in the denominator is $12.4$. Also, out of coincidence, the term $\ln{\left( \frac{a_3}{2} + b_3 \cdot \frac{m}{r} \right)}$ and the term $ \left( 4. \cdot \ln{T_{3\text{,crit}}} \right)$ roughly cancel each other out in magnitude. Therefore, a crude approximation gives:

\begin{equation}
    T_{3\text{,crit}} \sim \frac{a_3/2+b_3 \cdot (m/r)}{12.4}
\end{equation}

plug in the numbers: $a_3  = 54 $ and $b_3 =7.6$, we have

\begin{equation}
    T_{3\text{,crit}} \sim \frac{27+7.6 \cdot (m/r)}{12.4}
\end{equation}

or,

\begin{equation}
    T_{3\text{,crit}} \sim 2.18 + \underbrace{0.61 \cdot \frac{m}{r}}_{\text{due to gravity potential}}
\end{equation}

{\color{black}Now, the physical intuition is that the second term due to gravity potential can be dropped when it approaches the critical scenario, because as the escape process occurs it may \emph{absorb} more energy in various modes along the way of escape, and the escape process may be accelerated by a \emph{shock wave front}, that is, there is a \emph{phase transition} in the flow. We will re-visit this point when we discuss the \emph{tropopause}. Thus, 

\begin{equation}
    T_{3\text{,crit}} \sim 2.18
\end{equation}

Therefore, we can assume that the critical scenario occurs right around the transition from the \emph{entropy-limited} regime into the \emph{energy-limited} regime. This coincides with the observed cut-off temperature for large exoplanets at $T_{3\text{,crit}} \sim 2$. On the other hand, the second term $\left(\sim \frac{m}{r} \right)$ may be invoked to explain the ordinary positive-slope boundary bordering \emph{hot Saturns} in lg-lg mass-radius plot under normal non-limiting non-critical thermal escape scenario. Thus, in this respect of the critical \emph{phase transition} of flow status, these two terms can be separated.}

{\color{black}
\section{Tropopause}\label{tropopause}

Tropopause is the level which separates the troposphere, where convective upwelling dominates the energy transport mode, and the stratosphere, where radiative cooling dominates the energy transport mode. Because the visible opacity of H$_2$ gas goes like $\sim P^2$.

The tropopause is typically located at ten percent the pressure level of photosphere. Say, if photosphere is at 1 bar level, then tropopause is typically at 100 mb level. The thermal flux involved in the stratosphere and all the atmospheric mass above the tropopause is typically on the order of one percent level ($0.1^2=0.01$) of the total thermal flux received from the host star.

So, one should also estimate at what $T_{3\text{,crit2}}$ term (A) equals one percent (1\%) of term (B) in magnitude:

\begin{equation}
    \begin{split}
    &\left( 1.38\times 10^{10}\text{~W}\text{~m}^{-2} \right) \cdot \left( \frac{a_3}{2} + b_3 \cdot \frac{m}{r} \right) \cdot \exp{\left( -\frac{a_3/2+b_3 \cdot (m/r)}{T_{3\text{,crit2}}} \right)} \\ &\sim 0.01 \times \left( 5.67\times10^{4} \text{~W}\text{~m}^{-2} \right) \cdot T^4_{3\text{,crit2}}
    \end{split}
\end{equation}

so,

\begin{equation}
     \left( \frac{a_3}{2} + b_3 \cdot \frac{m}{r} \right) \cdot \exp{\left( -\frac{a_3/2+b_3 \cdot (m/r)}{T_{3\text{,crit2}}} \right) \sim \left( 4.1\times10^{-8} \right) \cdot T^{4}_{3\text{,crit2}}}
\end{equation}

or, equivalently,

\begin{equation}
    \begin{split}
     &\left( \frac{a_3/2+b_3 \cdot (m/r)}{T_{3\text{,crit2}}} \right) \cdot \exp{\left( -\frac{a_3/2+b_3 \cdot (m/r)}{T_{3\text{,crit2}}} \right)} \\ &\sim \left( 4.1\times10^{-8} \right) \cdot T^{3}_{3\text{,crit2}}
    \end{split}
\end{equation}

if we take natural logarithm on both sides:

\begin{equation}
     \ln{\left( \frac{a_3}{2} + b_3 \cdot \frac{m}{r} \right)}  -\frac{a_3/2 +b_3 \cdot (m/r)}{T_{3\text{,crit2}}} \sim -17.0 + 4 \cdot \ln{T_{3\text{,crit2}}}
\end{equation}

re-arranging the terms, we have:

\begin{equation}
      \frac{a_3/2+b_3 \cdot (m/r)}{T_{3\text{,crit2}}} \sim \ln{\left( \frac{a_3}{2} + b_3 \cdot \frac{m}{r} \right)}  + 17.0 - 4 \cdot \ln{T_{3\text{,crit2}}}
\end{equation}

finally,

\begin{equation}
    T_{3\text{,crit2}} \sim \frac{a_3/2+b_3 \cdot (m/r)}{\ln{\left( \frac{a_3}{2} + b_3 \cdot \frac{m}{r} \right)}  + 17.0 - 4 \cdot \ln{T_{3\text{,crit2}}}}
\end{equation}

again, by simple inspection, one can see that $T_{3\text{,crit2}} \sim 2$ because the dominating term in the numerator is $a_3/2 \approx 27.$, and the dominating term in the denominator is $17.0$. 

Also, out of coincidence, the term $\ln{\left( \frac{a_3}{2} + b_3 \cdot \frac{m}{r} \right)}$ and the term $ \left( 4. \cdot \ln{T_{3\text{,crit2}}} \right)$ roughly cancel each other out in magnitude. Therefore, a crude approximation gives:

\begin{equation}
    T_{3\text{,crit2}} \sim \frac{a_3/2+b_3 \cdot (m/r)}{17.0}
\end{equation}

plug in the numbers: $a_3  = 54.0 $ and $b_3 =7.6$, we have

\begin{equation}
    T_{3\text{,crit2}} \sim \frac{27+7.6 \cdot (m/r)}{17.0}
\end{equation}

or,

\begin{equation}
    T_{3\text{,crit2}} \sim 1.58 + 0.44 \cdot \frac{m}{r}
\end{equation}

Similar argument as in the  previous discussion can be invoked to drop the second term so that,

\begin{equation}
    T_{3\text{,crit2}} \sim 1.58
\end{equation}

Thus, this $T_{3\text{,crit2}}$ is close to the observed actual cut-off flux/temperature of small exoplanets at $T_{3\text{,crit2}} \sim 1.5$, which means that their limiting scenario is that their escaping matter flux is about one-percent ($\sim1\%$) level of their total energy flux.

So the rate of rise of temperature in stratosphere if without cooling is:

\begin{equation}
    \begin{split}
    \Delta T/\Delta t &= (\Delta q/\Delta t)/(m C_p) = 0.01 \cdot f /(C_p \cdot \rho_{\text{tropopause}} \cdot \mathcal{H}) \\ &= 0.01 \cdot f /(C_p \cdot P_{\text{tropopause}} / g )
    \end{split}
\end{equation}

Taking a stratospheric temperature of $\sim1000$K, $C_p=3.5$ R, $P=0.1\text{bar}=10^{5}\text{dyn cm}^{-2}$, $f=100 f_{\oplus}=1.37\cdot10^{8}\text{erg/s cm}^{-2}$, $g=g_{\oplus} \approx 10^{3}\text{cm s}^{-2}$, R$=8.314\cdot 10^{7}\text{erg/K mol}^{-1}$, so, $\Delta T/\Delta t \approx 0.5\times10^{-4}$K/s.

Given these rates of absorption and emission of heat above the tropopause, the time required for the temperature to change by one-degree Kelvin would be $1\text{K}/(\Delta T/ \Delta t)  \sim \Delta t \sim 10^{4}$ seconds. Thus, we say the \emph{radiative time constant} in this region, $\tau_{\text{rad}}$, is $10^{4}$ s (3 hours), for the 100-mbar level in the stratosphere assumed.

Therefore, in order for the temperature to be controlled radiatively in this layer, the time constant for turbulent mixing and convective overturn, $\tau_{\text{dyn}}$, the \emph{dynamic time constant}, must be longer than $\tau_{\text{rad}}$.

At the tropopause, a typical sample of a cloudtop material may travel a significant portion of the circumference of the planet $\sim10^{4}$km before encountering a downdraft, which requires a horizon mean wind velocity on order of $10^{4}\text{km}/10^{4}\text{s}\sim1$ km/s, which is comparable to the sound speed $c_s$ in a solar-composition gas in $500\sim2000$ Kelvin temperature range.

\begin{equation}
    c_s = \sqrt{\left( \frac{\partial P}{\partial \ln{\rho}} \right)_{\text{adiabatic}}} = \sqrt{\frac{\gamma R T}{\mu}} = \sqrt{\frac{C_p R T}{\mu (C_p - R)}} \sim 1.5-3 \text{~km/s}
\end{equation}

Therefore, we expect to encounter sub-sonic (Mach$<1$), trans-sonic (Mach$\sim1$), and then super-sonic (Mach$>1$), and then even hyper-sonic (Mach$\gg1$) flow regimes with a \emph{shock-wave front} present in certain part of the exoplanetary atmosphere, which are generated and powered by the strong host-stellar irradiation absorbed in the exoplanetary atmosphere. 

This scenario should be explored with detailed hydrodynamic calculation in future. Indeed this is a planetary-scale jet engine and then a rocket engine! We expect that as the irradiation power is increased, the atmosphere is trying to catch up with higher speed wind to carry away absorbed energy from sub-stellar host spot. However, at a critical point, the flow can no longer catch up and a \emph{shock front} emerges, and this \emph{shock front} is understood as a \emph{phase transition} of the flow structure which accelerates the exiting outflow from the ordinary thermal escape scenario into strong hydrodynamic jet with very high-speed shooting towards zenith. 

Our physical intuition is that for small exoplanets, the efficiency of this \emph{planetary-scale rocket engine} is on the order of one-percent ($1\%$) in terms of converting the absorbed host-stellar flux into the jet outflow. So this shock front is expected to occur right around \emph{tropopause}. 

On the other hand, for large exoplanets the efficiency of this \emph{planetary-scale rocket engine} is approaching the order of unity in terms of converting the absorbed host-stellar flux into super-sonic outflow accelerated by the \emph{shock wave front}. This shock front is expected to occur then near the level of \emph{photosphere}.}

\section{Mass Loss Rate Estimate under non-limiting scenario}
Turning back to the mass loss rate equation:

\begin{equation}
    \dot{m} = \bigg( 1.66\times10^{2}\text{~g} \cdot \text{cm}^{-2}\text{s}^{-1} \bigg) \cdot \exp{\left( -\frac{a_3/2 +b_3 \cdot (m/r)}{T_3} \right)} \cdot \frac{R_\oplus^2}{M_\oplus} \cdot \eta \cdot r^2
\end{equation}

plug in the number for $M_\oplus$ and $R_\oplus$, we have,

\begin{equation}
    \begin{split}
    \dot{m} &= \bigg( 1.66\times10^{2}\text{~g} \cdot \text{cm}^{-2}\text{s}^{-1} \bigg) \cdot \exp{\left( -\frac{a_3/2 +b_3 \cdot (m/r)}{T_3} \right)} \cdot \\& \frac{(6.371\times10^{8}\text{cm})^2}{6\times10^{27}\text{g}} \cdot \eta \cdot r^2
    \end{split}
\end{equation}

so that,

\begin{equation}
    \dot{m} = \bigg( 1.1 \times 10^{-8}\text{~s}^{-1} \bigg) \cdot \exp{\left( -\frac{a_3/2+b_3 \cdot (m/r)}{T_3} \right)} \cdot \eta \cdot r^2
\end{equation}


\begin{equation}
    \dot{m} = \left( \frac{1}{3.0\text{~year}} \right) \cdot \exp{\left( -\frac{a_3/2+b_3 \cdot (m/r)}{T_3} \right)} \cdot \eta \cdot r^2
\end{equation}


\begin{equation}
    \frac{\dot{m}}{m} \approx \left( \frac{1}{3.0\text{~year}} \right) \cdot \exp{\left( -\frac{a_3/2 +b_3 \cdot (m/r)}{T_3} \right)} \cdot \frac{\eta \cdot r^2}{m}
\end{equation}

now the characteristic lifetime $\tau$ of such a planet can be estimated as $m/\dot{m}$:

\begin{equation}
    \tau \sim \frac{m}{\dot{m}} \sim  3 \text{~yr} \cdot \frac{1}{\eta} \cdot \exp{\left( \frac{a_3/2 +b_3 \cdot (m/r)}{T_3} \right)} \cdot \frac{m}{r^2}
\end{equation}

\begin{equation}
    \tau \sim \frac{m}{\dot{m}} \sim  3 \text{~yr} \cdot \frac{1}{\eta} \cdot 10~\widehat{} \left[ \frac{a_3/2 +b_3 \cdot (m/r)}{2.303 \cdot T_3} \right] \cdot \frac{m}{r^2}
\end{equation}

\begin{equation}
    \tau \sim \frac{m}{\dot{m}} \sim  3 \text{~yr} \cdot \frac{1}{\eta} \cdot 10~\widehat{} \left[ \frac{27. + 7.6 \cdot (m/r)}{2.303 \cdot T_3} \right] \cdot \frac{m}{r^2}
\end{equation}

Therefore, the escape process and the lifetime of such a planet is highly sensitive to the planet temperature.


\section{On Transit Depth}
\label{sec:OnTransitDepth}

\begin{figure}[!ht]
    \centering
    \includegraphics[width=0.45\textwidth, angle=0]{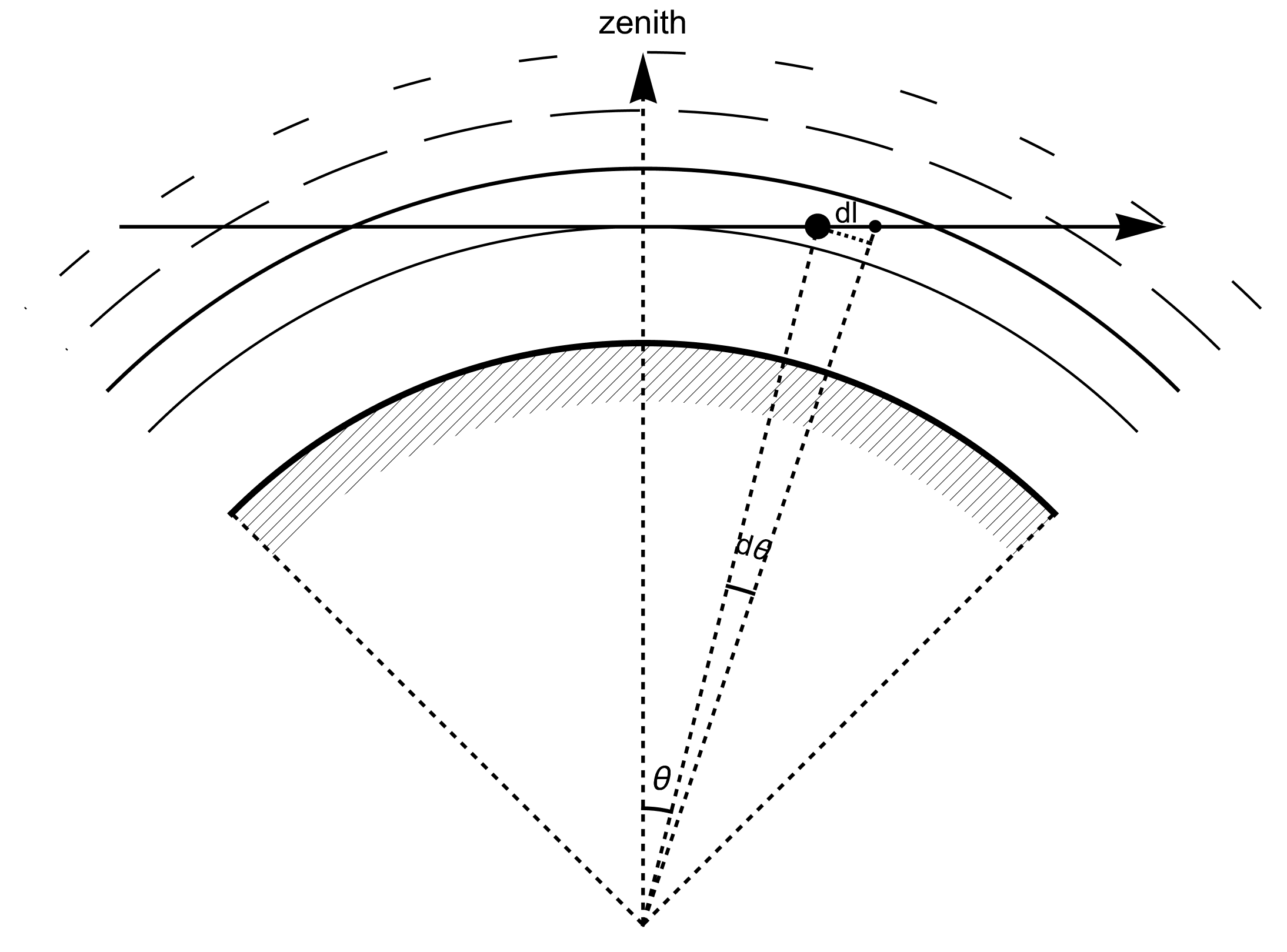}
    \caption{Schematic Drawing of the Grazing Transit Light Ray Path. Typical transit depth probes 1-10 mbar level, which is somewhat below the von Karman line so a uniform scale height $\mathcal{H}$ of the atmospheric constituents and mixing ratio may be assumed. Above the von Karman line the atmosphere becomes very tenuous so contributes little to the total opcacity encountered by this transit light ray beam. The critical transit light beam occurs at $\int n \cdot \sigma \cdot dl \sim 1$.}
    \label{fig:TransitDepth}
\end{figure}

{\color{black}
We assume that the grazing ray of light mostly probes the depth below the von Karman line where the atmospheric constituents are well-mixed with a common scale height $\mathcal{H}$, which is determined by the mean molecular weight of the atmospheric gas mixture $\bar{\mu}$, the local temperature $T$, and the local gravity $g$ as:

\begin{equation}
    \mathcal{H} = \frac{k_B \cdot T}{\bar{\mu} \cdot g} = \frac{k_B \cdot T}{\bar{\mu} \cdot \left( G M_p /r^2 \right)}
\end{equation}


\begin{equation}
    \frac{n}{n_0} = \exp{\left( -\frac{z}{\mathcal{H}} \right)} = \exp{ \left[ \frac{G \cdot M_p \cdot \bar{\mu}}{k_B \cdot T}  \cdot \left( \frac{1}{r}-\frac{1}{r_0} \right)  \right] }
\end{equation}

where

\begin{equation}
    r = r_0 \cdot \sec \theta = r_0 \cdot \frac{1}{\cos \theta}
\end{equation}

and the differential path length $dl$ is:

\begin{equation}
    dl = r \cdot d\theta \cdot \frac{1}{\cos \theta} = r_0 \cdot \frac{d\theta}{\cos^2 \theta}
\end{equation}

then, the critical ray path, which is reflected in the wavelength-dependent transit depth measurement is (please see Fig.~\ref{fig:TransitDepth}):

\begin{equation}
\begin{split}
1 &\sim  \int_{\text{along ray path}} n \cdot \sigma \cdot dl \\ 
  &=  \int \frac{d\theta}{\cos^2 \theta} \cdot r_0 \cdot \sigma \cdot n_0 \cdot \exp{ \left[ \frac{G \cdot M_p \cdot \bar{\mu}}{k_B \cdot T}  \cdot \left( \frac{1}{r}-\frac{1}{r_0} \right)  \right] } \\
  &=  \int \frac{d\theta}{\cos^2 \theta} \cdot r_0 \cdot \sigma \cdot n_0 \cdot \exp{ \left[ \frac{G \cdot M_p \cdot \bar{\mu}}{k_B \cdot T \cdot r_0}  \cdot \left( \cos \theta - 1 \right)  \right] } \\
  &= n_0 \cdot \sigma \cdot r_0 \cdot \int \frac{d\theta}{\cos^2 \theta} \cdot \exp{ \left[ \frac{G \cdot M_p \cdot \bar{\mu}}{k_B \cdot T \cdot r_0}  \cdot \left( \cos \theta - 1 \right)  \right] } \\
  &\approx n_0 \cdot \sigma \cdot r_0 \cdot 2 \cdot \int_{0}^{\pi/2} \frac{d\theta}{\cos^2 \theta} \cdot \exp{ \left[ \frac{G \cdot M_p \cdot \bar{\mu}}{k_B \cdot T \cdot r_0}  \cdot \left( \cos \theta - 1 \right)  \right] } \\
  &\approx n_0 \cdot \sigma \cdot r_0 \cdot 2 \cdot \int_{0}^{\pi/2} \frac{d\theta}{\cos^2 \theta} \cdot \exp{ \left[ \frac{G \cdot M_p \cdot \bar{\mu}}{k_B \cdot T \cdot r_0}  \cdot  \underbrace{ \left( \cos \theta - 1\right)}_{-2\cdot \sin^2 \frac{\theta}{2}}  \right] } \\
  &\approx n_0 \cdot \sigma \cdot r_0 \cdot 2 \cdot \int_{0}^{\pi/2} \frac{d\theta}{\cos^2 \theta} \cdot \exp{ \left[ \frac{G \cdot M_p \cdot \bar{\mu}}{k_B \cdot T \cdot r_0}  \cdot \left( -2\cdot \sin^2 \frac{\theta}{2} \right) \right] } \\
  &\approx n_0 \cdot \sigma \cdot r_0 \cdot 2 \cdot \int_{0}^{\pi/2} \frac{d\theta}{\cos^2 \theta} \cdot \exp{ \left[ \frac{G \cdot M_p \cdot \bar{\mu}}{k_B \cdot T \cdot r_0}  \cdot \left( -2\cdot \left(\frac{\theta}{2} \right)^2 \right)  \right] } \\
  &\approx n_0 \cdot \sigma \cdot r_0 \cdot 2 \cdot \int_{0}^{\pi/2} \frac{d\theta}{\cos^2 \theta} \cdot \exp{ \left[ \frac{r_0}{\mathcal{H}}  \cdot \left( -2\cdot \left(\frac{\theta}{2} \right)^2 \right)  \right] } \\
  &\approx n_0 \cdot \sigma \cdot r_0 \cdot 2 \cdot \int_{0}^{\pi/2} d\theta \cdot \exp{ \left[ \frac{r_0}{\mathcal{H}}  \cdot \left( -2\cdot \left(\frac{\theta}{2} \right)^2 \right)  \right] } \\
  &\approx n_0 \cdot \sigma \cdot r_0 \cdot 2 \cdot \int_{0}^{\infty} d\theta \cdot \exp{ \left[ \frac{r_0}{\mathcal{H}}  \cdot \left( -2\cdot \left(\frac{\theta}{2} \right)^2 \right)  \right] } \\
  &\approx n_0 \cdot \sigma \cdot r_0 \cdot 2 \cdot \int_{0}^{\infty} d\theta \cdot \exp{ \left[ -\frac{r_0}{2 \mathcal{H}} \cdot \theta^2  \right] } \\
  &\approx n_0 \cdot \sigma \cdot r_0 \cdot 2 \cdot \sqrt{\frac{2 \mathcal{H}}{r_0}} \cdot \int_{0}^{\infty} d\left( \sqrt{\frac{r_0}{2 \mathcal{H}}} \right) \theta  \cdot \exp{ \left[ -\frac{r_0}{2 \mathcal{H}} \cdot \theta^2  \right] } \\
  &\approx n_0 \cdot \sigma \cdot r_0 \cdot 2 \cdot \sqrt{\frac{2 \mathcal{H}}{r_0}} \cdot \frac{\sqrt{\pi}}{2} \\
  &\approx n_0 \cdot \sigma \cdot r_0 \cdot \sqrt{\frac{2\pi \mathcal{H}}{r_0}} \\
  &\approx n_0 \cdot \sigma \cdot \underbrace{\sqrt{2\pi r_0 \mathcal{H}}}_{\text{dimension of length, due to the geometry of transit grazing ray}}
\end{split}
\end{equation}

Now, we know that the light-matter interaction cross-section $\sigma$ (including both absorption and scattering) is frequency-dependent $\sigma_{\nu}$. On the other hand, $n_0$ depends both on the zenith height $z$ as well as mixing ratio $\xi$ of that particular species which contributes to the cross-section at that particular frequency $\nu$. So that,

\begin{equation}
    n_0 = \xi \cdot n_{\text{tot}} \cdot \exp{\left( -\frac{z}{\mathcal{H}} \right)}
\end{equation}

so, the frequency-dependent height $z_{\nu}$ (transit depth) in the atmosphere which the transit light ray probes is roughly:

\begin{equation}
    1 \sim \xi \cdot n_{\text{tot}} \cdot \exp{\left( -\frac{z_{\nu}}{\mathcal{H}} \right)} \cdot \sigma_{\nu} \cdot \sqrt{2\pi r_0 \mathcal{H}}
\end{equation}

so that, taking logarithm on both sides, we have the relation between transit depth and cross-section:

\begin{equation}
    0 \sim \underbrace{\ln{\left( \xi \cdot n_{\text{tot}} \cdot \sqrt{2\pi r_0 \mathcal{H}} \right)}}_{\sim\text{const}}   \underbrace{-\frac{z_{\nu}}{\mathcal{H}}  + \ln{ \sigma_{\nu} }}_{\text{correlation between $z_{\nu}$ and $\sigma_{\nu}$}} 
\end{equation}

Therefore, one can invert $\sigma_{\nu}$ and infer the source of the opacity/cross-section and identify the molecular species responsible for it, from the transit depth variation $z_v$ measured from the transit light curves.}



\clearpage





\end{document}